\RequirePackage{silence}
\ActivateWarningFilters[pdftoc,rerun,labels,sfloat,dfloat,vspace]
\documentclass[modern]{aastex631}

\usepackage{amsmath}
\usepackage[all]{hypcap}
\usepackage{color}

\graphicspath{{figures/}}
\hypersetup{
    pdfcreator={Naor Movshovitz},
}

\begin{document}


\newcommand{\V}[1]{\mathbf{#1}}
\newcommand{\grad}{\nabla}
\newcommand{\roofs}{\rho(s)}
\newcommand{\roofz}{\rho(z)}
\newcommand{\roofr}{\rho(r)}
\newcommand{\pofs}{p(s)}
\newcommand{\bro}{\boldsymbol{\rho}}
\newcommand{\Jstar}{\mathbf{J}^\star}
\newcommand{\code}[1]{\texttt{#1}}
\newcommand{\bx}{\V{x}}
\newcommand{\bigo}[1]{\mathcal{O}(#1)}
\newcommand{\about}[1]{\sim\!\!{#1}}
\newcommand{\unit}[1]{\;\mathrm{#1}}
\newcommand{\sub}[1]{_{\text{#1}}} 
\newcommand{\obs}{^{\text{obs}}}
\newcommand{\rig}{^{\text{rigid}}}
\newcommand{\wind}{^{\text{wind}}}
\newcommand{\smp}{^{\mathrm{s}}}
\newcommand{\req}{R\sub{eq}}
\newcommand{\rem}{R\sub{m}}
\newcommand{\arr}{\V{r}}
\newcommand{\arp}{\V{r'}}
\newcommand{\robg}{\rho\sub{bg}}
\newcommand{\rofg}{\rho\sub{fg}}
\newcommand{\mU}{M_{\mathrm{U}}}
\newcommand{\rU}{R_{\mathrm{U}}}
\newcommand{\mN}{M_{\mathrm{N}}}
\newcommand{\rN}{R_{\mathrm{N}}}
\newcommand{\mE}{M_{\oplus}}
\newcommand{\rE}{R_{\oplus}}
\newcommand{\sigloss}[1]{\sigma_{J_{#1}}/J_{#1}}
\newcommand{\logit}{\mathrm{logit}}
\newcommand{\romax}{\rho\sub{max}}
\newcommand{\note}[1]{\textbf{\textcolor{red}{#1}}}


\title{The Promise and Limitations of Precision Gravity:\\ Application to the
Interior Structure of Uranus and Neptune}
\author{Naor Movshovitz}
\author{Jonathan J. Fortney}
\affiliation{Department of Astronomy and Astrophysics, University of California,
Santa Cruz, California, 95064 USA}

\correspondingauthor{Naor Movshovitz}
\email{nmovshov@ucsc.edu}


\begin{abstract}
We study the constraining power of a high-precision measurement of the gravity
field for Uranus and Neptune, as could be delivered by a low periapse orbiter.
Our study is practical, assessing the possible deliverables and limitations of
such a mission with respect to the structure of the planets. Our study is also
academic, assessing in a general way the relative importance of the low order
gravity, high order gravity, rotation rate, and moment of inertia (MOI) in
constraining planetary structure. We attempt to explore all possible interior
density structures of a planet that are consistent with hypothetical gravity
data, via MCMC sampling of parameterized density profiles. When the gravity
field is poorly known, as it is today, uncertainties in the rotation rate on the
order of 10 minutes are unimportant, as they are interchangeable with
uncertainties in the gravity coefficients. By the same token, when the gravity
field is precisely determined the rotation rate must be known to comparable
precision. When gravity and rotation are well known the MOI becomes
well-constrained, limiting the usefulness of independent MOI determinations
unless they are extraordinarily precise. For Uranus and Neptune, density
profiles can be well-constrained. However, the non-uniqueness of the relative
roles of H/He, watery volatiles, and rock in the deep interior will still
persist with high-precision gravity data. Nevertheless, the locations and
magnitudes (in pressure-space) of any large-scale composition gradient regions
can likely be identified, offering a crucially better picture of the interiors
of Uranus or Neptune.
\end{abstract}


\section{Introduction}\label{sec:intro}
The gravity field of a giant planet is perhaps our best window into its interior
structure and composition. Obtaining gravity data is no simple task though.
Rough estimates can be deduced with ground-based observations of natural
satellites, but precise determinations must come from tracking radio signals
from probes sent to the outer solar system to orbit, or at least fly by these
worlds. Turning the hard-won gravity data into inferences about a planet's
interior is also not a straightforward task. There is no simple inversion; the
usual process involves creating models of the planet's interior and deducing
features of the planet we are interested in, for example the existence and size
of a heavy element core, based on how well the calculated gravity of model
planets with such features matches the observed gravity of the real planet.

The past decade has seen the \emph{Juno} Mission to Jupiter and the
\emph{Cassini} Grand Finale orbits at Saturn revolutionize our understanding of
these planets as the precision of the gravity fields of these giant planets have
improved by two orders of magnitude
\citep{Bolton2017,Iess2018,Iess2019,Durante2020}. The bulk compositions of these
planets are dominated by hydrogen and helium, and the typical questions involve
the mass and distribution of heavy elements within the planets, to better
understand structure, formation, and evolution. Since these data have
revolutionized our views of Jupiter and Saturn
\citep{Stevenson2020,Mankovich2021}, it is a natural question to wonder how
similar data may be able be alter our views of Uranus and Neptune. A variety of
orbiter mission concepts have been put forward with comprehensive science cases,
including improving our view of the interior structure of these planets
\citep{Hofstadter2019,Fletcher2020,Rymer2021}.

The advances in giant planet gravity field data have necessitated advances in
modeling efforts, first in terms of new methods of ultra-precise gravity field
calculations \citep{Hubbard2012,Hubbard2013a,Nettelmann2017a,Nettelmann2021a}.
A second effort has been in re-examining assumptions that go into models. What
may not be well appreciated are the host of assumptions that typically go into
models, making virtually every inference about a giant planet's structure
strongly model dependent.

{\subsection{The role of gravity in interior models}} Physical interior
models try to create a self-consistent description of the composition, density,
pressure, and temperature, at every point inside the planet. Gravity
{enters the problem} from the requirement of hydrostatic equilibrium,
connecting the pressure $p$ (really its gradient) and density $\rho$ at every
point. Hydrostatic equilibrium also allows us to use a one dimensional
structure, $\roofs$ and $\pofs$, where $s$ is the mean (volume equivalent)
radius of a surface of constant density and pressure\footnote{The existence of
surfaces of constant density and pressure is also guaranteed by the condition of
hydrostatic equilibrium.}. Once a self-consistent $\roofs$ is determined, it can
be integrated to yield a total planetary mass, radius, and external gravity
field which can be compared with observed values to judge the model's overall
likelihood. For the planets Uranus and Neptune, a recent reference in the
physical picture is \citet{Nettelmann2013b}.

The gravitational potential exterior to the planet, $V\sub{e}(\V{r})$, is
described by the expansion
\begin{equation}\label{eq:gravpot}
V\sub{e}(\V{r}) = -\frac{GM}{r}\left(1 -
\sum_{n=1}^{\infty}(a_0/r)^{2n}J_{2n}P_{2n}(\cos\theta)\right),
\end{equation}
the connection with the interior mass distribution is in the coefficients
\begin{equation}\label{eq:Jdef}
J_n = -\frac{1}{Ma_0^n}\int{}\rho(\V{r'})(r')^nP_n(\cos\theta')\,d\tau'.
\end{equation}
In this expansion $M$ is the planet's mass, $a_0$ is a normalizing constant with
dimensions of length, commonly a whole number of kilometers close to the
planet's equatorial radius, and $P_n$ are the Legendre polynomials. The
integrals are carried over the volume of the planet.

Equation~(\ref{eq:gravpot}) depends on the co-latitude angle $\theta$ but not on
azimuth, and is also strictly north-south symmetric (even $n$ only). Clearly,
this is an approximate, partial description of the full gravity field.
Furthermore, the integrals~(\ref{eq:Jdef}) require knowledge of the equilibrium
shape of the model. planet. Methods for deriving this shape (sec.~\ref{sec:tof})
rely on equilibrium between gravitational and centrifugal \emph{potential},
which greatly restricts the rotation state under consideration. In fact rigid
rotation at a constant rate $\omega$ is almost always assumed. Real planets, on
the other hand, exhibit more complex gravity. Non-zero odd gravity harmonics
($J_{3,5,7,9}$) have been measured for Jupiter \citep{Iess2018} and used to
infer the depth of the observed asymmetric surface flow \citep{Kaspi2018a}. And
in Saturn evidence for strong differential rotation was found not in the odd Js
but rather in the unexpected magnitude of even harmonics higher than $J_6$
\citep{Iess2019,Galanti2019}.

There is no doubt that such non-hydrostatic effects can be expected to exist in
Uranus and Neptune too. Nevertheless it is appropriate to focus on the simpler
gravity field with the implied uniform rotation as these are the most important
for determining the bulk structure of the interior. We discuss nonuniform
rotation again in sec.~\ref{sec:windtalk} but in the rest of the text
``gravity'' implies the zonal, north-south symmetric gravity of
eq.~(\ref{eq:gravpot}).

\subsection{The role of the equation of state}
Pressure and density are also related via a thermodynamic equation of state
(EOS). A self-consistent solution is found, by iteration, such that the value of
pressure satisfying hydrostatic equilibrium, which depends on gravity, matches
everywhere the value given by the EOS. The EOS requires knowledge, at every
point, of temperature and composition. Temperature can be calculated by
equations of heat transfer, perhaps with input from \emph{cooling models} that
follow a long-term evolution of the planet. In detailed \emph{structure models}
a prescribed, static temperature structure (usually adiabatic) or, equivalently,
entropy structure is used. Composition cannot be calculated by an equation; it
must be stipulated.

And {that fact regarding composition is a limitation}: the model must
assume the very thing it is supposed to infer. To be sure, there are some very
good assumptions that one can make. For example, if the target planet is a gas
giant we may assume that, in a large fraction of the volume of the interior, the
dominant species is a mix of hydrogen and helium, although the relative
proportion of hydrogen to helium should be allowed to change with depth, given
helium phase separation \citep{Stevenson1977,Mankovich2020a}.

In the case of the ice giants, on the other hand, the dominant species is not so
easy to guess. Relatively recent models within the physical picture
\citep{Nettelmann2013b} suggest important roles for H/He, water and other
volatiles, as well as rock/iron. Some models even suggest more rock that
volatiles \citep{Teanby2020}. Some of the most important questions about planet
formation (that are perhaps the main motivation for the model in the first
place) depend most strongly on the inferred content, composition, and
distribution of heavier elements, with this ``metals'' mass fraction referred to
as $Z(s)$. Recent reviews can be found in \citet{Helled2020a} and
\citet{Helled2020b}. Whereas for Jupiter and Saturn the outcome of the inferred
$Z(s)$ is not overly sensitive to exactly \emph{which} heavy element (i.e.,
volatiles vs.~rocks) one uses in the model, the same is \emph{not} true for
Uranus and Neptune.

However, such a hypothetical model is already much too complex. To resolve the
interior to a meaningful degree requires discretizing the continuous variables
on a fine grid in $s$, with at least hundreds and preferably thousands of grid
points, resulting in thousands of model parameters and an impracticable task.
Some very strong simplifying assumptions are needed. The most important one is
the assumption of some kind of layering.

\begin{figure}[tb!]
\centering
\includegraphics[width=1.0\textwidth]{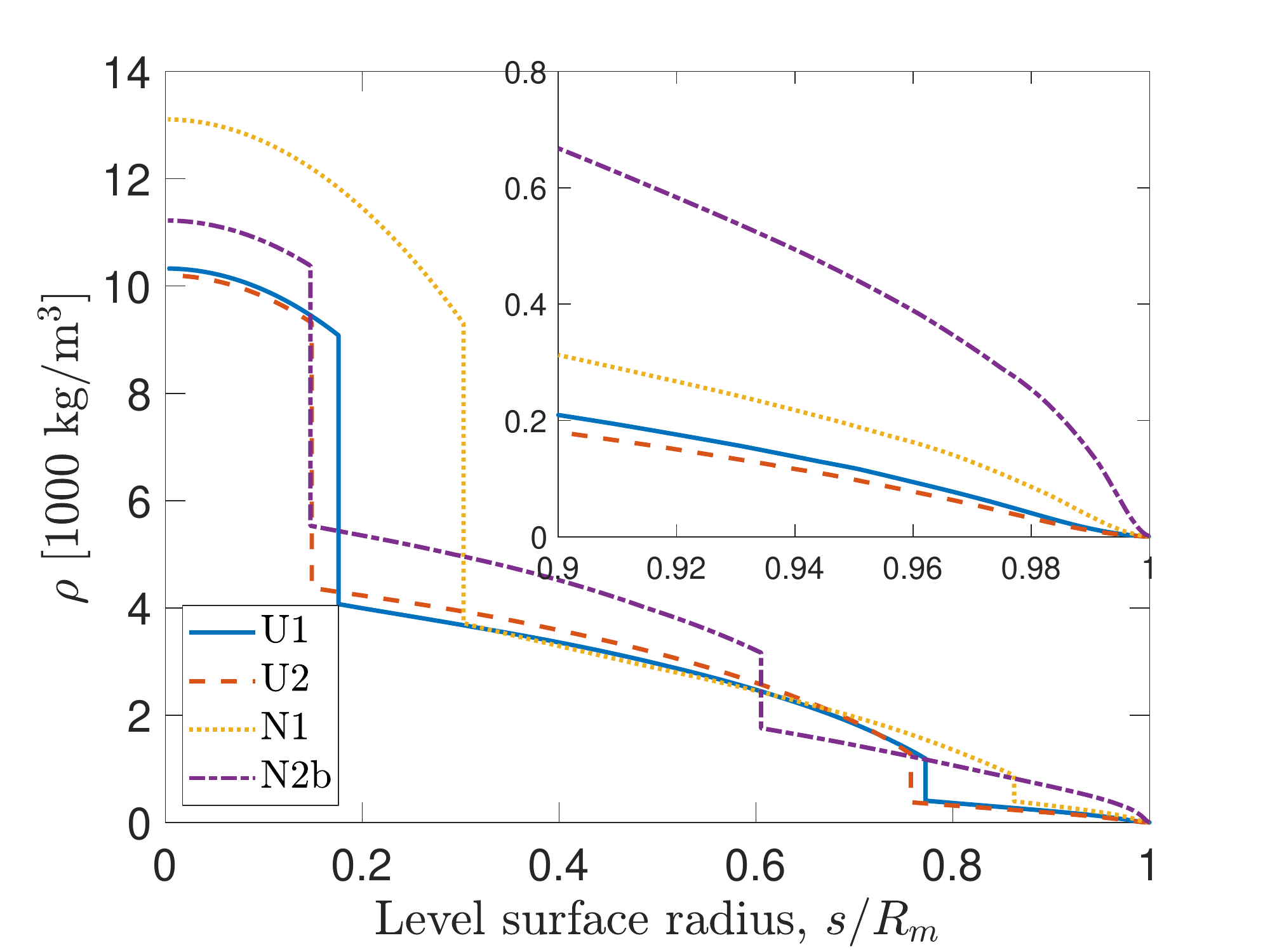}
\caption{Density profiles $\rho(s)$, of the three-layer models of 
\citet{Nettelmann2013b}, re-plotted from their data. $\rem$ is the mean (volume
equivalent) radius of a surface of constant density, which by hydrostatic
equilibrium must also be a surface of constant pressure and potential; $s_0$ is
the mean radius of the 1-bar surface.}
\label{fig:N13_profs}
\end{figure}

The most common class of models, for both the gas giants and ice giants, has for
a long time been the three-layer model. The planet is assumed to consist of
radial regions, each of homogeneous composition. This assumption is made
typically for computational expediency and potentially out of physical
reasoning, the latter being that thermodynamics may support, under some
conditions, the existence of fully convective regions, with boundaries between
them that resist mixing \citep{Bailey2021}. Of course, that thermodynamics may
support such configurations is no proof that these are the only possible ones.
Modelers have been gradually increasing the sophistication of
layered-composition models, {for example by adding layers of
composition gradients}, and will no doubt continue to do so, while still
retaining the basic paradigm. Figure~\ref{fig:N13_profs} shows what density
profiles deriving from such models of Uranus and Neptune \citep{Nettelmann2013b}
may look like; the three-layer structure is clearly visible.

\subsection{Composition agnostic models}

An alternative approach that has been used is to create so-called
\emph{empirical} models. Rather than parameterize the planet's composition and
then solve for the pressure and density structure, these models parameterize the
structure directly and make inferences about the composition from the resulting
models \citep{Helled2009,Helled2011b,Movshovitz2020}. For example, some models
may assign a synthetic (and simple) pressure-density relation to one or more
regions of the planet. A popular choice is a combination of one or more
polytropic regions, where $p(s)\propto\rho(s)^{1+\frac{1}{n}}$ for polytropic
index $n$, and perhaps a region of constant density approximating a core
\citep{Neuenschwander2021}. A synthetic pressure-density relation can be used
just like a physical EOS to derive self-consistent equilibrium shape and density
profiles. A three-layer structure like those shown in fig.~\ref{fig:N13_profs}
can be approximated by using three different polytropes in three radial regions.

The density profile can also be parameterized directly, and this is our
preferred approach here. A parametric mathematical representation of a curve is
chosen, mapping a vector of parameter values $\V{x}\in{\mathbb{R}^n}$ (hopefully
$n$ is not too large) to a density profile $\roofs$ and thereby, through
hydrostatic equilibrium, to a self-consistent interior structure. The advantage
of being entirely divorced from an assumed equation of state, real or synthetic,
is that density profiles representing both layered composition as well as
continuous composition can be represented with the same model. The disadvantage
is that without an EOS, real or synthetic, to guide us, it is harder to know how
to interpret the resulting density profile. Our works builds on pioneering work
in this area, of ``random'' interior models of Uranus and Neptune
\citep{Marley1995a,Podolak2000}, that were motivated by an exploration to move
past the 3-layer model and explore {the widest possible range} of density
distributions allowed by the gravity field.

{\subsection{Nonuniqueness of models}}
Regardless of how a model planet is constructed, with a physical EOS, a
synthetic EOS, or a synthetic $\roofs$, there remains the problem of degeneracy
of solutions. The problem is, simply, that multiple models can be constructed
that all match the measured gravity field to within a specified uncertainty. In
fact there are infinitely many such models, covering an unknown region of
parameter space. When modelers claim to constrain some desired property of the
planet, for example if they say that the heavy element content of the planet is
found to be between some upper and lower bounds, what they mean is that, given
their chosen model design, no solutions were found with heavy element content
outside those bounds, that match the gravity measurement to withing some
tolerance. It is typically not known how much of this constraint is due to the
measured or observed properties of the planet and how much is due to the
assumptions built into the model, or even to incomplete exploration of the model
parameter space.

That is the main {deficiency we aim} to address in this work. Focusing on
Uranus and Neptune as the target planets, we choose a parameterization designed
for maximum flexibility and a sampling method designed for maximum coverage, and
generate representative random samples of $\roofs$ from several different
distributions, each constrained by a different combination of observable
planetary properties. Concretely, we constrain the $\roofs$ distributions by
{applying progressively stronger information as follows}: (1) The
planet's mass and radius only; (2) Mass, radius, and rotation period; (3) Also
adding the planet's $J_2$ and $J_4$ gravity coefficients, with uncertainty
varying from that of currently available data to precision limited only by
computational considerations; (4) With nominal values of gravity coefficients of
increasing order (up to $J_{12}$), hypothetically, as if known to great
precision.

The difference between distributions of density profiles obtained with the
different constraints illustrates the ``constraining power'' of the different
observables.

Our focus on Neptune and Uranus here is motivated by a number of factors.
Compared with the gas giants, Jupiter and Saturn, the vital observables of the
ice giants, including and especially their gravity fields and rotation periods,
are much less well known. A future mission to these worlds with one of the goals
being to obtain more precise measurements of their gravity fields would be
extremely valuable, this is not in doubt. What is still unknown, however, is
what would be the expected relationship between the precision of such
measurements and our ability to translate them into better constraints of the
interior structure of Uranus and Neptune. Our work here provides motivation for
such missions and also a framework to understand their limitations, such that we
can better assess {whether and} how a precision gravity field could
revolutionize our view of these planets.

We explain in detail the methods of this experiment in sec.~\ref{sec:method}.
Theoretical results are illustrated in sec.~\ref{sec:results}. Results specific
to Uranus and Neptune, demonstrating the difference between models constrained
with currently available data for these planets and data improved by a plausible
mission scenario are shown in sec.~\ref{sec:uncons}. Our conclusions are
summarized in sec.~\ref{sec:discussion}.

\section{Experimental method}\label{sec:method}

\subsection{Overview}\label{sec:overview}
In order to evaluate the potential of precise measurements of the gravity fields
to improve our knowledge of the interiors of the ice giants we generate samples
of interior density profiles that are constrained by successively higher-order
and/or more precisely known gravity coefficients. By comparing the range of
density values, and variety of density profiles attained by each sample we get
an idea of the ``constraining power'' of gravity as a measured observable.

We start with a sample of density profiles guided by knowledge of the mass and
radius of the planet only, to use as a baseline for comparison. The oblate shape
of the rotating planet is calculated {because it affects the mass,} but
the associated gravity field is not expected to match observation. Even so, the
space of allowed $\roofs$ curves is already constrained by basic physics.
Trivially, $\roofs$ must be monotonic strictly decreasing with $s$ and its
integral from the center to the surface should match the planet's mass, within
{observational} uncertainty. Additionally, the relationship between
pressure and density in hydrostatic equilibrium implies
$\lim_{s\to{0+}}d\rho/ds=0$. By themselves these conditions only restrict the
shape of the $\roofs$ curve, not its scale, but we can constrain the scale by
putting reasonable limits on the density at both the surface and the center of
the planet.

{ On the surface, which in our models we take to mean the 1-bar
surface, we get a rough estimate of density by applying the ideal gas law to a
protosolar mix of hydrogen and helium (mean molecular weight $2.319\unit{amu}$)
at the observed temperature, $T\sub{1bar}$. This leads to a nominal value of
$\rho\sub{1bar,U}=0.367\unit{kg/m^3}$ for Uranus, and for Neptune
$\rho\sub{1bar,N}=0.387\unit{kg/m^3}$. The 1-bar temperature itself is not a
direct measurement. It is inferred from analysis of radio occultation data from
the \emph{Voyager 2} mission \citep{Lindal1992} and includes uncertainty of
several percent. And of course the atmospheric composition is unknown. Combining
the uncertainties from temperature and composition (mean molecular weight), we
should allow some ``play'' in the 1-bar surface density, treating it as a
sampled variable with an appropriate prior. However in preliminary samples we
found that the resulting $\Delta\rho\sub{1bar}$ is too small to have any effect
on the sample, and so we keep $\rho\sub{1bar}$ fixed in all models to speed up
the sampling process.}

The density at the center of the planet is, of course, not directly available.
We can estimate a reasonable upper bound value by considering the order of
magnitude of the central pressure. We set the upper limit for both planets at
$20,000\unit{kg/m^3}$, comfortably above the density of pure rock at
${50}\unit{Mbar}$.

The rotation period of the planets is also not precisely known.
\citep{Jacobson2009,Jacobson2014,Helled2010,Podolak2012,Nettelmann2013b}. For a
given density profile, changing the rotation period will change the equilibrium
shape and, {as the equatorial radius is held fixed, also} the mass and
gravity. We therefore include the rotation period as one of the sampled
parameters, with an estimated prior, allowing the MCMC procedure to sample from
the space of density profiles and rotation period simultaneously. The effect of
rotation period uncertainty is discussed further in sec.~\ref{sec:results}.

{ To summarize: the baseline sample, for each planet, is drawn from the
space of one dimensional profiles restricted to monotonically decreasing
$\roofs$ with vanishing $d\rho/ds$ at the center and satisfying
$\rho(s_0)=\rho\sub{1bar}$ and $\rho(s)\le\rho\sub{max}$ everywhere.}
The prior includes information about the imperfectly known rotation period, and
the likelihood function driving the sampling procedure compares the integral of
the density profile with the planet's reference mass $M$. More detail about
the exact sampling procedure is given in the appendix. The values used for these
baseline models are summarized in Table~\ref{tab:observables}.


\begin{deluxetable}{LCC}\label{tab:observables}

\tablecaption{Reference planetary values}

\tablehead{\colhead{} & \colhead{Uranus} & \colhead{Neptune}}

\startdata
G\;(10^{-11}\unit{m^3\;kg^{-1}\;s^{-2}}) & 6.67430(15) & 6.67430(15) \\
M\;(10^{24}\unit{kg}) & 86.8127(40) & 102.4126(48) \\
a_0\;(\mathrm{km}) & 25559 & 24764 \\
T\sub{1bar}\;(\mathrm{K}) & 76(2) & 72(2) \\
P\;(\mathrm{sec}) & 62064(600) & 57996(600) \\
J_2\;(\times{10}^6) & 3510.7(7) & 3536.5(47) \\
J_4\;(\times{10}^6) & -34.2(13) & -36.0(31) \\
\enddata

\tablecomments{Reference values from \citet[][2018 CODATA]{CODATA2018} and
\url{http://ssd.jpl.nasa.gov}, analysis of Voyager data from \citet{Lindal1992},
rotation period from \citet{Podolak2012}, and $J_n$ values from
\citet{Jacobson2009,Jacobson2014} renormalized to fixed value of $a_0$.}

\end{deluxetable}

\vspace*{-1.0\baselineskip}
We then proceed to constrain the interior density profiles further by using
additional information about the planet's gravity field. Currently, approximate
determination of the gravity fields of Neptune and Uranus comes form a
combination of ground based astrometry and radio data from both \emph{Voyager}
missions \citep[][and references therein]{Jacobson2009,Jacobson2014}. Gravity
harmonics $J_2$ and $J_4$ are available, with significant uncertainty, for both
planets. We use the nominal values from Table~\ref{tab:observables} to center
the multidimensional Gaussian we use in the sampling likelihood function,

\begin{equation}\label{eq:Jloss}
D_J^2 = \sum_{\text{even }n}^{n\sub{max}}\left(\frac{J_{n}-J\obs_{n}}
{\sigma_{J_{n}}}\right)^2.
\end{equation}

But for the $\sigma_{J_n}$ uncertainties we are less interested in the precision
of the \emph{Voyager} missions data; our question is the theoretical
constraining power of increasingly precise gravity. We assign, in different
samples, $\sigma_{J_n}/J_n=10^{-\mu_n}$ with $\mu_n$ ranging from 2 to 6.
Concretely, we obtained samples where $n\sub{max}$ was 2, 4, 6, and 12. For
$n\sub{max}>4$ we do not have measured nominal values for the Gaussian centers;
we use the mean values from the sample with $n\sub{max}=4$.

Comparing samples obtained with $n\sub{max}=2$ but with varying values of
uncertainty, $\sigma_{J_2}$, will show the expected benefit of more precise
measurements of the low-order coefficients. Comparing samples obtained with
increasing $n\sub{max}$ and fixed $\sigma_{J_n}$ will show the expected benefit
of measuring higher-order coefficients, even crudely. Comparing samples obtained
with the same $n\sub{max}$ and $\sigma_{J_n}$ but different priors set on
rotation period will show the expected benefit of more precise determination of
rotation. In a real mission, of course, these are not distinct, independent
measurements. Nevertheless, knowing which of the above options provides the most
benefit, in terms of constraining models of interior structure, may help
optimize the mission design.

Before we look at the resulting sample distributions we need to define the
density profile parameterization used in this work.

\subsection{Parameterization of the density profiles}\label{sec:params}
Consider the density profiles of the three-layer Uranus and Neptune models of
\citet{Nettelmann2013b}, shown in Figure~\ref{fig:N13_profs}. These profiles are
derived from models that solve the planetary structure equations for an assumed
{layered} composition, using physical equations of state. Our goal is to
find a suitable direct parameterization of density as a function of radius,
$\rho(s)$, that is capable of capturing the profiles of Fig.~\ref{fig:N13_profs}
as a subset and is otherwise as flexible as possible. It is also important to
keep the number of parameters small and, even more important, to minimize
parameter correlations. The more parameters we use and the stronger the
correlations between them, the longer it will take any sampling algorithm to
adequately cover the sample space.

With these requirements in mind, what are the important features of the curves
in Fig.~\ref{fig:N13_profs} that we need to allow for? Each $\rho(s)$ curve is
smooth and monotonic, except for two sharp discontinuities. In the physical
models these discontinuities are ``built-in''; the models assume layers of
homogeneous composition with sharp boundaries between them. Three-layer models
thus have two density discontinuities, the implicit assumption being that
regions of composition gradients of length scale below the model's resolution
separate the layers. The smoothness of the $\rho(s)$ curve between these density
``jumps'' suggests that the entire profile can be well-represented by a
piecewise-polynomial function. Such a parameterization was used successfully in
\citet{Movshovitz2020} to generate density profiles for Saturn, but here we
utilize an alternative parameterization that is superior to the
piecewise-polynomial in two significant ways, both having to do with
representing the density jumps.

There are two important limitations to representing $\roofs$ with
piecewise-polynomials. The first and more obvious is that this parameterization
only allows for sharp density discontinuities, not gradual ones. There are two
mathematical discontinuities (of the first kind) in every density profile, with
parameters controlling their location, say fractions $z_1$ and $z_2$ of the
planet's radius, and jump magnitude, say $\delta_1$ and $\delta_2$ in density
units\footnote{For reasons of efficiency one may choose to use some one-to-one
transformation of these parameters \citep[e.g.][Appendix B]{Movshovitz2020}, but
their physical meaning remains.}. The jumps can merge ($z_1\to{}z_2$) or one or
both may vanish ($\delta_i\to{0}$), but they cannot approximate a more gradual
transition, a gradient region detectable on the scale of the model. These sharp
density jumps are features of layered composition models and we want our
parameterization to be able to reproduce them, but we would like it to have the
flexibility to capture gradient regions as well, something that
composition-based models have difficulty with.

The second limitation of the piecewise-polynomial parameterization is a
technical one, that becomes apparent when we try to use MCMC algorithms to
sample from the parameters' joint posterior. It turns out that the parameters
are very strongly correlated, leading to impractically long convergence time.
One way to mitigate this problem is to fix values of the parameters $z_1$ and
$z_2$, obtain their marginal distributions by sampling the other parameter
values under their conditional probability, and repeat the process for a range
of reasonable values for $z_i$. A sample from the full posterior can be created
by drawing from the marginal probabilities, in proportion to their relative
likelihoods. While this approach provides a working method, it is more
cumbersome and time consuming. Worse, the additional step of combining the
marginals into a single posterior involves the difficult task of calculating, at
least approximately, the \emph{posterior odds} ratio (also called the Bayes
factor or the evidence integral or, simply, the evidence). While this is a
common and well-studied task, it still has no generally agreed upon best method
\citep{Nelson2018}.

Our alternative parameterization represents $\roofs$ with a single continuous
and continuously differentiable function:
\begin{equation}\label{eq:ppwd}
\roofz = \rho(s/\rem) = \sum_{n=2}^{8}a_n(z^n - 1) + \rho_0 +
\sum_{n=1}^{2}\frac{\sigma_n}{\pi}\Bigl(\frac{\pi}{2} +
\arctan\bigl(-\nu_n(z - z_n)\bigr)\Bigr).
\end{equation}
The first line is a degree-8 polynomial in $z=s/\rem$. It is constrained to have
a vanishing derivative at $z=0$ and to pass through the point $(1,\rho_0)$.
Since we reference our $J$ values to the 1-bar surface we take
$\rho_0=\rho\sub{1bar}$ to be the 1-bar density.

\begin{figure}[tb!]
\centering
\includegraphics[width=1.0\textwidth]{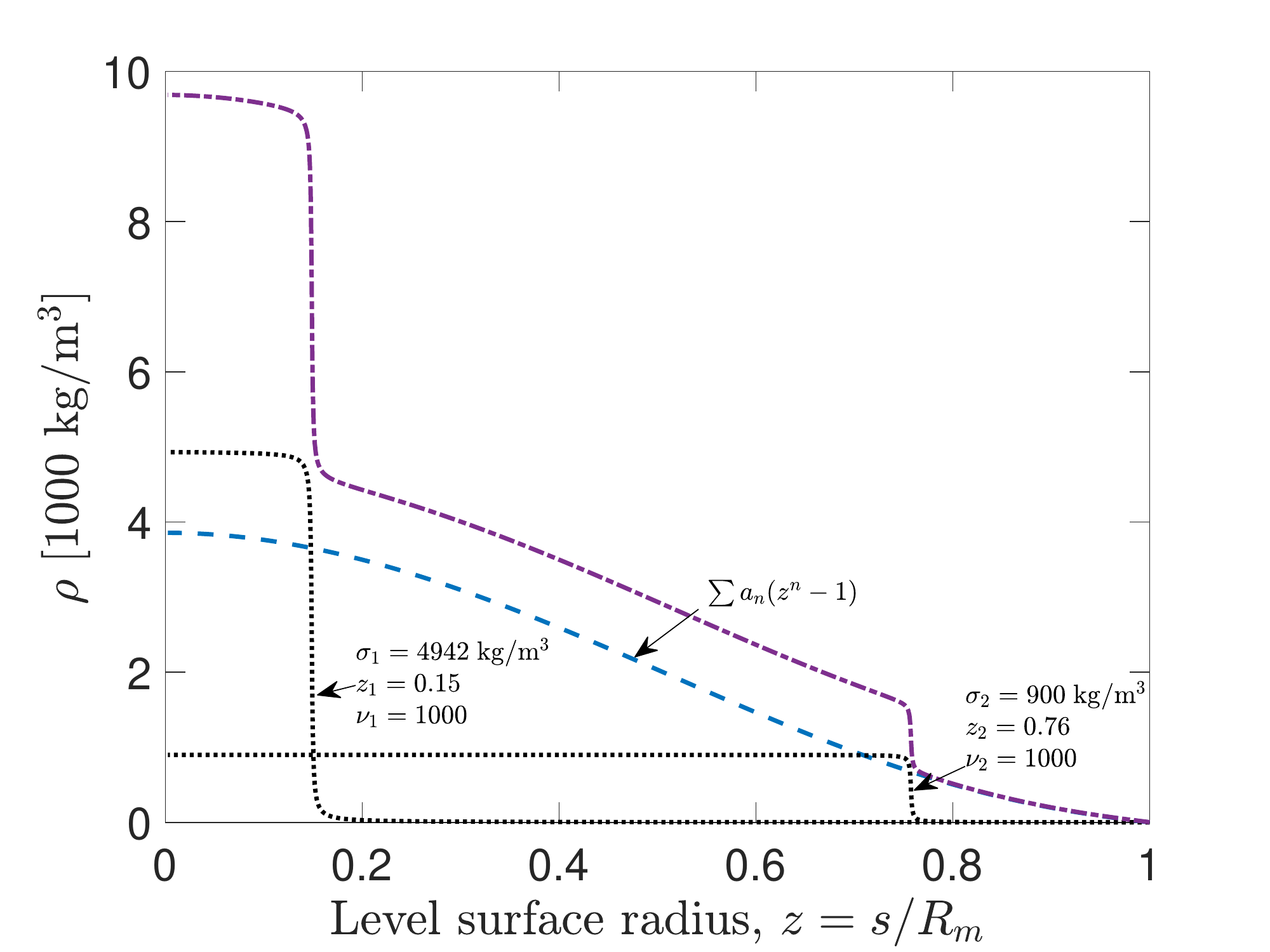}
\caption{Illustration of the density parameterization used in this work.
The dashed line is a degree-8 polynomial, the first part of eq.~\eqref{eq:ppwd},
constructed to match the surface density $\rho_0$ (a small but non-zero value)
and have vanishing derivative near the center. The two dotted lines are the
inner ($z_1=0.15$) and outer ($z_2=0.76$) Sigmoid functions of the second part
of eq.~\eqref{eq:ppwd}. In this example both are approximating a sharp
discontinuity ($\nu_1=\nu_2=1000$), meaning the increase in density
($\sigma_1,\sigma_2$) happens over a short span. The dot-dashed line is the sum
of the other three lines resulting in a density profile that matches the mass,
radius, and low-order gravity of Uranus.}
\label{fig:ppwd_sketch}
\end{figure}

The second line in eq.~\eqref{eq:ppwd} is a Sigmoid-function parameterization of
possible density jumps, overlain on top of the polynomial. In this version of
the parameterization we allow up to two such jumps. The parameter $z_1$ defines
the location (in normalized radius) of the center of the inner of these. A
density increase of $\sigma_1$ (in density units) is applied asymptotically and
symmetrically around this point, the width being controlled by the
non-dimensional sharpness parameter $\nu_1$. The jump can be made arbitrarily
sharp, to resemble a discontinuity like the ones in Fig.~\ref{fig:N13_profs}, by
increasing the value of $\nu_1$. Conversely, small values of $\nu_1$ result in
smooth, gradual density increase, indistinguishable from the background
polynomial. The location, scale, and sharpness of the outer density jump are
similarly determined by $z_2$, $\sigma_2$, and $\nu_2$, respectively. An example
is illustrated in Figure~\ref{fig:ppwd_sketch}, using parameter values tailored
to approximate the shape of the Uranus model of \citet{Nettelmann2013b}.

{
\subsection{Solving for equilibrium shape and gravity}\label{sec:tof}
There are several methods of calculating $J_n$ for a model planet. They vary in
theoretical precision and, more importantly, the precision they achieve \emph{in
practice} when working with manageable resolution
\citep{Hubbard2014,Wisdom2016,Nettelmann2017a,Debras2017,Nettelmann2021a}. In
this work we use the Theory of Figures (ToF) algorithm, taken to 4th-order
(ToF4) where possible and to 7th-order (ToF7) where necessary\footnote{Simply
because ToF7 is slower to run than ToF4.}. We estimate the practical precision
of this method, when calculating the gravity of a given density profile (as
opposed to a given density--pressure relation) with a given rotation period to
have fractional error,
$\sigma_{J_n}/J_n\approx{}10^{-6},10^{-5},10^{-4},10^{-4},10^{-2},10^0$, for
$n=2, 4, 6, 8, 10, 12$, respectively. This estimate is derived by comparison
with the extremely precise benchmark $n=1$ polytrope gravity solutions of
\citet{Wisdom2016}; they are the residual differences that remain when the ToF7
algorithm is applied to very high resolution models. We then choose the
resolution of our models (the number, $N$, of discretized density levels) such
that a doubling of $N$ results in a relative difference to the resulting $J_2$
less than a part per million. We find that $N=4098$ is always sufficient, and
for consistency we use this resolution in all models in every sample, even when
high precision is not required. }

\subsection{Sampling}
As given in eq.~\eqref{eq:ppwd} the number of parameters required to completely
define the density profile is 13 (since we take $\rho_0$ to be constant). A
degree-8 polynomial with two boundary conditions takes 7 parameters, and two
potential density jumps take three parameters each (location, scale, and
sharpness). Of course this polynomial-plus-Sigmoid parameterization can be made
even more flexible, by increasing the polynomial degree and/or adding
more Sigmoid terms. But the advantages of added flexibility must be carefully
weighted against the cost of increasing the dimensionality of the sample space.

As it is, sampling from the 13-dimensional parameter space is a computationally
expensive operation, although much less so than with the piecewise-polynomial
parameterization. The advantage comes not from the number of parameters, which
is similar in both, but from the weaker parameter correlations. The
polynomial-plus-Sigmoid parameters are less strongly correlated, meaning a small
change in one parameter while keeping the other parameters fixed results in a
smaller overall change in the density profile. This significantly improves the
behavior of the sampling algorithm. Even so, ensuring a large enough draw of
independent samples with adequate coverage of the parameter space is not
straightforward. We utilize the ensemble sampler algorithm of
\citet{Goodman2010} implemented in the \code{emcee} package
\citep{Foreman-Mackey2013} but we find it necessary to employ a ``tempering''
procedure, whereby sampling initially follows a modified loss function with
arbitrarily lower sensitivity to parameter values to {facilitate
the algorithm moving through parameter space},
followed by sampling with the full loss function. Repeated application of this
procedure results in a set of independent draws from what is hopefully the
static but unknown posterior.

The full details of the sampling procedure are given in
appendix~(\ref{sec:chains}).

\subsection{Some caveats\label{sec:windtalk}}
The methodology described above was designed to minimize the impact of implicit
assumptions and model limitations. Nevertheless, some necessary assumptions
remain. Perhaps the strongest of these is the assumption, made throughout, of
rigid rotation, in which every part of the planet is assumed to be moving at a
single angular rotation rate $\omega$. Many planets, including Uranus and
Neptune, exhibit signs of non-rigid rotation but this does not always invalidate
rigid-rotation models. Surface phenomena that involve only a small
fraction of a planet's mass would have negligible effect on the gravity field.
But deeper rooted {latitude- and depth-dependent motion} will
sufficiently change the equilibrium state at the time of measurement so that the
observed gravity field's coefficients are a mix of the {rigid}
equilibrium gravity and non-rigid \emph{wind} perturbations:
\begin{equation}\label{eq:J_correction}
J_n\obs{} = J_n\rig{} + J_n\wind{}.
\end{equation}
We say perturbation, but for high order $J_{n>6}$ the {non-rigid}
component {(also referred to as the dynamic component)} may be
significant or even dominant. { Models that assume rigid rotation
to calculate $J_n$ from a given density profile} should be made to match
$J_n\rig{}=J_n\obs-J_n\wind{}$ instead of $J_n\obs$. Thus getting a solid handle
on the non-rigid rotation's contribution to the gravity field, at least its
expected magnitude if not a nominal correction, should be considered a
prerequisite to using $J_n\obs$ to constrain interior models.

On both Uranus and Neptune, fast, east-to-west atmospheric streams are observed
at cloud level; the crucial question being the depth of these winds. Analysis of
the difference between $J_4\obs{}$ and the $J_4\rig{}$ of relatively
simple interior models suggests that these are relatively shallow jets,
involving a small fraction of the planets' mass \citep{Kaspi2013}, but the
possibility remains that a better characterization of the gravity field will
require addressing the correction due to dynamics. Work on this problem
continues \citep{Kaspi2017,Kaspi2018a,Galanti2017,Iess2018} and it is clear that
without a good understanding of dynamic effects the usefulness of high-order
gravity measurements would be much reduced.

A much more minor limitation comes from the use of composition-agnostic
parameterization, which can theoretically generate $\roofs{}$ profiles that
would be ``unphysical'' for one reasons or another, including characteristics
that are difficult to predict or detect. It is not impossible, for example, that
some density profile exist in the sample that if used with realistic equations
of state to back out a temperature profile, would imply unrealistic temperature
gradients somewhere in the interior. Any additional restriction of the
parameterization that would guard against such features would be useful in
further constraining the allowable models, but would necessarily rely on
imperfectly known equations of state and/or additional model assumptions.

{
Conversely, the specific functional form~(\ref{eq:ppwd}), in addition to being
possibly too permissive physically, may not be flexible enough mathematically.
Certainly, it is \emph{not} a complete basis for the space of all functions in
$(0,1)$ nor even for the more restricted space of functions representing a valid
density profile. Just how much of that space may be ``missing'' is difficult to
analyze. In section~\ref{sec:baseline} we offer some reassurance, by examining a
baseline sample, that the chosen functional form can indeed capture a very wide
variety of density profiles.
}

Finally, any work that makes use of MCMC sampling must acknowledge that
convergence to the correct static probability distribution is never assured. We
describe in the appendix our sampling procedure and our reasons for accepting
the resulting samples as correct.


\section{Constraining power of gravity}\label{sec:results}
\subsection{A baseline sample\label{sec:baseline}}
\begin{figure*}[tb!]
\gridline{
\fig{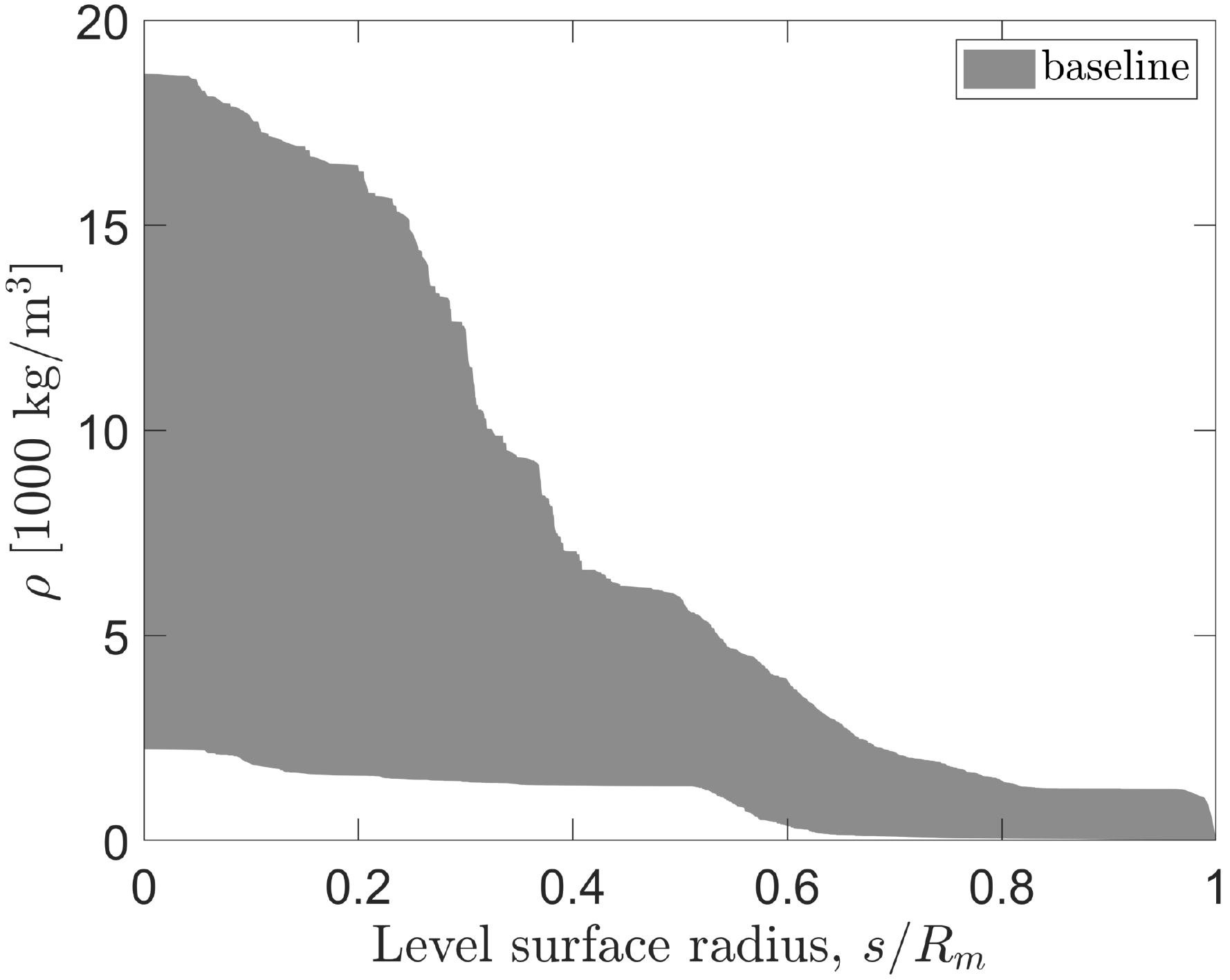}{0.33\textwidth}{}
\fig{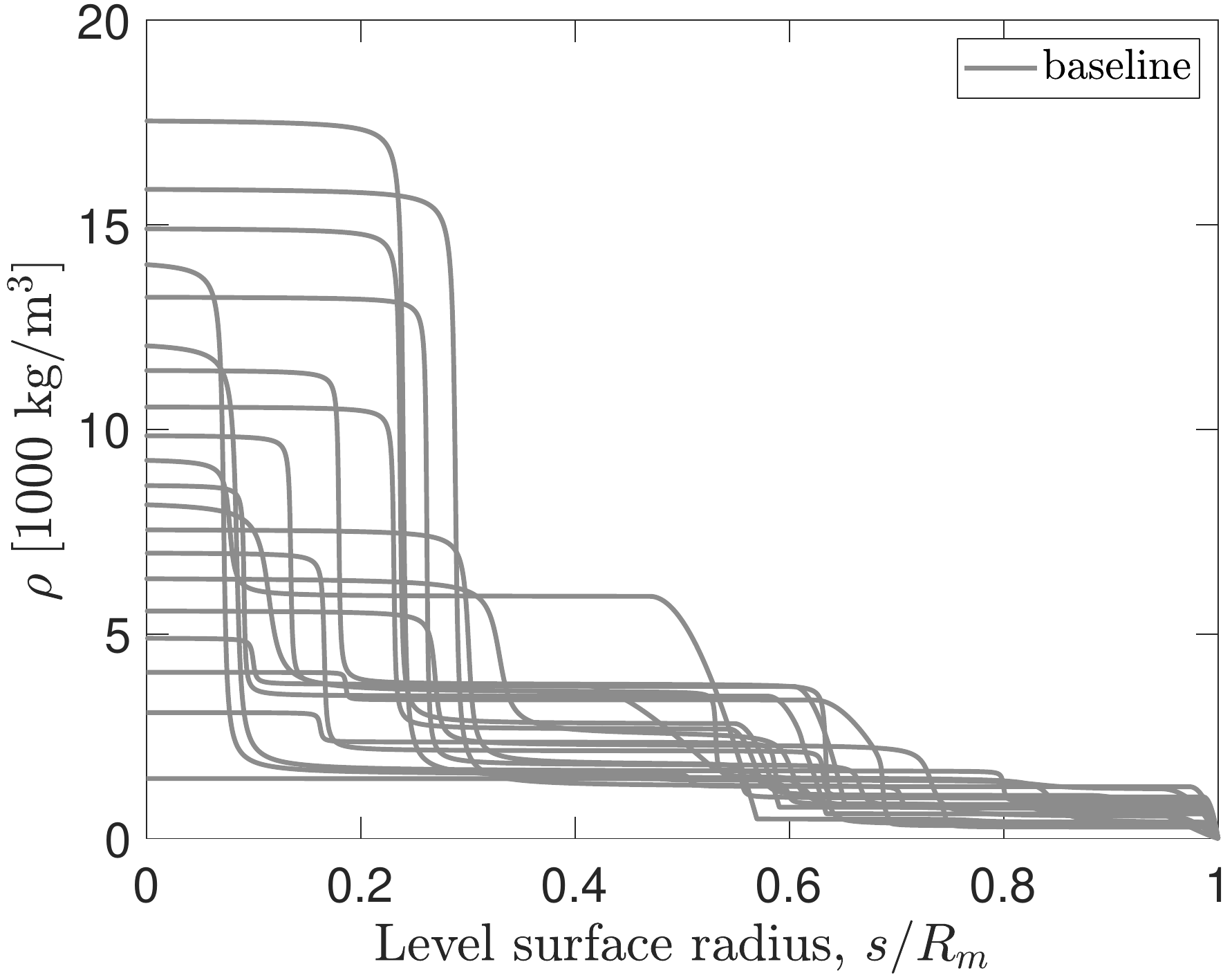}{0.33\textwidth}{}
\fig{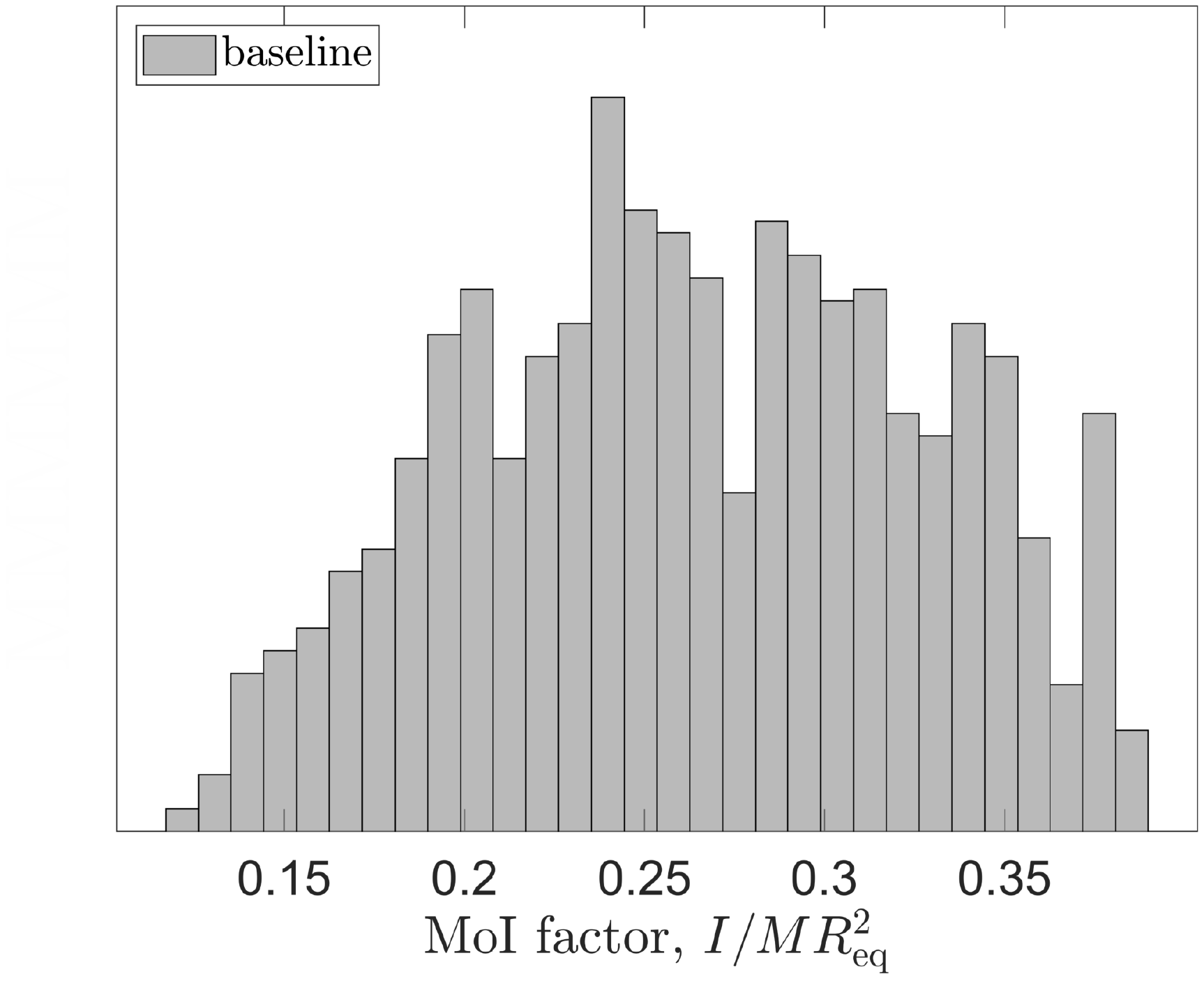}{0.32\textwidth}{}
}
\vspace{-2.0\baselineskip}
\caption{Three views of the baseline sample. Left: shaded region showing the
range of density values, for each level surface mean radius $s$, between the 2nd
and 98th percentiles of the sample; in effect the extent of the sample minus the
occasional outlier. Middle: a subset of 20 density profiles from the sample,
taken at equal intervals of percentiles of the central $(s=0)$ density values.
Right: histogram of the moment of inertia factor, integrated while accounting
for the equilibrium shape of level surfaces.}
\label{fig:baseline}
\end{figure*}

We begin by looking at the baseline sample. We will look at Uranus samples
first, because we find it more instructive to examine different views of the
same sample side-by-side, rather than compare the same view for both planets.
Figure~\ref{fig:baseline} shows three different views of the baseline Uranus
sample. The left panel shows what we call the \emph{envelope} view; a shaded
area covering, for each radius, roughly a ``two-sigma'' spread in density values
obtained there by the sample. This view is helpful in providing, at a glance, a
sense of the overall extent of the density values reachable under the relevant
constraint. Recall that for the baseline sample the only constraints were that
each density profile integrates to the correct mass, with the boundary
conditions $\rho(1)=\rho_0$ and $\rho(0)\le\rho\sub{max}$, and the Gaussian
prior set on the rotation period (Table~\ref{tab:observables}).

The middle panel of Figure~\ref{fig:baseline} is what we call the
\emph{ensemble} view. A subset of $20$ density profiles from the sample is
shown, where we attempted to pick ones that show the variety of possible density
profiles reachable under the given constraint. This view is more helpful for
seeing features like the location and scale of density jumps and identifying
regions of possible layering.

Lastly, the right panel is a histogram of the moment of inertia (MoI) values
from the sample. As an integrated, scalar value, it serves as a quantitative
measure of the ``width'' of the distribution.

The distribution of density profiles evident in the baseline sample is not
surprising and does not contain much information about Uranus in particular or
planetary interiors in general. Figure~\ref{fig:baseline} is nevertheless
important because, first, every other sample will be compared against the
baseline in order to gauge the effectiveness of the employed constraints, and
second, because it serves to validate the combination of the chosen
parameterization scheme and sampling procedure. Recall that the goal was to have
a parameterization flexible enough, and a sampling procedure robust enough, so
that any restriction of the resulting density profiles evident in the sample can
be attributed to the constraints that were deliberately used. The baseline
sample was expected to, and in fact did cover essentially the entire range of
plausible interior profiles. From panels (b) and, especially, (c) it is clear
that both nearly ``flat'' profiles as well as very centrally condensed ones are
reachable with this sampling method. Even if we cannot guarantee the exact shape
of the probability distribution (it is possible, for example, that regions of
apparent low probability would fill up gradually if much larger samples were
drawn, see also appendix~\ref{sec:chains}) it is at least clear that the
parameterization and/or sampling do not by themselves restrict the resulting
samples.

\subsection{Rotation period priors\label{sec:rotation}}
\begin{figure*}[tb!]
\gridline{
\fig{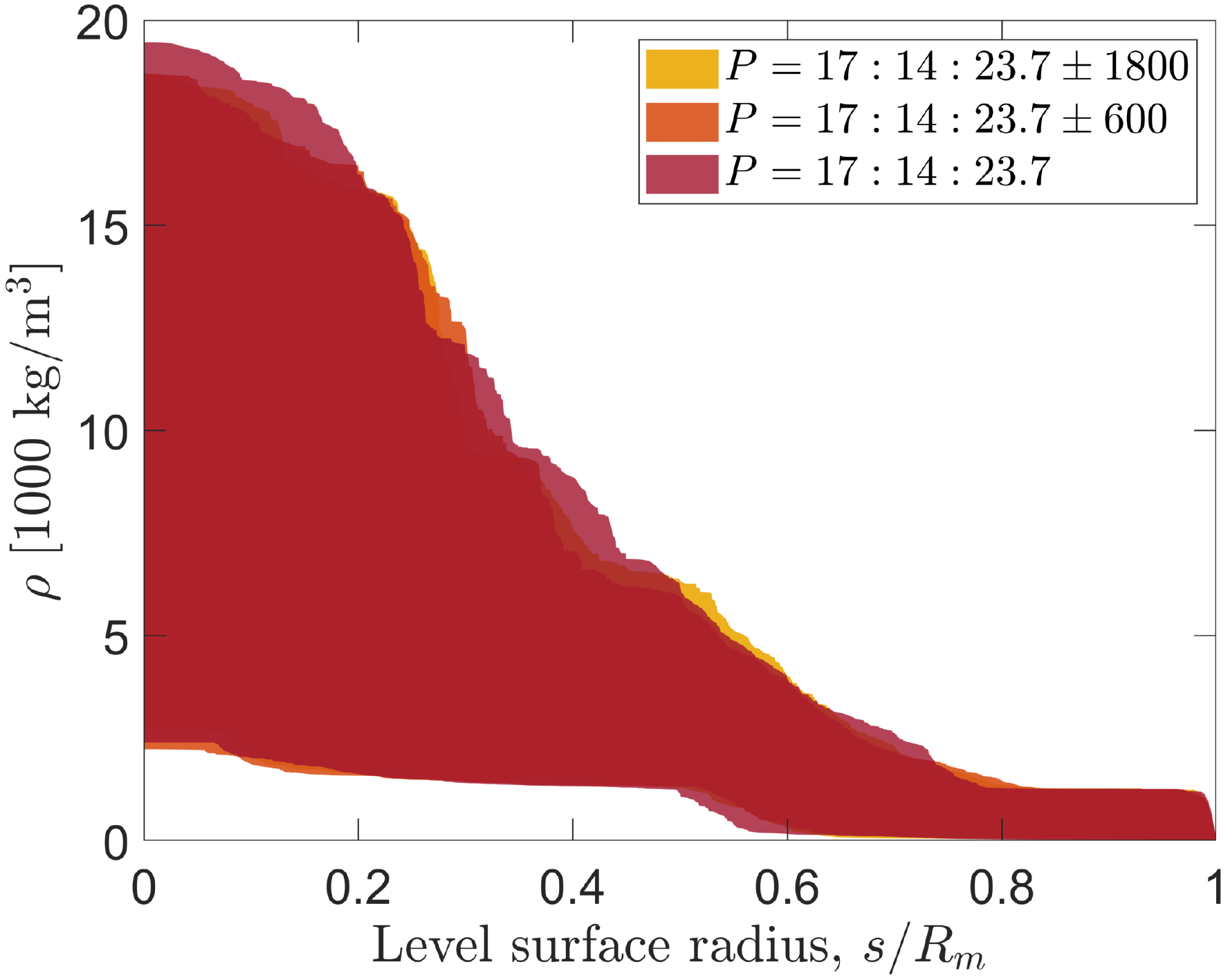}{0.33\textwidth}{}
\fig{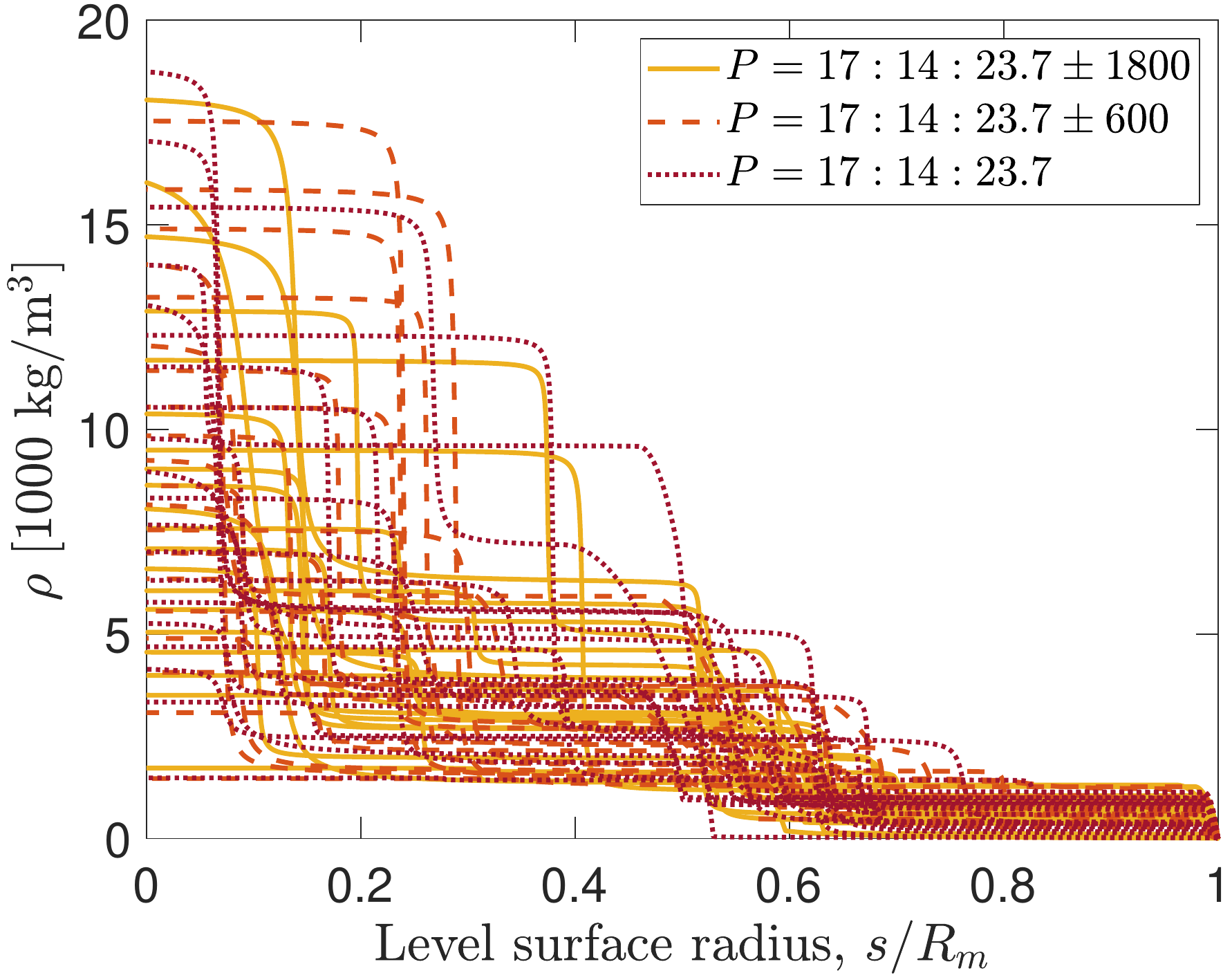}{0.33\textwidth}{}
\fig{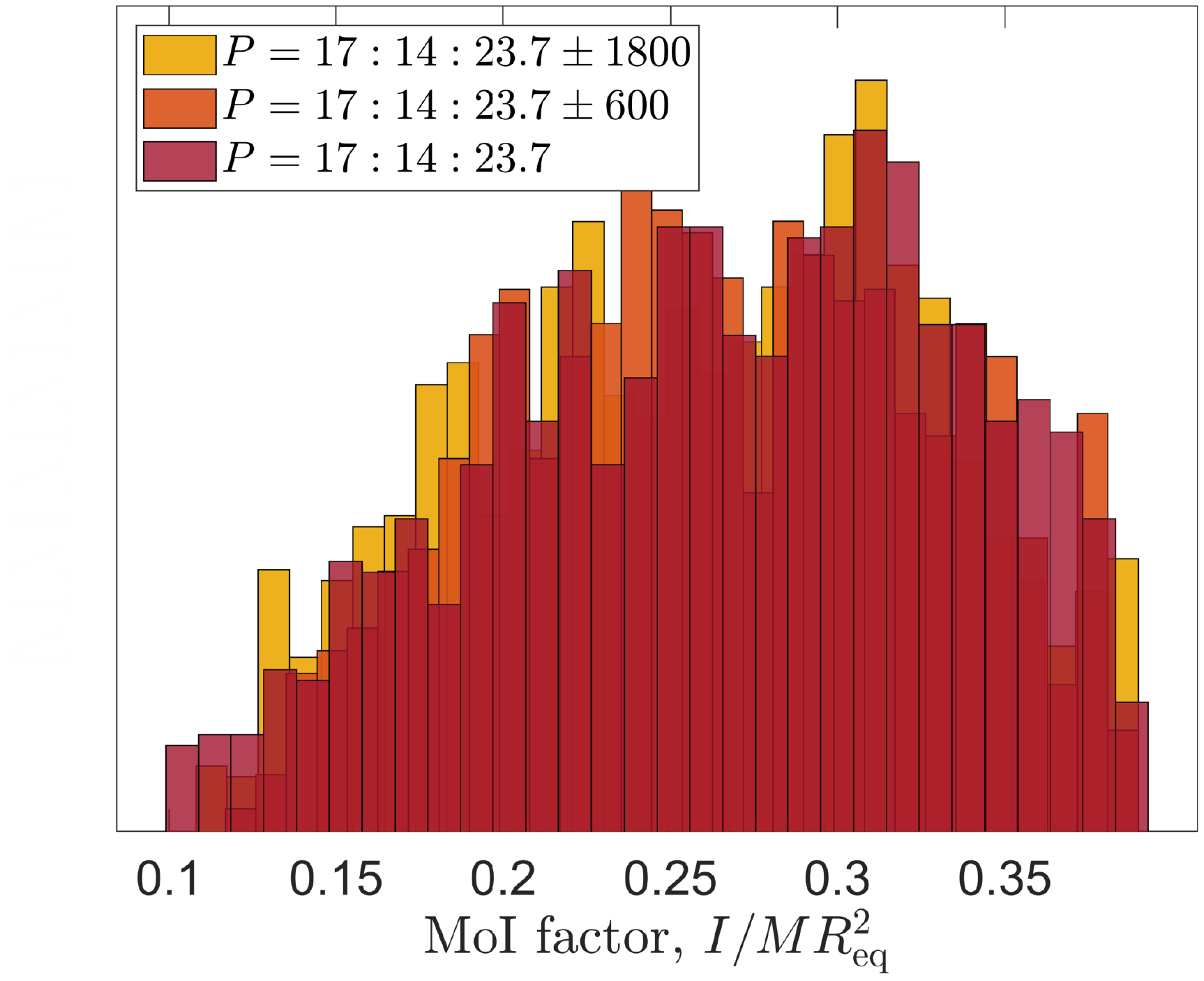}{0.32\textwidth}{}
}
\vspace{-2.0\baselineskip}
\caption{Same views as Fig.~\ref{fig:baseline} showing samples obtained with
large uncertainty (30 minute, light shade, solid lines), realistic uncertainty
(10 minute, medium shade, dashed lines), and no uncertainty (dark shade, dotted
lines) on the rotation period. The samples are essentially the same.}
\label{fig:rotation}
\end{figure*}

A more precise determination of the rotation, by itself, is not a helpful
constraint. In Figure~\ref{fig:rotation} the baseline sample (rotation period a
Gaussian prior with $\sigma=10\unit{min}$, \citep{Podolak2012}) is shown again,
overlain with two more samples. One obtained assuming much higher uncertainty in
the rotation period (Gaussian prior, $\sigma=30\unit{min}$), and the second
obtained with a perfectly known rotation period: the relevant parameter kept
constant instead of being sampled.

It is perhaps surprising that these samples are essentially identical; that the
rotation period can be varied by as much as an hour with apparently no effect on
the shape of the allowable density profiles. This is because density and
rotation are in a sense interchangeable, when the only observable use to
constrain the sample is total mass. While faster rotation increases the planet's
oblateness which would result in a smaller mass for the same density profile
(the outer mean radius is decreased since the equatorial radius is fixed), this
is easily compensated by a small increase in density in the upper layers. It is
possible that much stricter interior boundary conditions, i.e., much smaller
$\rho\sub{max}$ and applying a central $\rho\sub{min}$, would reveal the limits
of allowable rotation period. We did not test this because there is no realistic
prospect of independently estimating the central density.

We stress that even though the rotation period \emph{by itself} is not a helpful
constraint, it will become important when high-precision gravity is considered,
in sec.~\ref{sec:rotredux}.

\subsection{Low order gravity\label{sec:j24}}
\begin{figure*}[tb!]
\gridline{
\fig{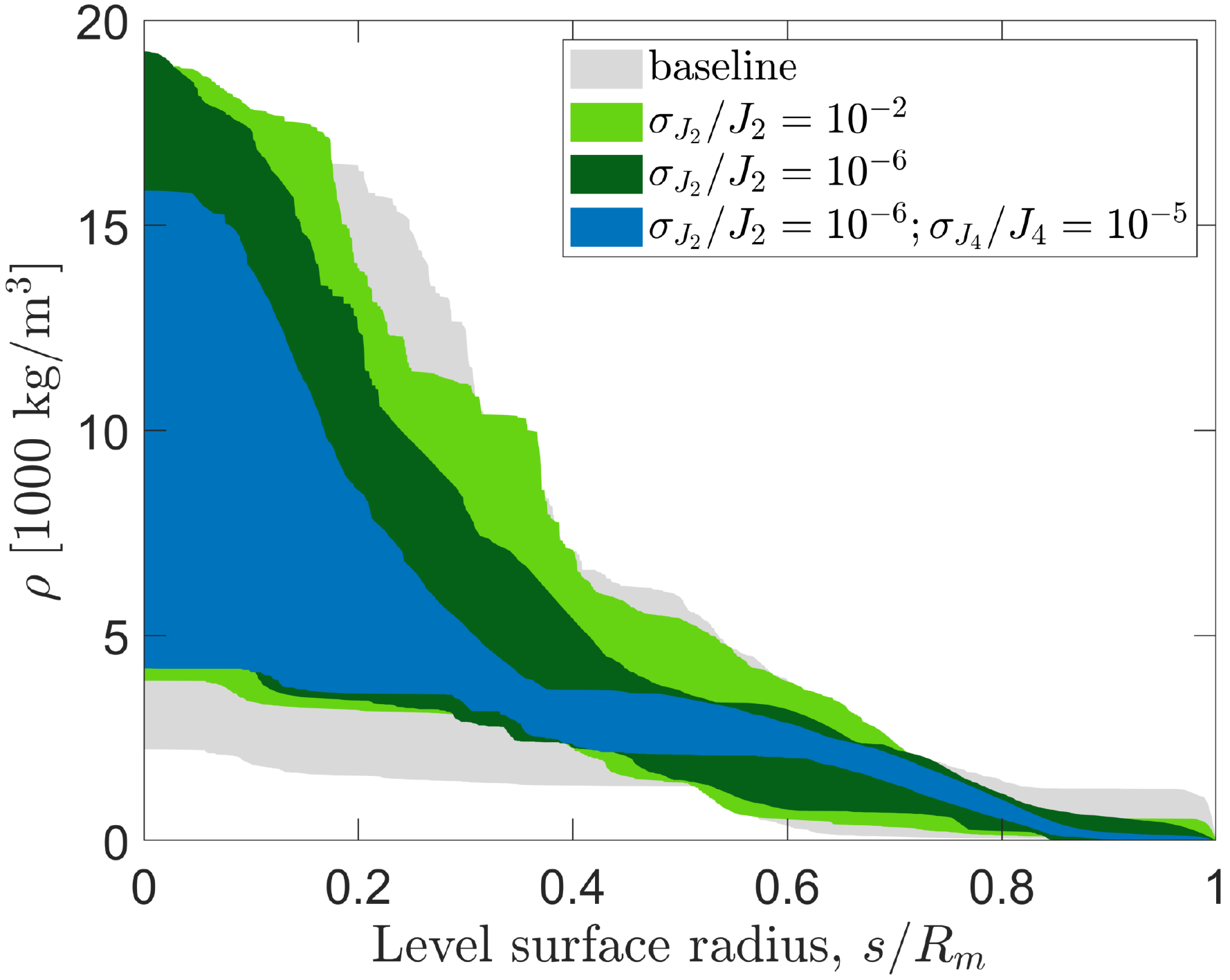}{0.33\textwidth}{}
\fig{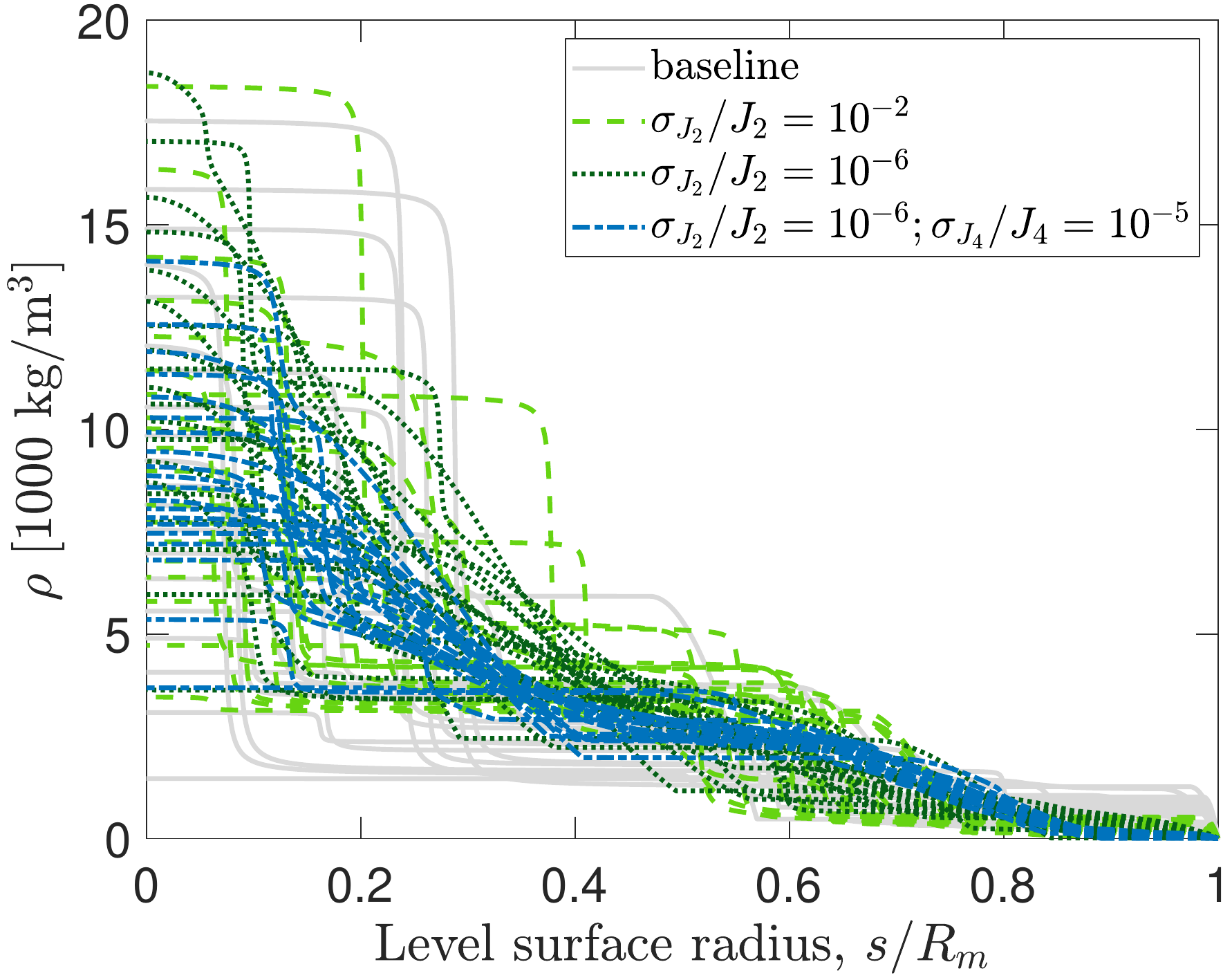}{0.33\textwidth}{}
\fig{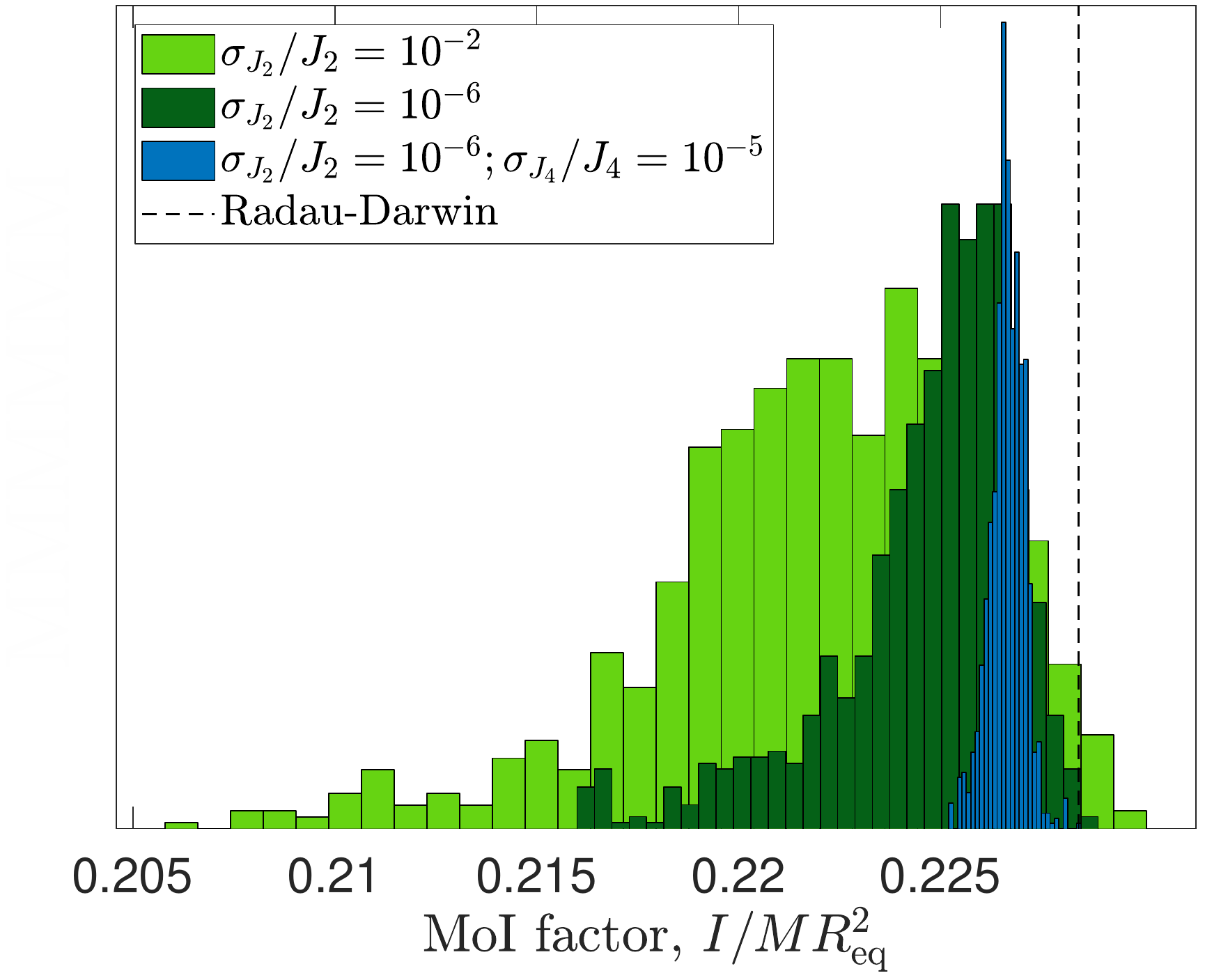}{0.32\textwidth}{}
}
\vspace{-2.0\baselineskip}
\caption{Same views as Fig.~\ref{fig:baseline}, showing samples constrained by
$J_2$, poorly known, $J_2$ precisely known, and $J_2$ and $J_4$ precisely known,
in addition to the previous constraints of mass, radius, and rotation period.}
\label{fig:lowj}
\end{figure*}

Adding information about the gravity field, even just a crudely estimated $J_2$
value, significantly shrinks the extent of allowable density profiles. Looking
at Figure~\ref{fig:lowj} we can deduce two important, if unsurprising lessons
about the constraining power of gravity, and a third that was perhaps not as
obvious \emph{a priori}. The first is that gravity is most sensitive to the
density structure at the outer regions of the planet, gradually becoming less
sensitive the deeper we look, and almost indifferent to the density near the
center of the planet. This behavior is well understood and easily predicted, at
least qualitatively, from the definition of the gravity coefficients as
integrals over the radius. Nevertheless we find it instructive to see a direct,
quantitative illustration. This is most clearly seen in the left panel of
Figure~\ref{fig:lowj}, while the middle and right panels better illustrates the
second unsurprising fact, that gravity restricts the \emph{shape} of the density
better than its overall extent. For example, even the most restrictive sample,
the one assuming very precisely known $J_2$ and $J_4$ (blue dash-dot lines in
the figure), allows the central density to reach values almost as low or as high
as the less restrictive samples, or even the baseline sample. But there are many
ways in which a $\roofs{}$ curve can reach those values, and some of these
curves appear in the less restrictive samples and are notably missing from the
last one. In particular, large and sharp density increases at large
($0.3\lesssim{s/\rem}\lesssim{0.6}$) radii seem to be disfavored or even
disallowed by this nominal value of $J_4$, thereby telling us, in this case,
that Uranus is unlikely to have a huge rocky core.

The last, and least predictable lesson gleaned from Figure~\ref{fig:lowj} is
that, loosely speaking, it is better to know more $J$ values than to know the
same $J$ value more precisely. This is probably best seen in the right panel, as
well as in other samples, not shown here, with different combinations of $J_n$
and $\sigma_{J_n}$. While not exactly surprising it is nevertheless not obvious
why, for example, a four order of magnitude improvement in the precision of the
$J_2$ value would, by itself, amount to only a modest shrinking of the allowed
range of densities, density profiles, or MoI values.

\subsection{High order gravity}
There are at present no usable estimates of actual $J_n$ values for either
planet, for $n>4$. To continue investigating the potential constraining power of
high-precision, high-order gravity measurement we need to assign some
hypothetical yet reasonable values to $J_n$ and $\sigma_{J_n}$.

\begin{figure*}[tb!]
\includegraphics[width=1.0\textwidth]{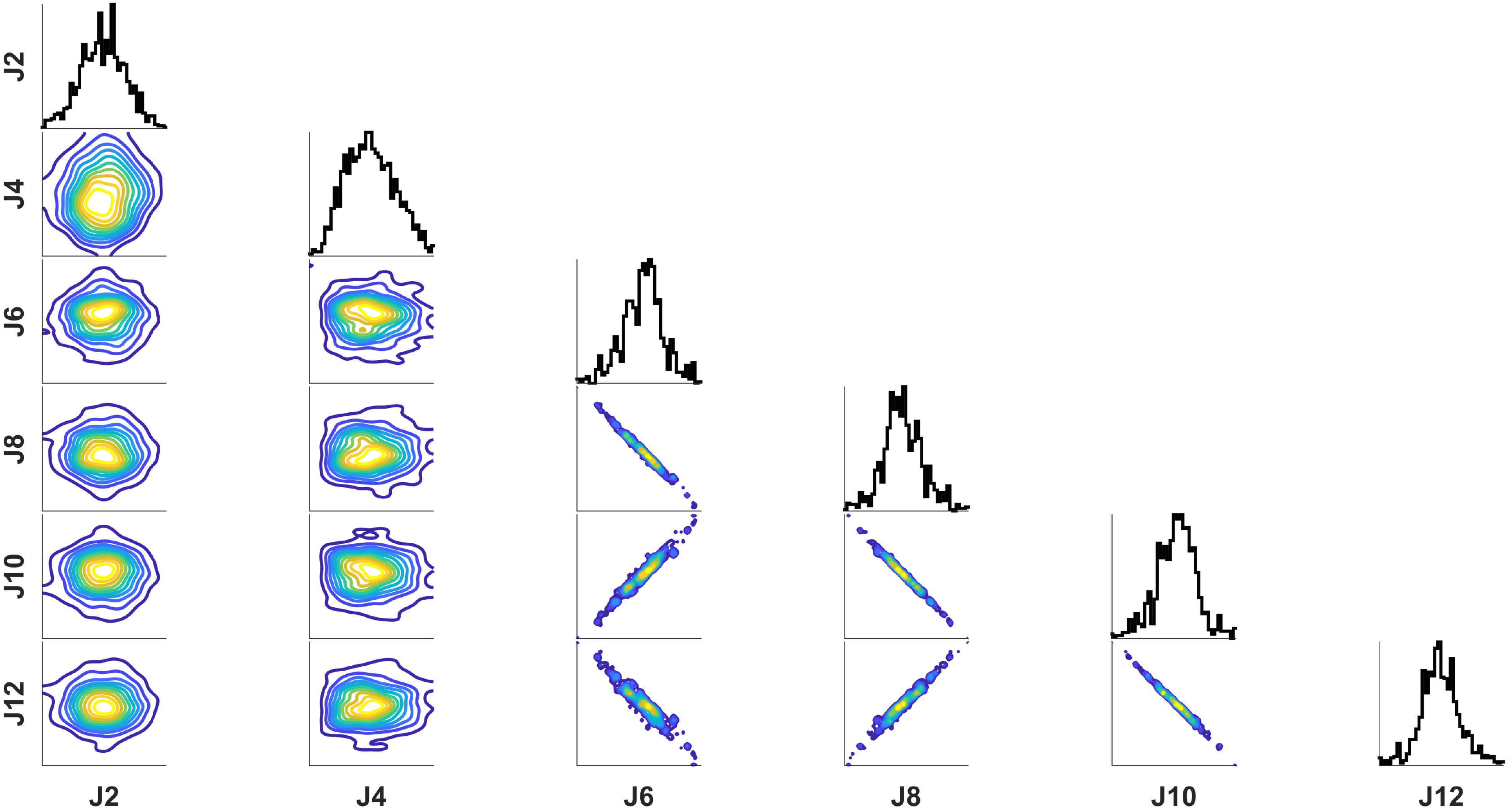}
\caption{$J$ values from an example MCMC run, illustrating correlations between
$J_n$.}
\label{fig:Jdistribution}
\end{figure*}

For nominal $J_n$ values the only sensible choice is to use the mean values from
the previous samples, specifically from the most constraining one obtained so
far (blue, dash-dot curves in fig.~\ref{fig:lowj}). Only $J_2$ and $J_4$ values
were used in the likelihood function driving the sampling algorithm, but all
values of $J_n$ are available after the fact. Figure~\ref{fig:Jdistribution}
reminds us that the distributions of $J_n$ values for different $n$ are highly
correlated, yet there is a range of allowable values that could be further
restricted.

The usable precision of gravity data, the $\sigma_{J_n}$, is limited by the
worst of three factors. The first is the expected accuracy of determination by a
hypothetical future orbiter mission, should it become available. For this we can
look to the recent successes of the radio science teams of the \emph{Cassini}
and \emph{Juno} missions in reconstructing the gravity fields of Saturn and
Jupiter, respectively, from the spacecraft accelerations deduced from Doppler
shifts detected in the radio link between the spacecraft and tracking stations
on Earth. We may take the exquisite precision obtained during those missions as
{best-case bound of} what might be expected from a future Uranus/Neptune
mission. But recall that this truly impressive precision applies to $J_n\obs{}$
and not to $J_n\rig{}$ (eq.~\ref{eq:J_correction}). The second factor therefore
is the expected accuracy of a correction term, $J_n\wind$. It is hard to
speculate what that correction would look like, although it's safe to assume it
will come with attached uncertainty greater than the formal
$\sigma_{J_n}\obs{}$.

{
The last limiting factor is the precision with which we are able to calculate
the equilibrium shape and gravity of a model planet. It is unusual for a numeric
computation to rival observational data in terms of poorer accuracy, but this is
in fact the case here. For the high-order gravity samples we solve for
equilibrium shape and gravity using the ToF7 method (sec.~\ref{sec:tof}), which
gives us usable values for up to $J_{12}$ and the associated $\sigma_{J_n}$ for
the likelihood function.
}

\begin{figure*}[tb!]
\gridline{
\fig{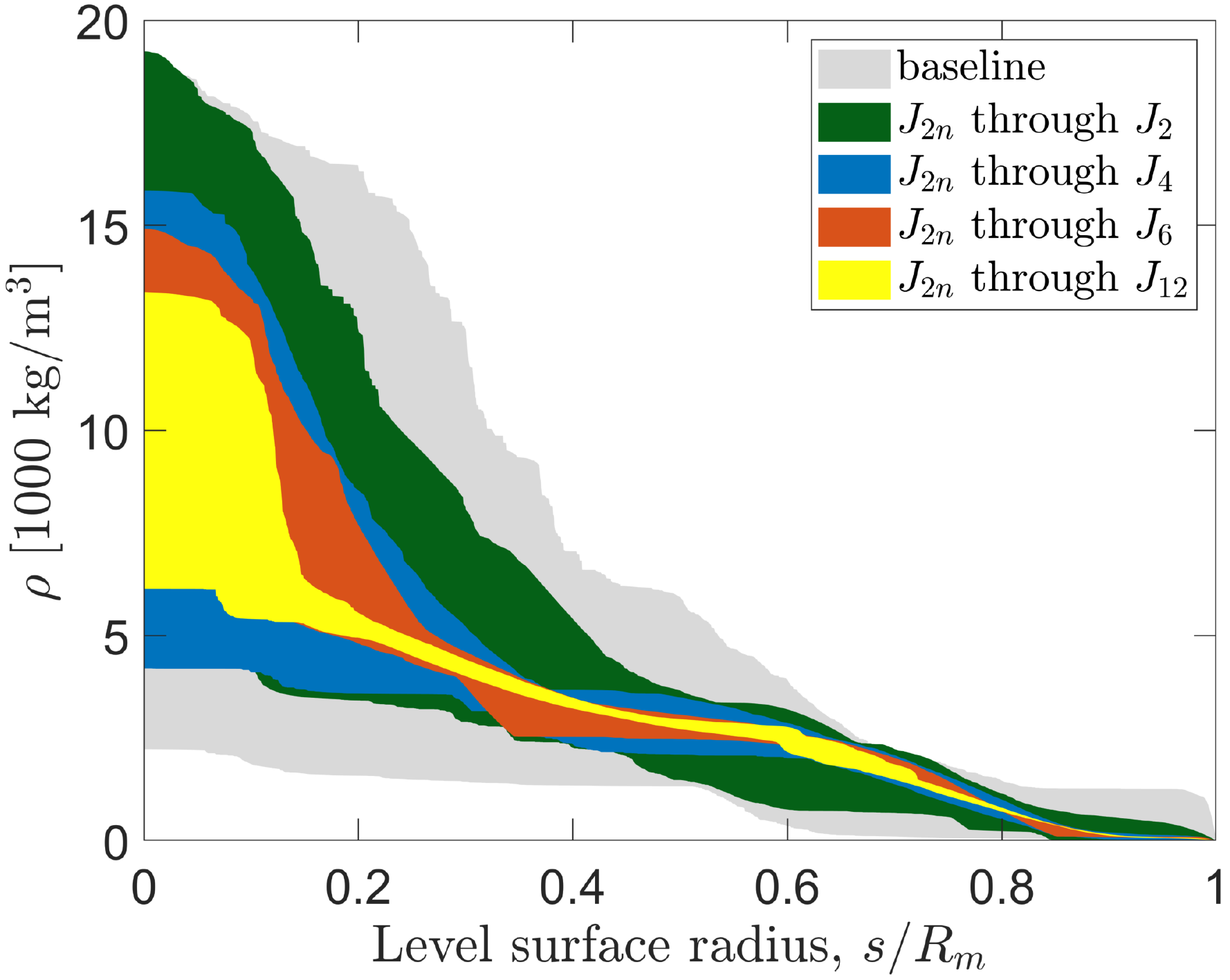}{0.33\textwidth}{}
\fig{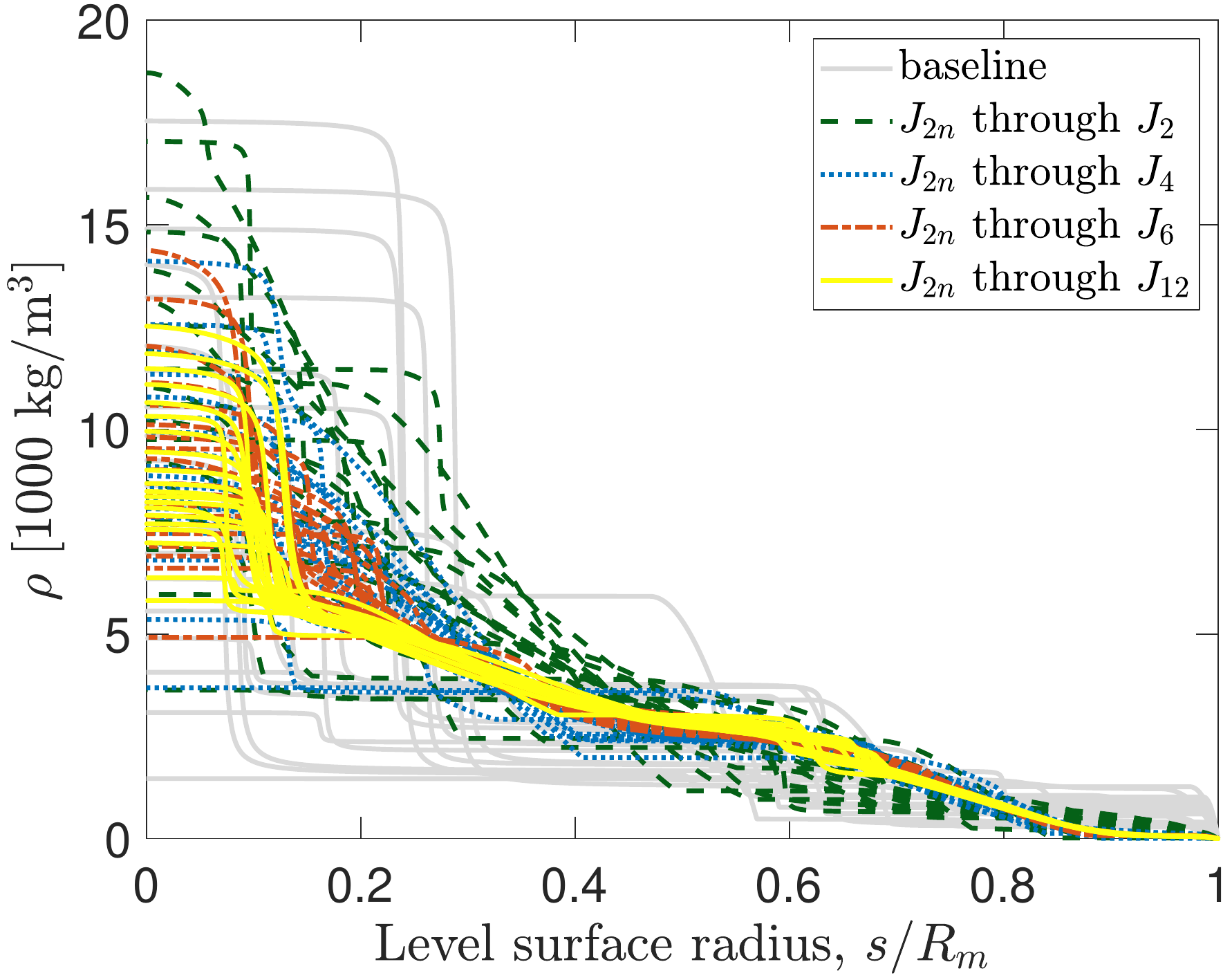}{0.33\textwidth}{}
\fig{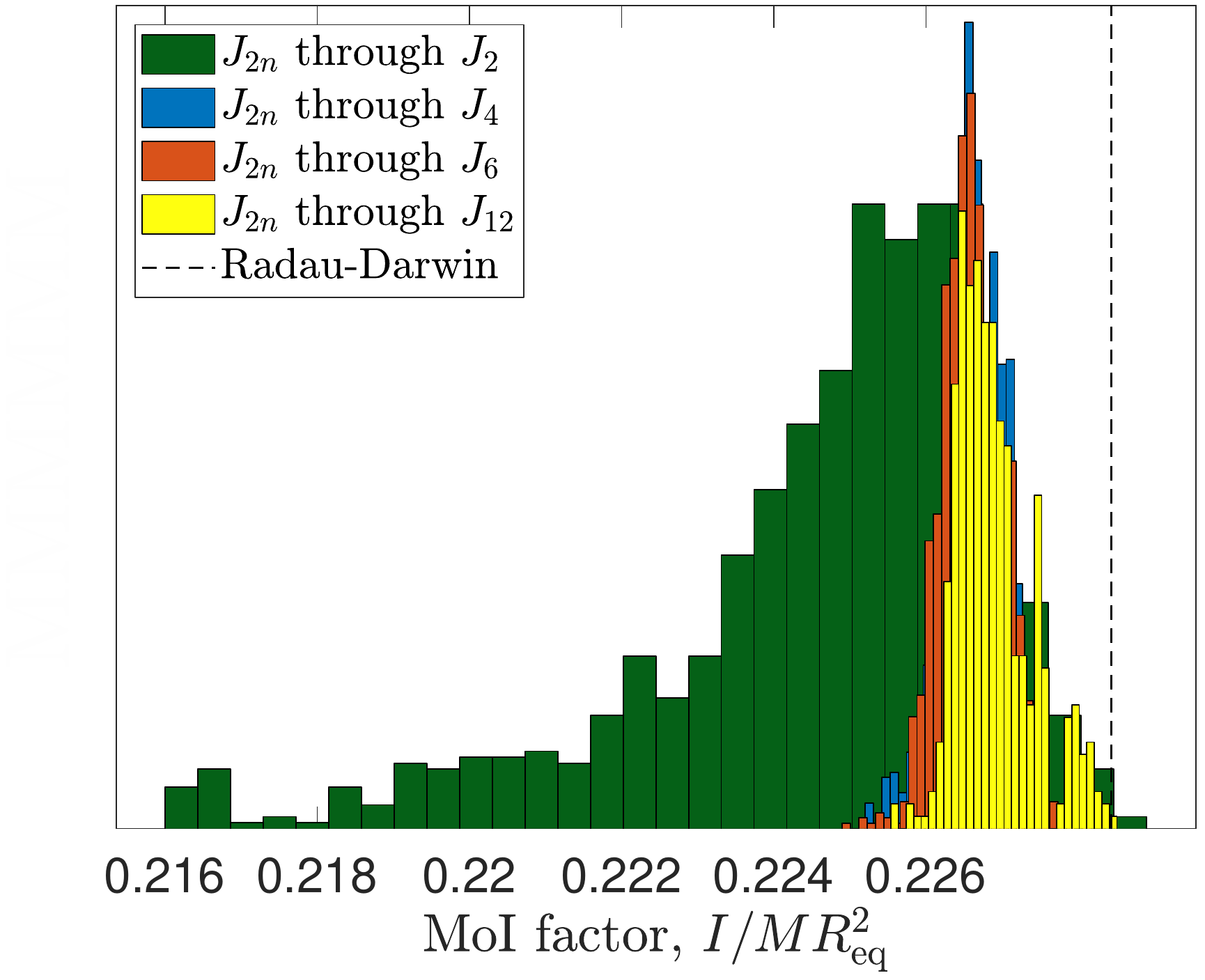}{0.32\textwidth}{}
}
\vspace{-2.0\baselineskip}
\caption{Same views as Fig.~\ref{fig:lowj} showing samples constrained with
successively higher order $J_n$ up to $J_{12}$ ($n\sub{max}$ in
eq.~\ref{eq:Jloss}), in addition to the constraints of mass, radius, and
rotation period. The likelihood function used $\sigloss{2}=10^{-6}$,
$\sigloss{4}=10^{-5}$, $\sigloss{6}=10^{-4}$, $\sigloss{8}=10^{-4}$,
$\sigloss{10}=10^{-2}$, and $\sigloss{12}=10^{0}$.}
\label{fig:hij}
\end{figure*}

With this framework, Figure~\ref{fig:hij} shows the distributions from samples
constrained by progressively higher-order gravity, up to $J_{12}$. Again we see
that high-precision gravity provides an excellent constraint on the density
overall but cannot, by itself, pin down the central density or even the
existence of a distinct central region of high density.

The moment of inertia factor seems to be very tightly correlated with the
gravity, as soon as we go beyond $J_2$. While correlation is expected, both
gravity and MoI being essentially different integrals of the same $\roofs{}$, it
has been suggested that an independent measurement of the MoI can provide new
information, not already contained in the gravity field
\citep{Helled2011a,Helled2011c}. It may be true that knowledge of $J_n$ to any
order and with full precision is equivalent to knowledge of $\roofs$, and
therefore of every other quantity derived from it. Whether or not this is
strictly true in a precise, mathematical sense is not all that relevant. The
relevant question is how much variability is still possible in the MoI value
once gravity is measured to realistic order and with realistic precision. Our
answer, if this sample is representative, is: about $0.1\%$ (but see below for
the importance of rotation period uncertainty).

\subsection{Rotation and gravity\label{sec:rotredux}}
A planet's rotation period may be estimated by a variety of methods
\citep[e.g.][and references
therein]{Read2009,Helled2015,Gurnett2007,Mankovich2019,Anderson2007a} and with
varying degrees of precision. In an example of a best case scenario, Jupiter's
deep interior rotation is tied to the precession of the polar axis of its strong
magnetic field which can be measured to sub-second precision
\citep{Seidelmann2007}. In other cases we are not so lucky; the rotation periods
of Uranus and Neptune are estimated with uncertainty of at least 10 minutes (in
each direction) \citep{Podolak2012,Helled2010} and perhaps much higher.

\begin{figure*}[tb!]
\gridline{
\fig{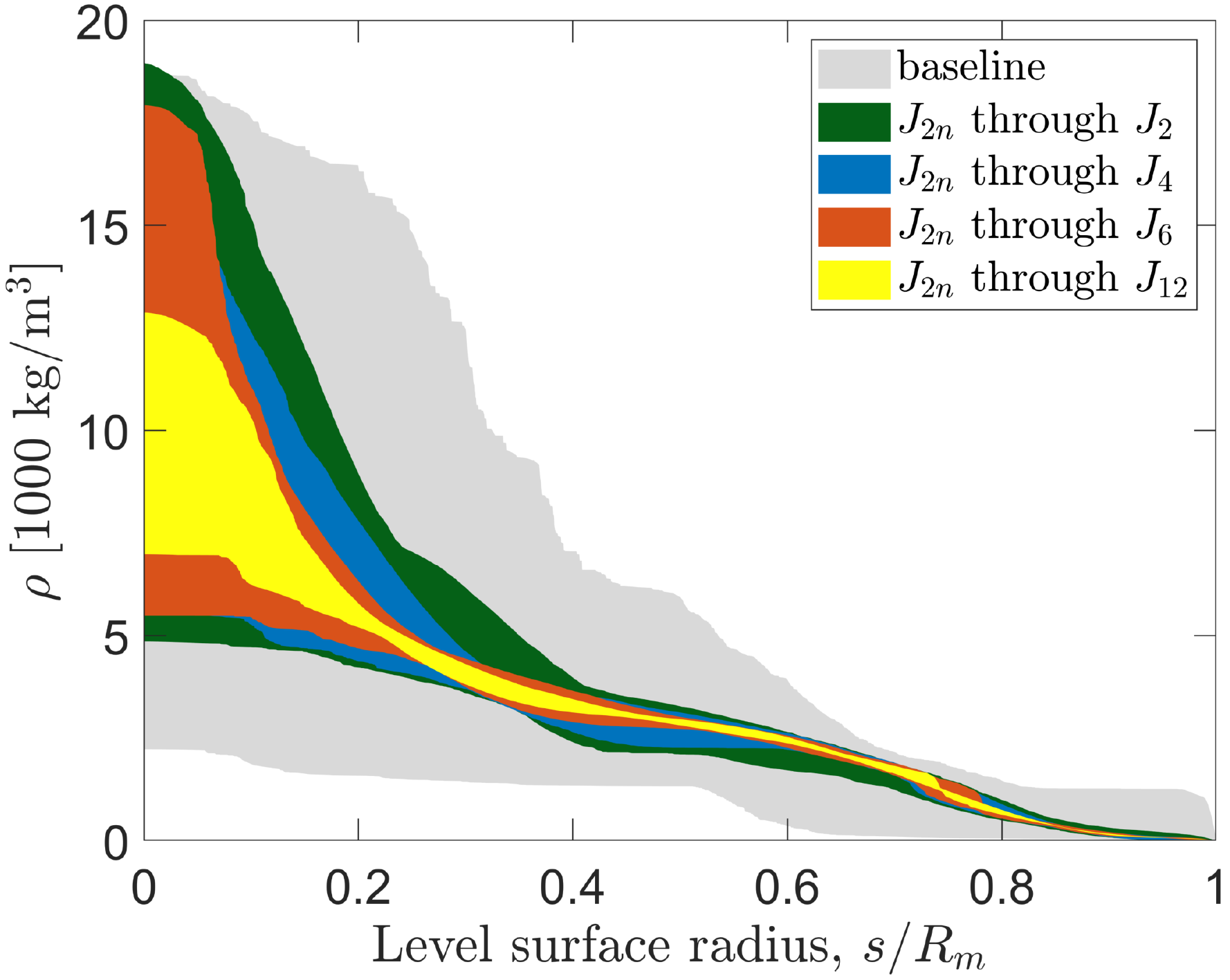}{0.33\textwidth}{}
\fig{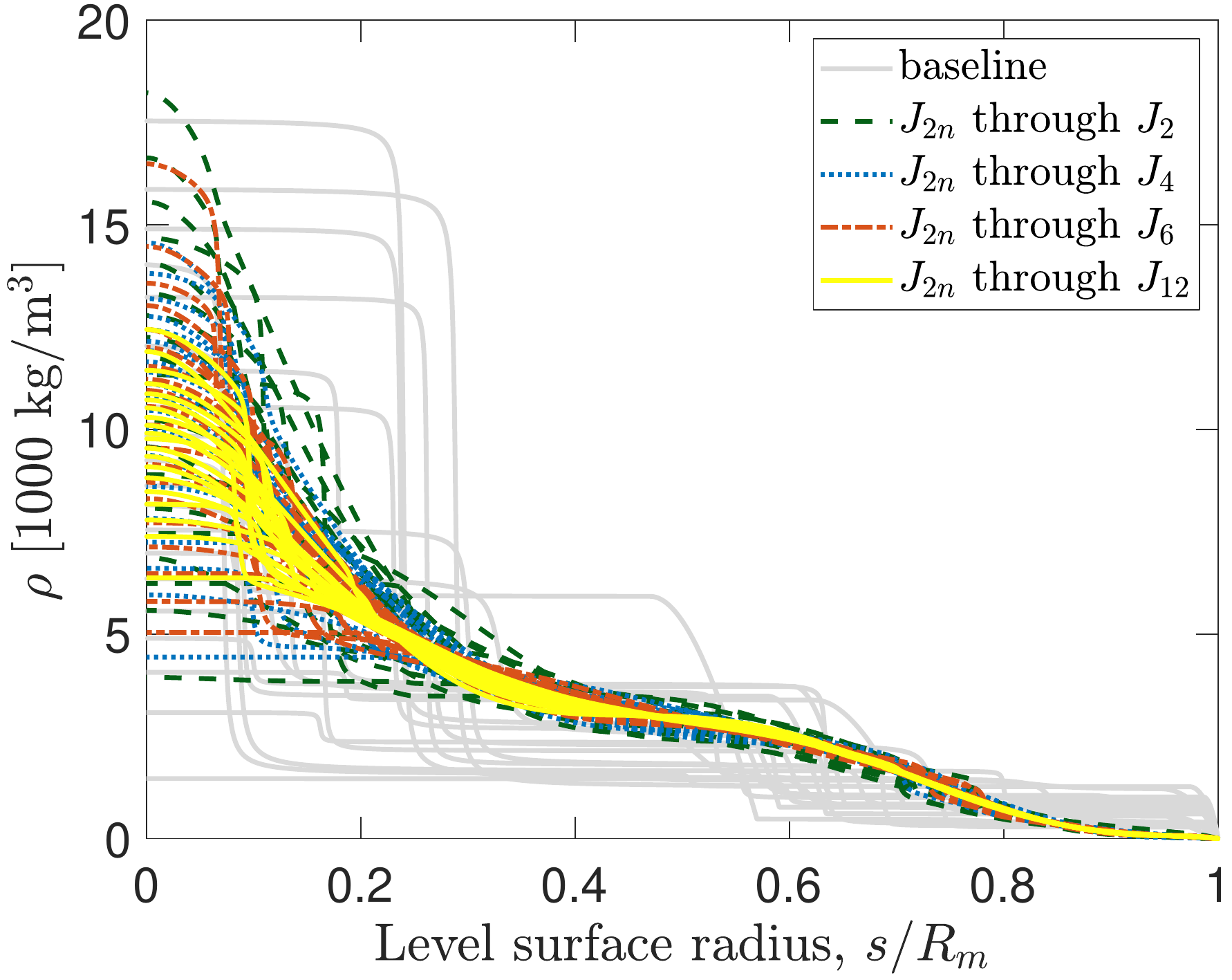}{0.33\textwidth}{}
\fig{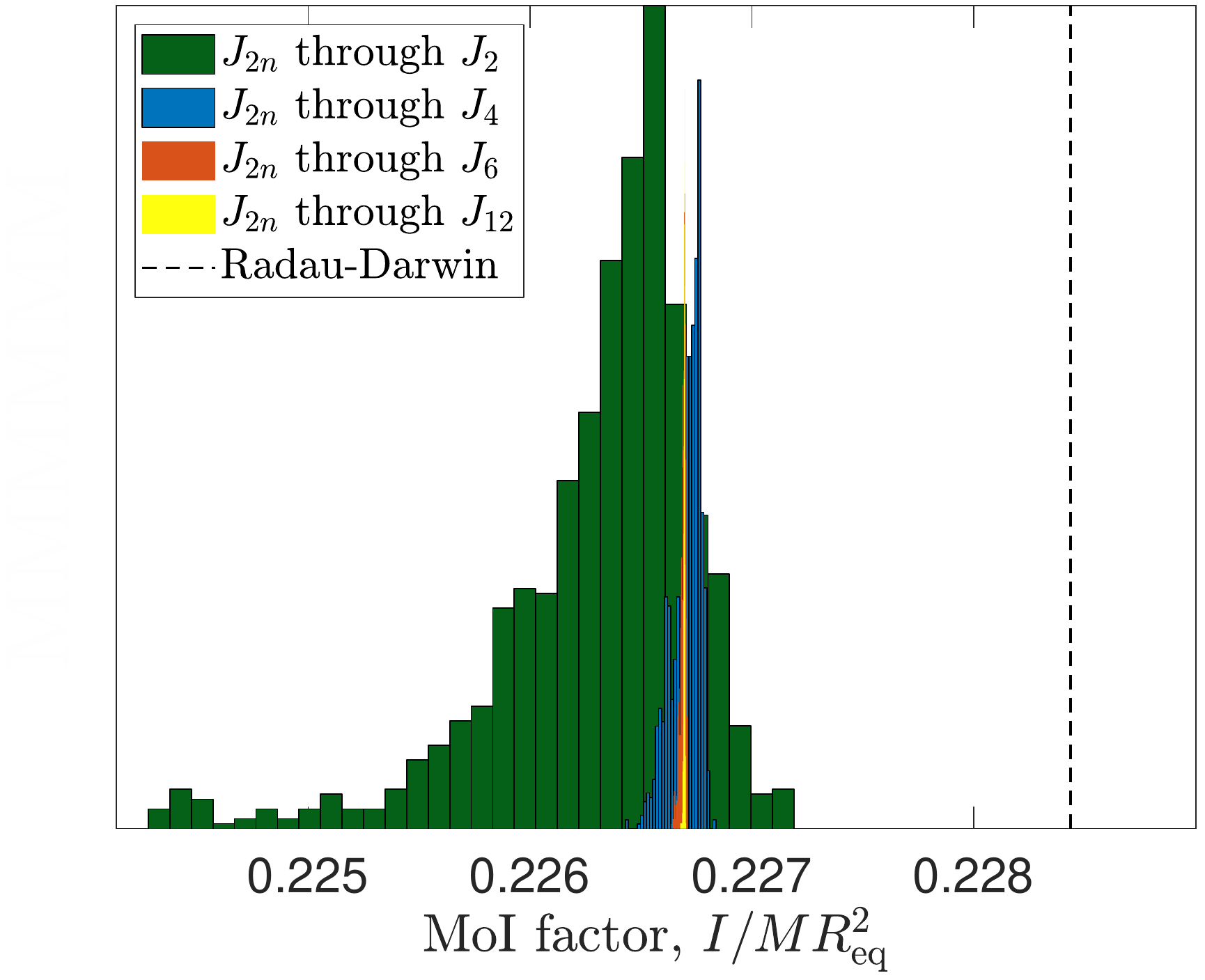}{0.32\textwidth}{}
}
\vspace{-2.0\baselineskip}
\caption{Same views as Fig.~\ref{fig:hij} showing samples obtained with a
constant, rather than sampled, rotation period. (Note changed horizontal axis
limits in MoI histogram.)}
\label{fig:hijfixrot}
\end{figure*}

In section \ref{sec:rotation} above we concluded that shrinking the uncertainty
in the rotation period by itself does little to further constrain the space of
allowable density profiles. However the same is not true when gravity is
considered also. Figure~\ref{fig:hijfixrot} shows a sample obtained assuming a
precisely known rotation period. Compare with Figure~\ref{fig:hij}. The
constraining power of gravity is significantly enhanced by precise knowledge of
rotation. Or, said another way, ignoring uncertainty in rotation may lead to
unjustifiably tight constraints being deduced from models. Assuming a precisely
known rotation period \footnote{Actually most models, including ours, prefer to
fix a dimensionless rotation parameter such as $m=\omega^2s_0^3/GM$ where
$\omega=2\pi/P$ and $s_0$ is the 1-bar level surface mean radius. This is
\emph{almost} equivalent to fixing the rotation period $P$ itself, but as the
equilibrium shape and with it $s_0$ are allowed to change, one cannot keep the
mass, equatorial radius, rotation period, and rotation parameter all fixed
simultaneously. Normally benign, this subtlety can cause confusion when directly
comparing models derived by different groups.} is a common, though not universal
simplification made by modelers. We caution that such a simplifying assumption
should be carefully justified or, better yet, avoided.

As evident from the right panel of Figure~\ref{fig:hijfixrot}, when both gravity
and rotation period are known to high precision, the remaining variation in MoI
value all but disappears (sample $\sigma/\mu\approx{10^{-5}}$). An independent
measurement of the MoI would have to be extremely precise if it is expected to
distinguish between different models already fitting the other constraints.

\subsection{Pressure-density relation}
The density profile, $\roofs$, was the focus of our attention because it is the
quantity that directly determines the gravity field, and is therefore directly
inferred by its measurement. But the density itself is not really what we are
most interested in. What we would like to know, ultimately, is what the planet
is made of, its composition, and how the various molecular species are
distributed inside the planet.

Gravity by itself can never give us this information, without additional
information and/or assumptions. But it can get us a step closer by noting that
the condition of hydrostatic equilibrium defines a one-to-one relationship
between density and pressure, once the gravity field is known. If everywhere in
the planet the weight of a layer of fluid is exactly balanced by the force due
to pressure gradient, $\grad{p}$, then
\begin{equation}\label{eq:HE}
\grad{p} = -\rho\grad{U}
\end{equation}
everywhere, where $U$ is the total potential, gravitational plus centrifugal.
Note that $U$ must be known in the interior of the planet, while the measured
$J_n$ values only relate to the external potential. Luckily, the process of
calculating the external potential from a given $\roofs$ and rotation period
also furnishes the potential on interior level surfaces as a useful byproduct.
We can integrate, numerically, eq.~(\ref{eq:HE}) starting from the 1-bar
reference level and obtain $p(s)$ and therefore $p(\rho)$ on every level
surface.

\begin{figure*}[tb!]
\gridline{
\leftfig{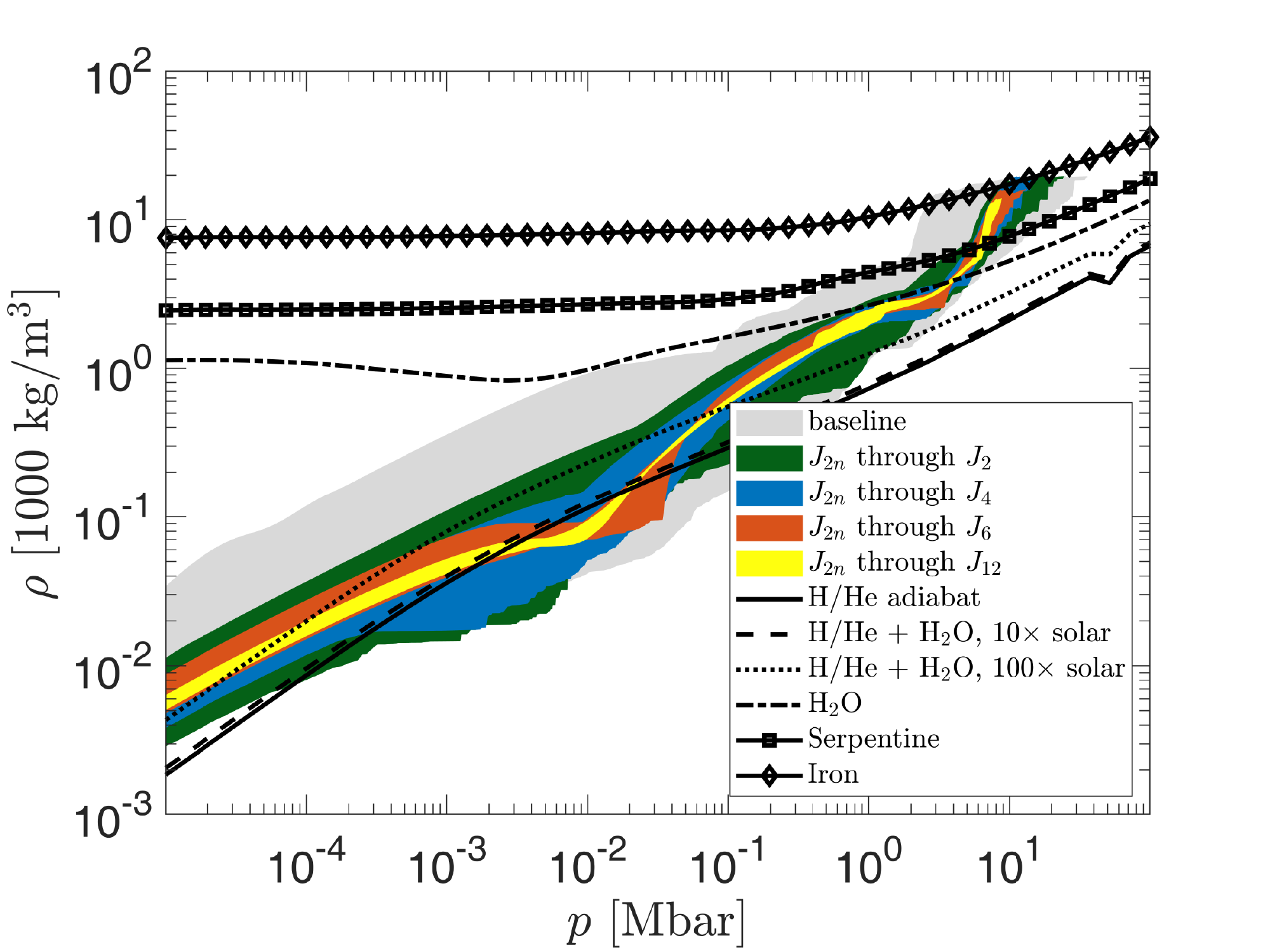}{0.5\textwidth}{}
\rightfig{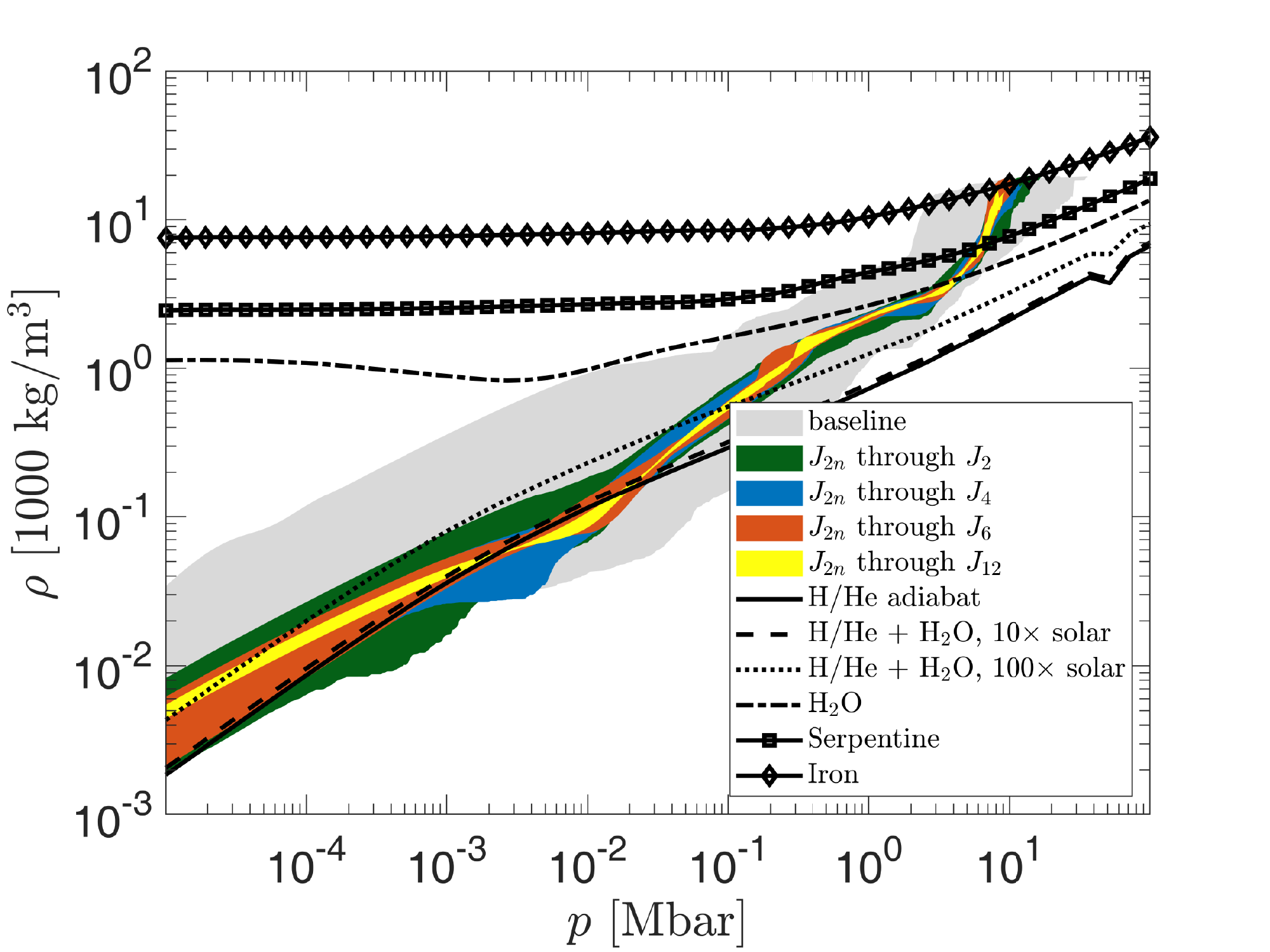}{0.5\textwidth}{}
}
\vspace{-2.0\baselineskip}
\caption{Pressure-density envelopes integrated from sampled $\roofs$ profiles
of Uranus.
\emph{Left}: samples obtained with Gaussian prior on rotation period;
\emph{Right}: samples obtained with known rotation period. Overlain isentropes
for hypothetical homogeneous compositions extend from a common
$T\sub{10bar}=150\unit{K}$.}
\label{fig:barotropes}
\end{figure*}

The pressure-density relation, also called a \emph{barotrope}, is not quite
enough to uniquely relate to composition (a true equation-of-state would require
the temperature profile as well) but it can already help by setting some bounds.
Figure~\ref{fig:barotropes} shows the distribution envelope of barotropes
integrated from the Uranus density profiles of Figures~\ref{fig:hij}
and~\ref{fig:hijfixrot}. The left panel corresponds to samples obtained with our
conservative, $\sigma=10\unit{min}$ prior on rotation period, and the right
panel corresponds to samples obtained with a precisely known rotation period.
Adiabats of several compositions, computed with the SCvH and ANEOS equations of
state \citep{Saumon1995,Thompson1990} are overlain for
comparison.\footnote{ The plots begin at 10 bars because the SCvH
equation of state table does not extend down to the required temperature at 1
bar.} {(The dotted line approximates a high metallicity envelope by
adding water to the H/He adiabat at $Z=0.57$ mass fraction, or about 100 times
the solar O:H abundance.)}

We see in Figure~\ref{fig:barotropes} how progressively higher-order gravity is
able to shrink the allowable region in $\rho$--$p$ space. The mass and radius of
the planet (grey, baseline region) already tell us something about the possible
composition; {for example that Uranus is not dense enough to have a
significant iron core}. The low-order $J_2$ and $J_4$, if measured with higher
precision, can be used to narrow down further possible configurations, and
higher-order $J_n$ would help even more. But a precise determination of the
underlying rotation period is necessary for maximum benefit.

\section{Realistic constraints on Uranus and Neptune\label{sec:uncons}}
\subsection{Constraints with presently known gravity}
In the previous section we investigated the ability of gravity field
measurements, in general, to constrain the interior density distribution of a
fluid planet. We used nominal values of mass, radius, and low-order $J$s for
Uranus to illustrate the results but the models presented above should not be
used for making predictions about the real planet Uranus, since they use
hypothetical uncertainty values. In this section we look at samples of interior
models obtained with presently available best values and uncertainties for both
Uranus and Neptune. We use the same views of the resulting distributions as in
section~\ref{sec:results} but showing side by side the same view for both
planets.

\begin{figure*}[tb!]
\gridline{
\leftfig{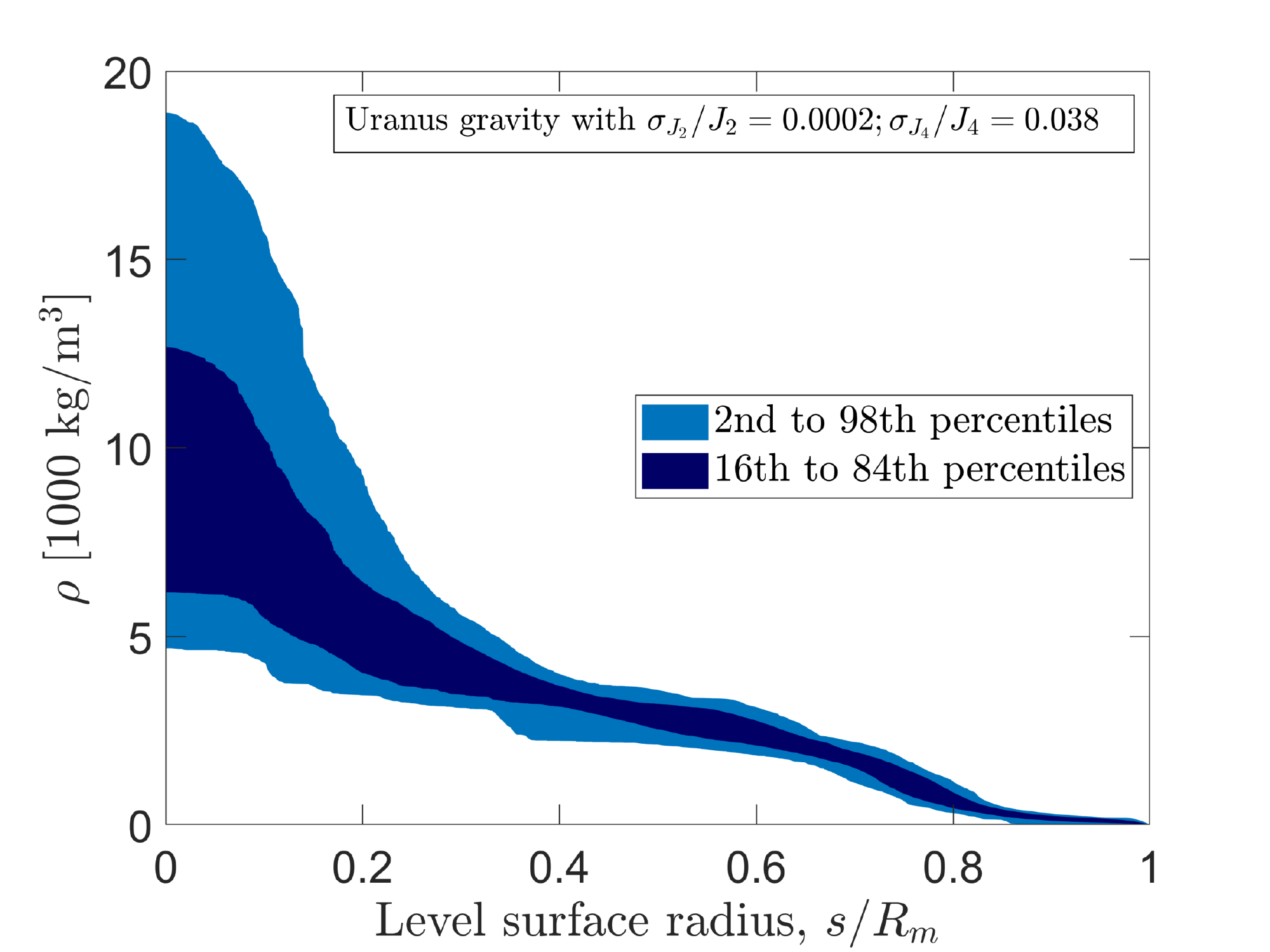}{0.5\textwidth}{}
\rightfig{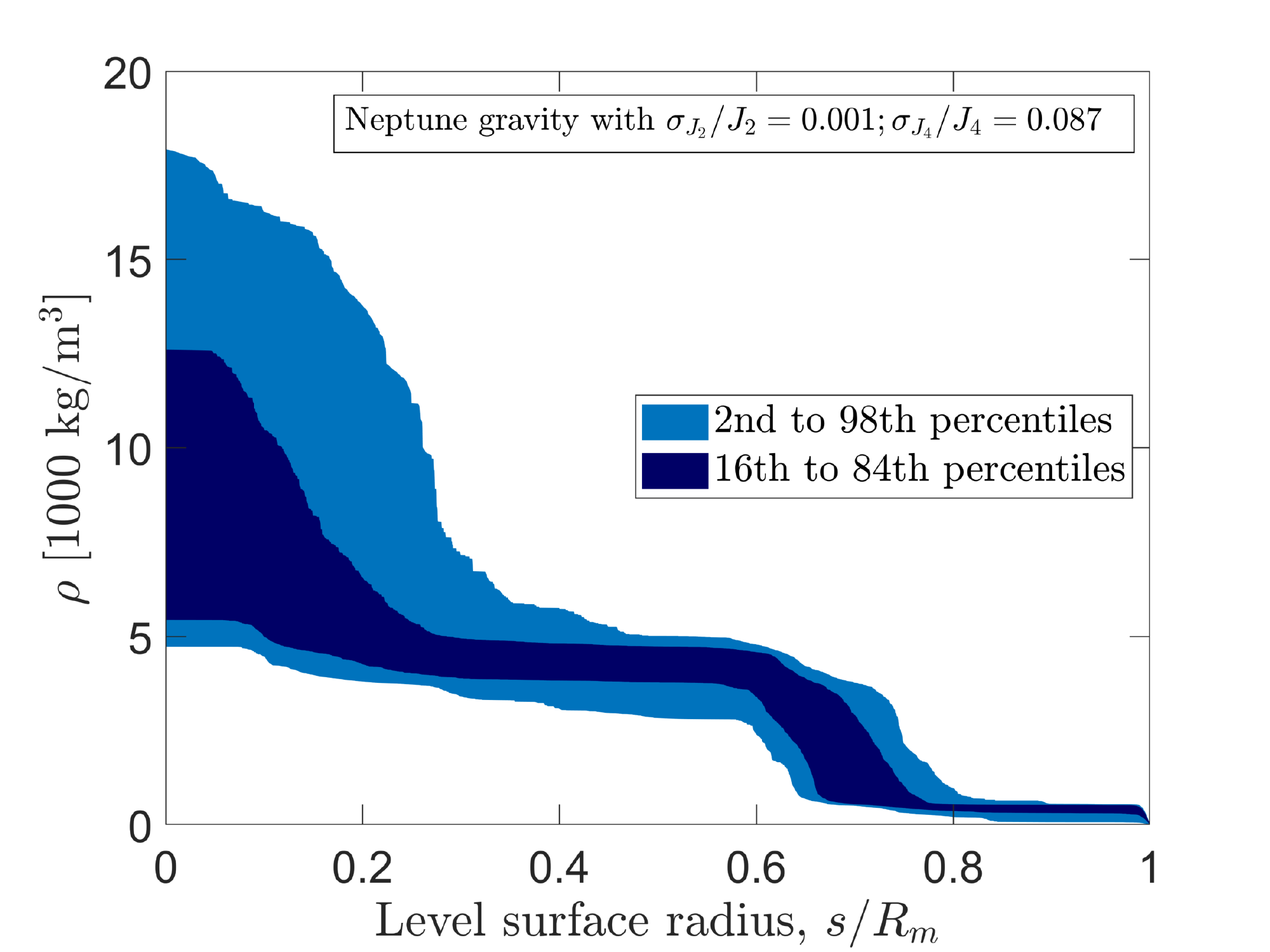}{0.5\textwidth}{}
}
\vspace{-2.0\baselineskip}
\caption{Envelope view of sampled $\roofs$ profiles of Uranus (left) and Neptune
(right) matching currently available observables
\citep{Jacobson2009,Jacobson2014}. The dark shaded regions show the range of
density value at every radius that lie between the 16th and 84th percentiles
(roughly the sample's ``one sigma'' spread) and the light shaded regions show
value between the 2nd and 98th percentiles, roughly the sample's ``two sigma''
spread.}
\label{fig:un_envelope}
\end{figure*}

\begin{figure*}[tb!]
\gridline{
\leftfig{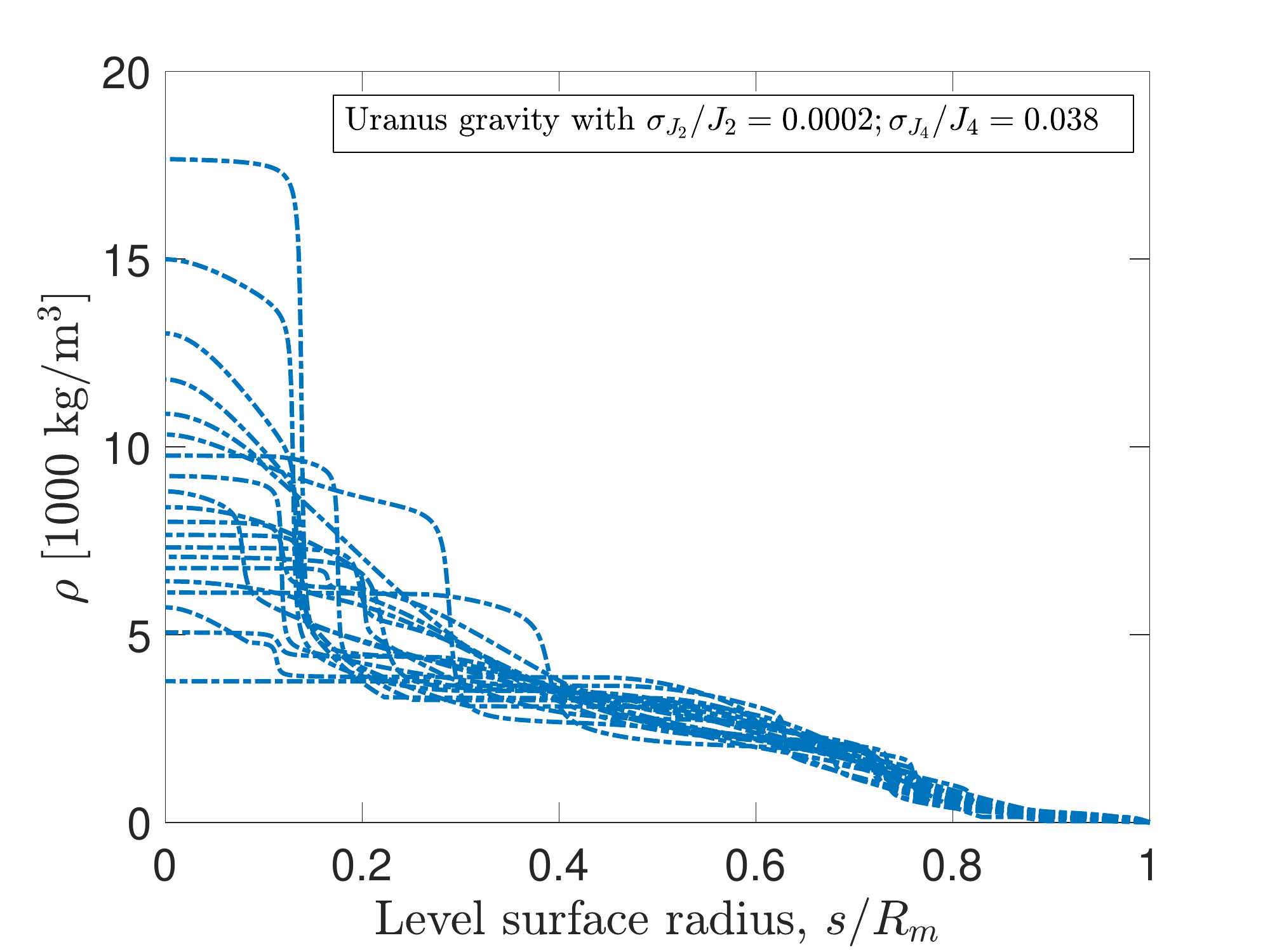}{0.5\textwidth}{}
\rightfig{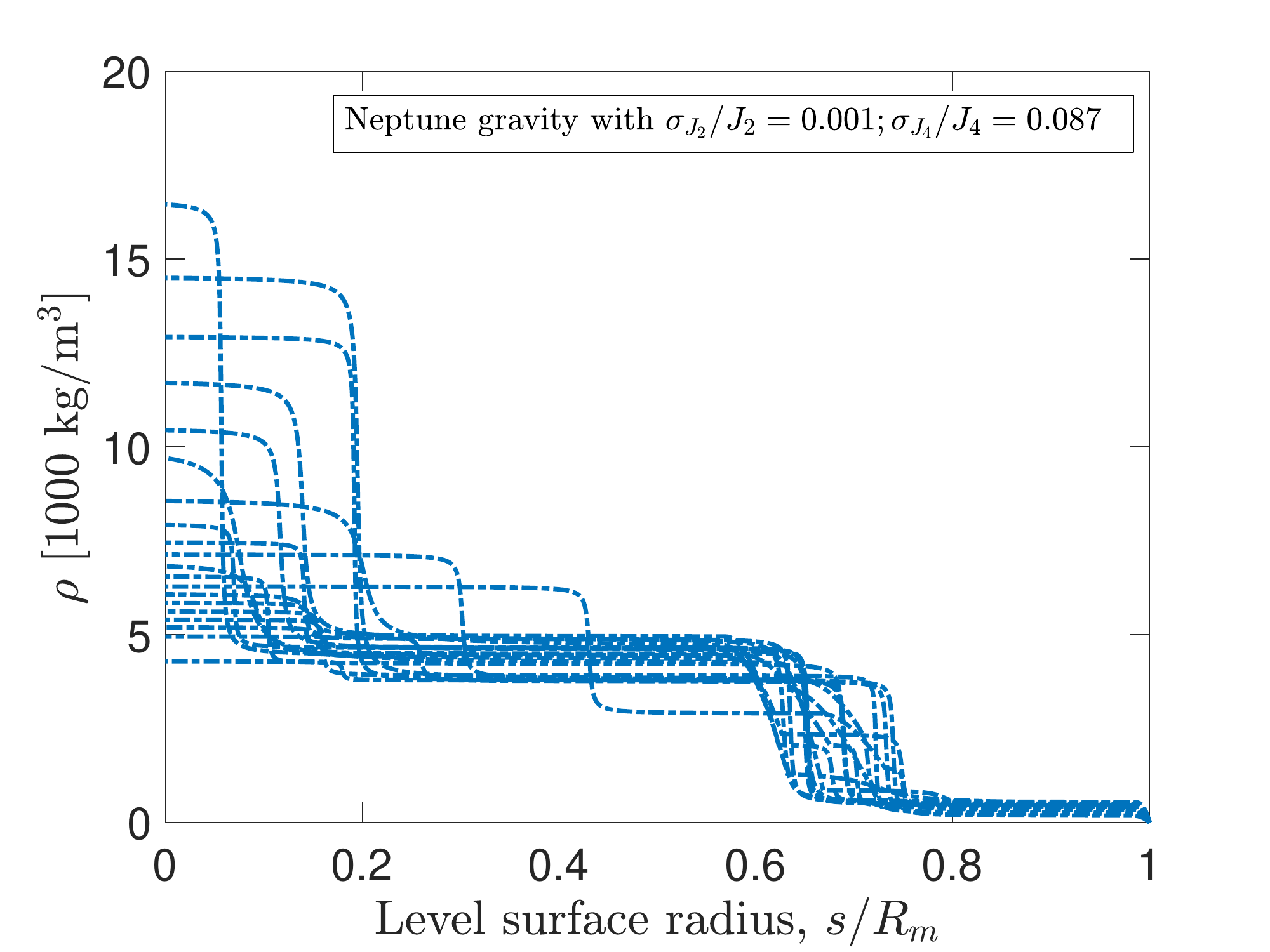}{0.5\textwidth}{}
}
\vspace{-2.0\baselineskip}
\caption{Ensemble view of sampled $\roofs$ profiles of Uranus (left) and
Neptune (right) matching currently available observables.}
\label{fig:un_ensemble}
\end{figure*}

\begin{figure*}[tb!]
\gridline{
\leftfig{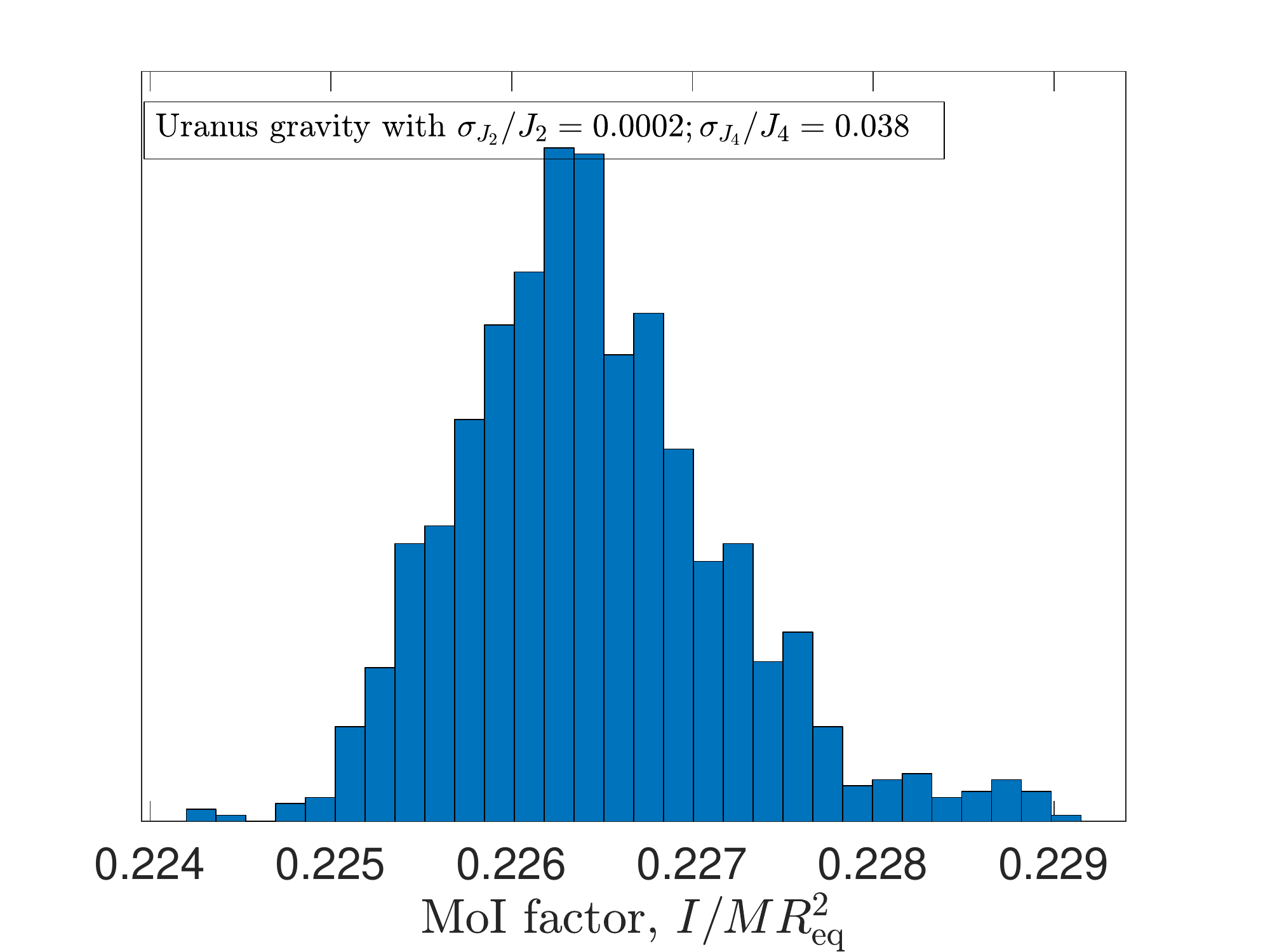}{0.5\textwidth}{}
\rightfig{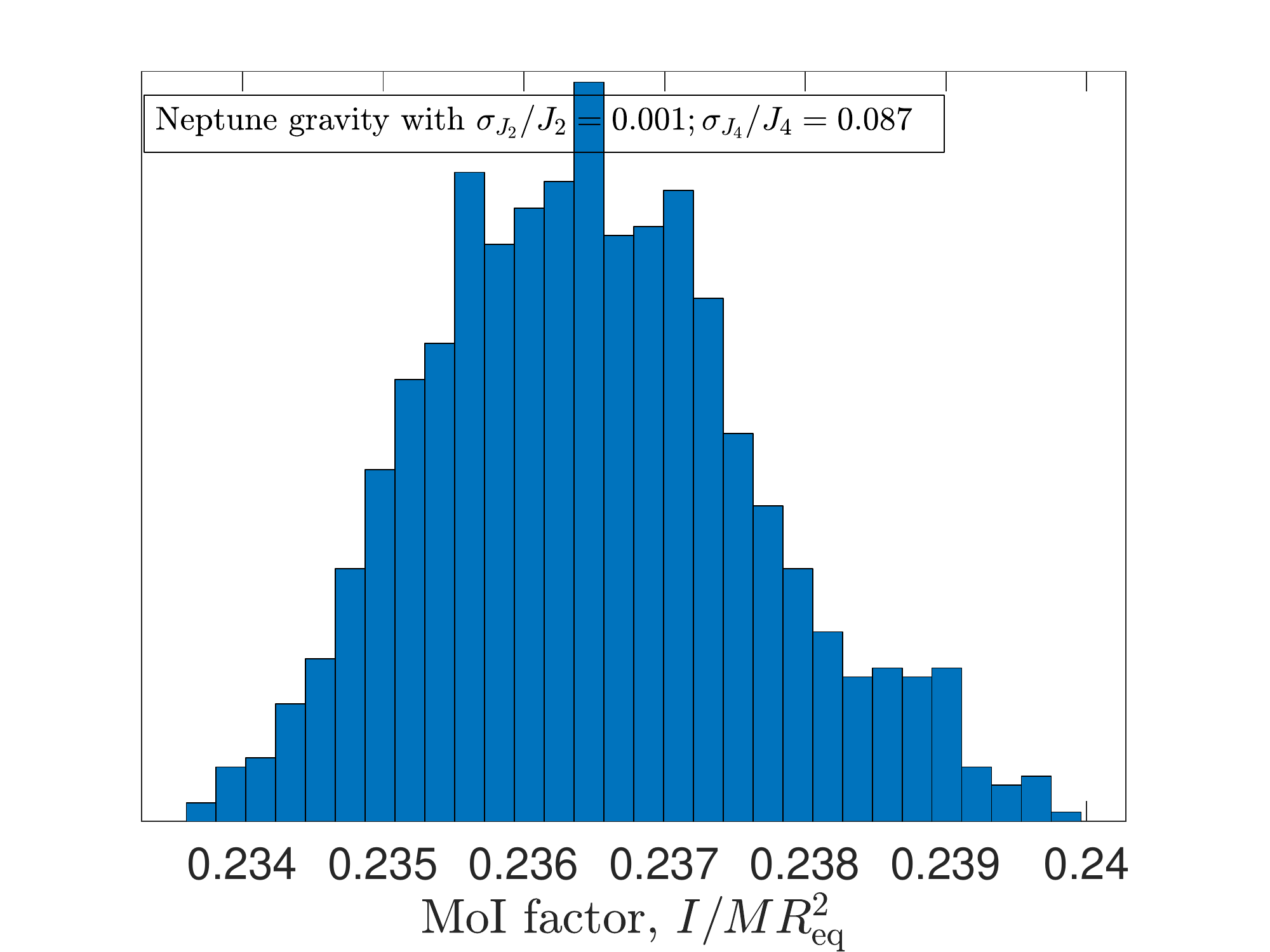}{0.5\textwidth}{}
}
\vspace{-2.0\baselineskip}
\caption{Moment of inertia factor histograms from sampled profiles of Uranus
(left) and Neptune (right) matching currently available observables.}
\label{fig:un_moi}
\end{figure*}

Figure~\ref{fig:un_envelope} compares the planets in envelope view. The more
accurately determined gravity of Uranus, compared with that of Neptune
(Table~\ref{tab:observables}) allows a much tighter constraint of density in the
upper envelope but, as expected, does not help with the deep interior. A similar
picture is evident with the ensemble view, Figure~\ref{fig:un_ensemble}, while
the MoI histograms in Figure~\ref{fig:un_moi} allow a prediction: the range of
possible values of moment of inertia factor, should it ever be independently
measured,{ is $0.225\lesssim{I/MR^2}\lesssim{0.229}$ for Uranus, and
$0.234\lesssim{I/MR^2}\lesssim{0.239}$ for Neptune.}

\begin{figure*}[tb!]
\gridline{
\leftfig{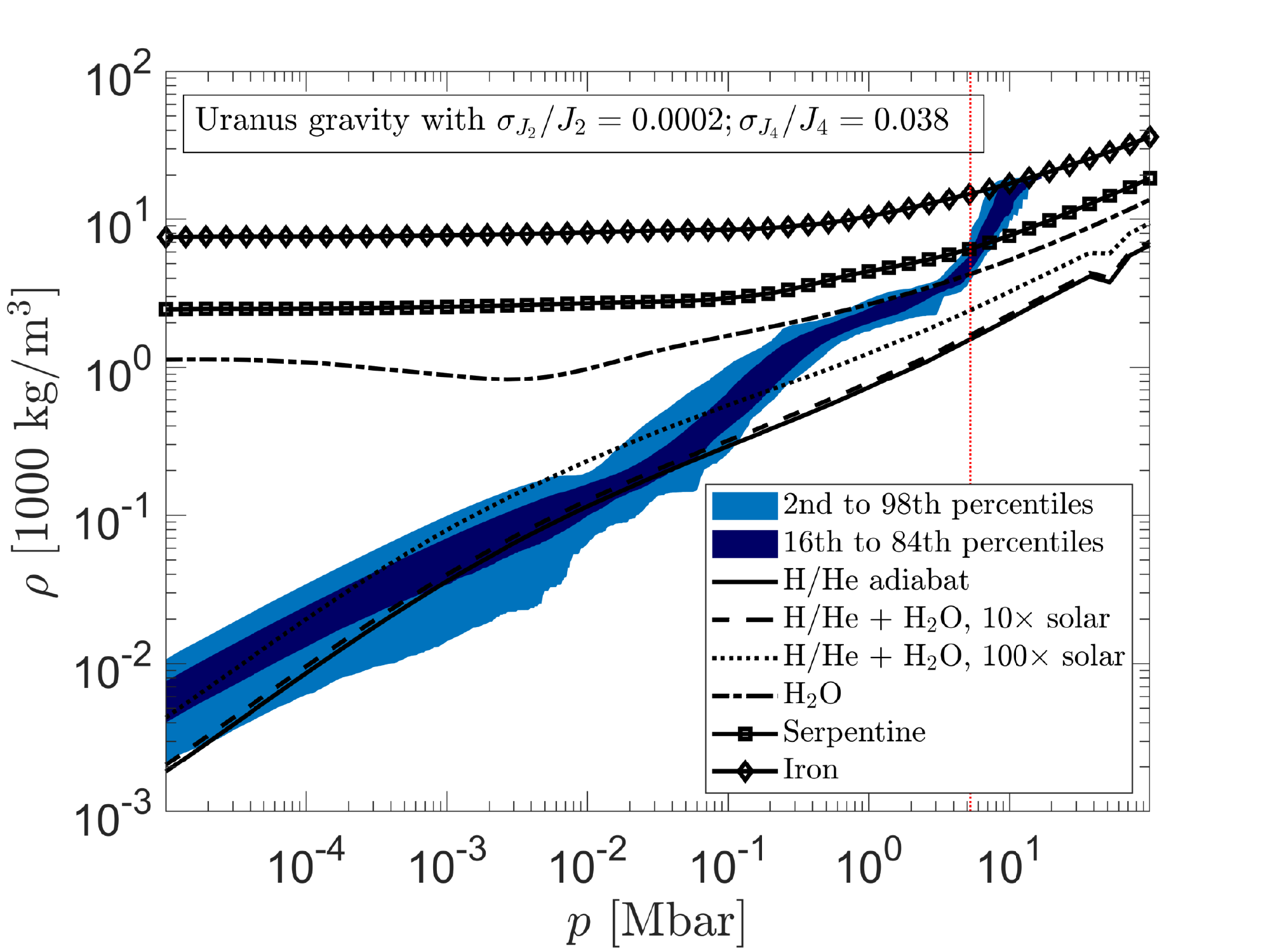}{0.5\textwidth}{}
\rightfig{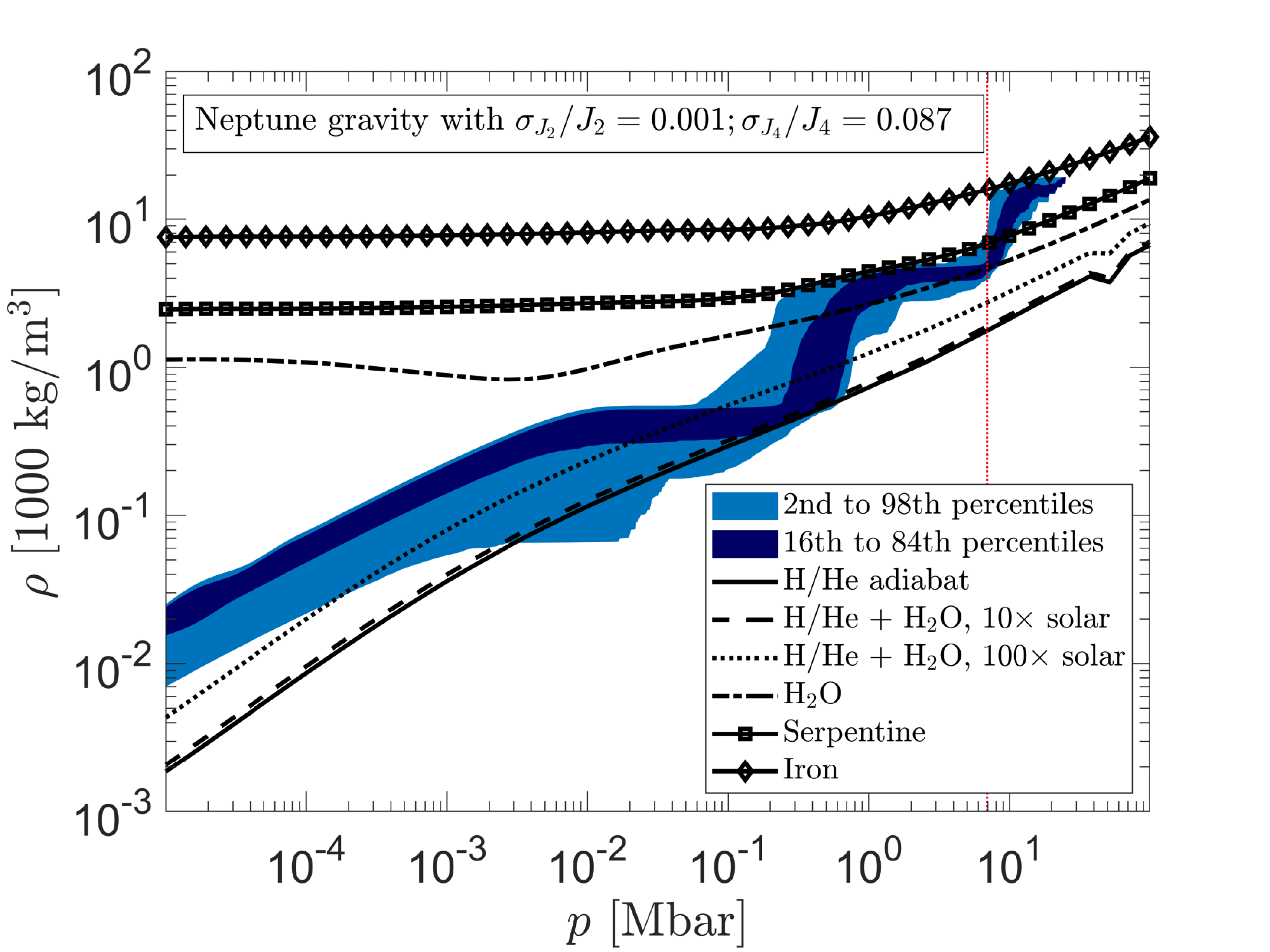}{0.5\textwidth}{}
}
\vspace{-2.0\baselineskip}
\caption{Barotrope view of sampled profiles of Uranus (left) and Neptune (right)
matching currently available observables. {The thin vertical line marks
the sample's minimum value of central pressure.}}
\label{fig:un_barotrope}
\end{figure*}

The barotrope view, Figure~\ref{fig:un_barotrope}, allows us to make some
statements about the planets' compositions, albeit only very generally. Both
planets are too dense to \emph{not} include significant amounts of heavy
elements. No surprise there. But it seems that Neptune's entire envelope must be
enriched with He or heavier elements, while a large (in pressure) fraction of
Uranus is consistent with a solar composition H/He mixture, or even a somewhat
helium poor atmosphere. This dichotomy is consistent with the one found in the
models of \citet[][their Table 2]{Nettelmann2013b}, where the envelope of both
planets was assumed to have a constant H/He ratio and much higher metallicity
was required in Neptune's envelope compared with Uranus. Observational
constraints on the atmospheric abundances in both planets, from spectral data,
are inconclusive, showing similar C:H ratios in both planets, for example, but
detecting signatures of CO and HCN in the atmosphere of Neptune but
not of Uranus \citep{Gautier1995}. Clearly, this would be one of the more
important observations to improve upon should the opportunity arise.

It also appears, from Figure~\ref{fig:un_barotrope}, that both planets
{allow for} significant amounts of molecules heavier than water in their
central regions, suggesting the existence of rocky cores. While the above
interpretations are admittedly loose, and can probably be made more robust by
comparisons with additional adiabatic and non-adiabatic barotropes of different
compositions, they have the benefit of not being strongly model dependent. No
assumptions at all about composition or temperature profile were made in
obtaining the samples, and hence the blue shaded areas in
Figure~\ref{fig:un_barotrope}. The generality of the solutions is limited only
by the flexibility of the parameterization, eq.~\eqref{sec:params}, and the
thoroughness of the sampling procedure (appendix~\ref{sec:chains}), and the
implicit assumptions of hydrostatic equilibrium and rigid rotation at a
well-known rotation rate.

\subsection{A realistic scenario for future tight gravity field
constraints\label{sec:j26}}

In sec.~\ref{sec:windtalk} we discussed how making use of high-order gravity,
$J_n\obs{}$, would be impossible without some robust estimate of the effect of
{non-rigid rotation}, $J_n\wind{}$. For this reason we obtained one more
sample, for each planet, with the assumption of well known $J_2$, $J_4$, $J_6$,
and rotation period $P$, but unknown $J_n$ for $n>6$. These constraints
correspond to a plausible scenario in which a future mission is able to obtain
high-precision gravity and rotation period but uncertainty about the effects of
deep zonal winds renders the higher order $J_n$ unusable. This is similar to the
current situation for Saturn where large differential rotation strongly affects
the even $J_n$ deduced from \emph{Cassini} Grand Finale data and dynamical wind
models must therefore be used to predict a $J_n\wind{}$ correction
\citep{Iess2019}.

\begin{figure*}[tb!]
\gridline{
\leftfig{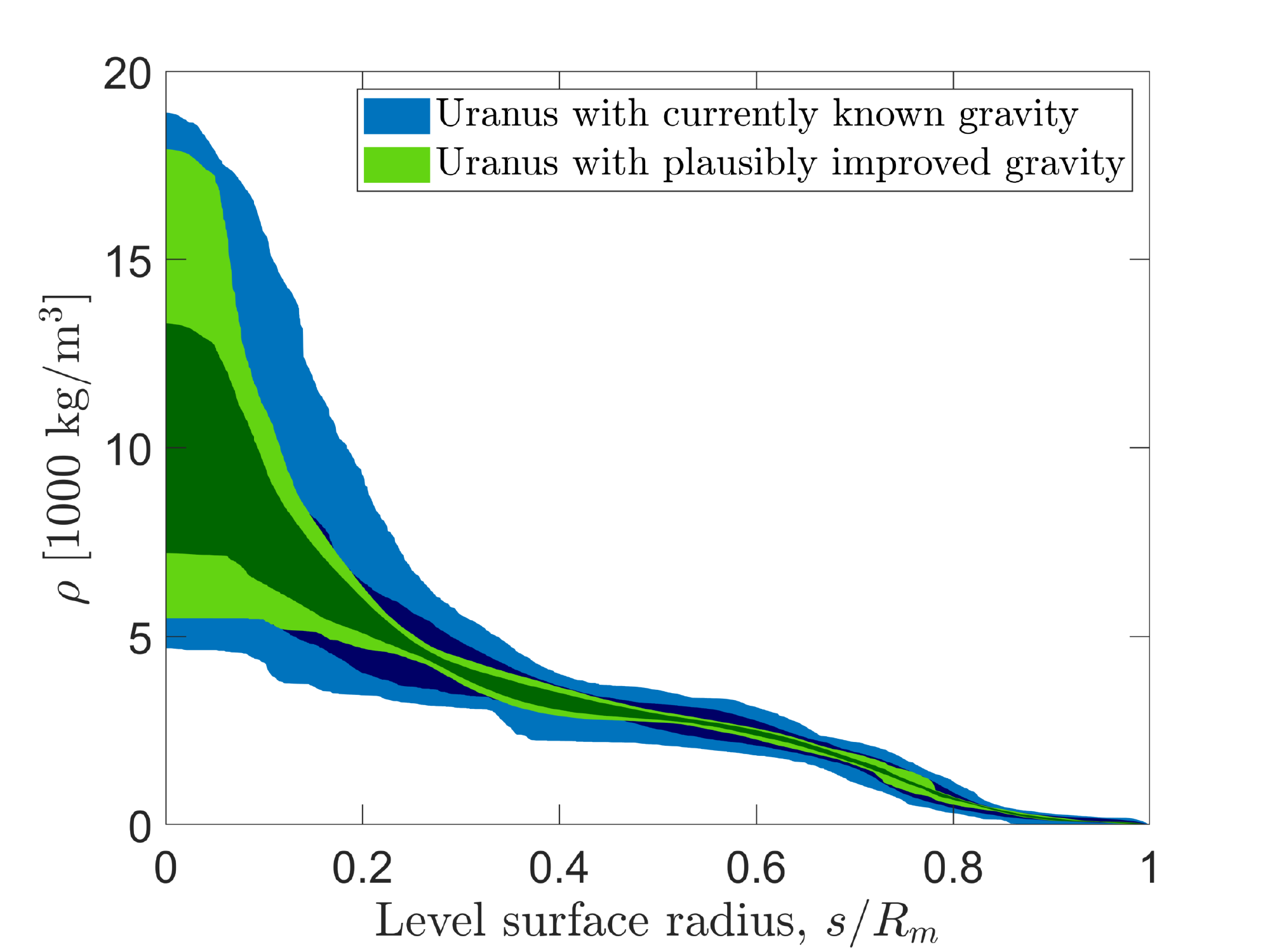}{0.5\textwidth}{}
\rightfig{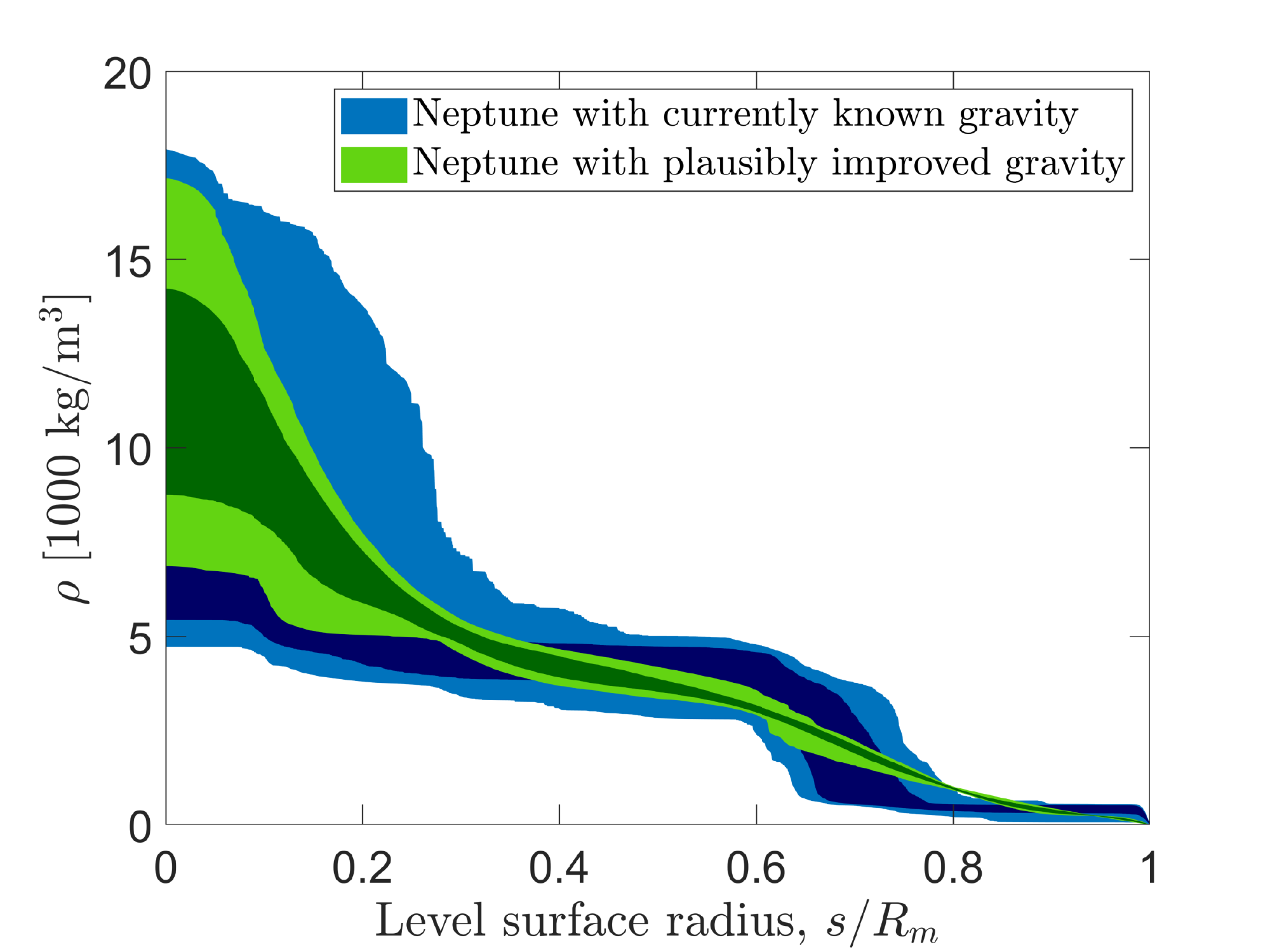}{0.5\textwidth}{}
}
\vspace{-2.0\baselineskip}
\caption{Envelope view of Uranus (left) and Neptune (right). Green shaded areas
are samples obtained with the constraints of the scenario of sec.~\ref{sec:j26},
showing the 1-sigma (dark green) and 2-sigma (light green) sample range, and
overlain on the samples of Fig.~\ref{fig:un_ensemble} (blue) for comparison.}
\label{fig:unj26_envelope}
\end{figure*}

\begin{figure*}[tb!]
\gridline{
\leftfig{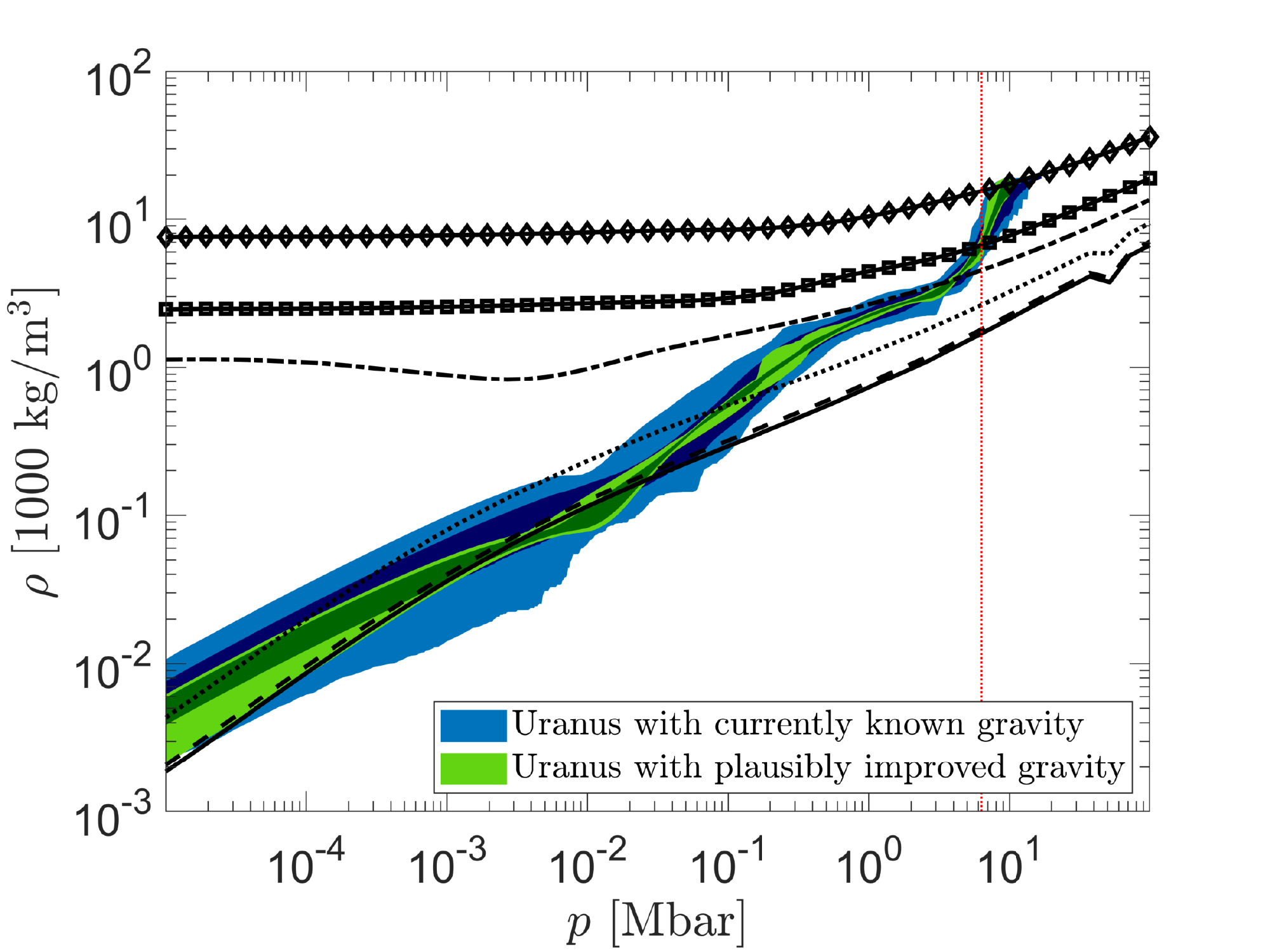}{0.5\textwidth}{}
\rightfig{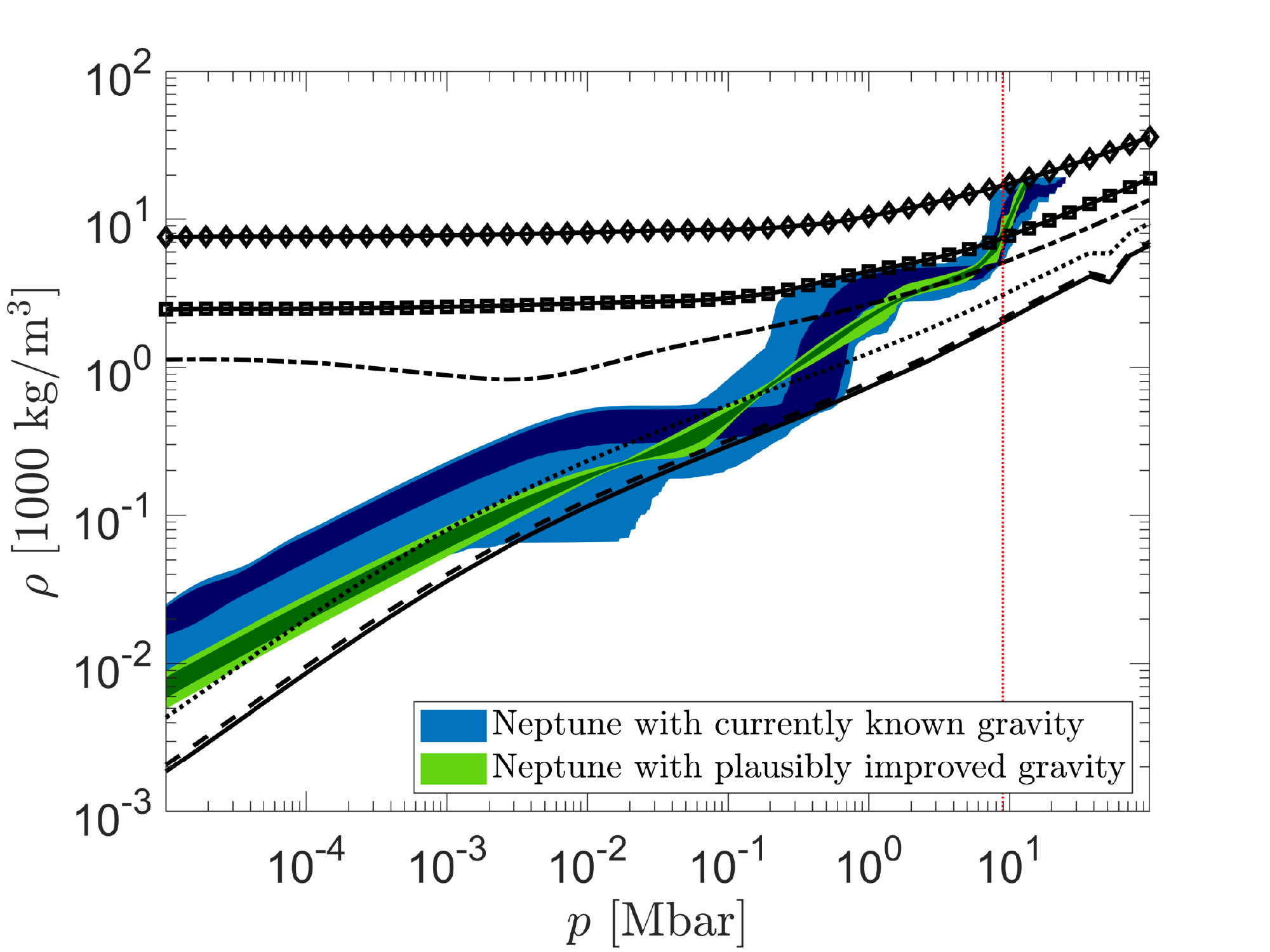}{0.5\textwidth}{}
}
\vspace{-2.0\baselineskip}
\caption{Barotrope view of Uranus (left) and Neptune (right). Green shaded area
are samples obtained with the constraints of the scenario of sec.~\ref{sec:j26},
showing the 1-sigma (dark green) and 2-sigma (light green) sample range, and
overlain on the samples of Fig.~\ref{fig:un_barotrope} (blue) for comparison.
{The thin vertical line marks the improved sample's minimum value of
central pressure.}}
\label{fig:unj26_barotrope}
\end{figure*}

With such a data set, how would our knowledge of the interior density profile
for either Uranus or Neptune be improved? We find the improvement would be quite
significant. Figure~\ref{fig:unj26_envelope} shows the sampled profiles of
density vs.~radius, while Figure~\ref{fig:unj26_barotrope} shows density
vs.~pressure, which is potentially the most illustrative.

There are several aspects to note. In particular, the metallicity of the outer
H/He envelope ($P<0.1$ Mbar) could potentially be reliably assessed, given any
consistency or inconsistency with a solar metallicity H/He adiabatic density
profile. In addition, a metallicity enhancement that would deviate from a
uniform enhanced metallicity could potentially be ``seen'' with gravity data.
This would be more readily done at higher metallicities that deviate strongly
from solar.

The long-standing question of whether the ``middle'' layers of Uranus and
Neptune are exclusively made of water and other ``ices'', or instead have less
water but more H/He and rocks to give a similar density, may not be answerable,
directly. However, the pressures at which changes in density structures occur,
and the ``slope'' in density vs.~pressure space of density profiles may well
allow for plausible explanations for any constrained density profiles to be
determined. Such profiles could be connected directly to predicted composition
profiles from planet formation \citep{Helled2017,Ormel2021} and thermal
evolution models \citep{Vazan2020,Scheibe2021,Stixrude2021}. Clearly, the
constraints on interior models would not only become much tighter than possible
with currently available data but also tight enough in absolute terms to
distinguish between different temperature/composition profiles.

\newpage
\section{Summary and conclusions}\label{sec:discussion}
In the previous sections we described an experiment designed to gauge the
ability of a precise measurement of a planet's gravity to constrain the possible
distribution of mass in its interior. Gravity is a long range force and the
planet's gravitational potential at any exterior point is determined by an
integral of the mass density over the entire planet so that, in principle,
knowledge of one should inform on the other. In practice the gravitational
potential can be measured with varying degrees of uncertainty: crudely, with
the aid of natural satellites; better by tracking a flying-by spacecraft; with
potentially exquisite precision by a dedicated orbiting mission. There is no
simple formula connecting the degree of precision of the observed gravity,
described by a set of expansion coefficients $J_n$ and associated uncertainty
$\sigma_{J_n}$, with the degree of constraint on the implied density
distribution $\roofs{}$. Our main goal was to provide this connection by
generating samples from the space of allowable $\roofs{}$ limited by
successively higher order and higher precision gravity coefficients.

Examining the samples presented in sec.~\ref{sec:results} led us to the
conclusions summarized below. Many of these are unsurprising, being predictable
at least qualitatively from the nature of the gravity integral. Nevertheless we
find useful the more direct and more quantitative demonstration made possible by
the sampling framework.

\begin{enumerate}
\item Even the most crude estimation of a planet's gravitational potential, say
a measurement of only the $J_2$ coefficient to within one percent
(Fig.~\ref{fig:lowj}) narrows down considerably the space of allowable density
profiles, compared with a baseline constrained only by the planet's mass,
radius, and boundary conditions. This narrowing down is most evident in the
upper ten to twenty percent (by radius) of the planet's interior, very quickly
disappearing with depth.

\item The degree of constraint applied to the interior $\roofs{}$ by a
gravity measurement (a set of $J_n, \sigma_{J_n}, {n\le{n\sub{max}}}$) can be
loosely quantified by the width of the distribution of associated moment of
inertia values, a scalar quantity integrated from (and sensitive to!) $\roofs{}$
over the equilibrium shape of the planet.

\item More precisely known values of the low-order gravity coefficients are not
nearly as useful as adding even crudely measured values of higher-order
coefficients. Compare figures~\ref{fig:lowj} and~\ref{fig:hij} and especially
the rightmost panel in each. This fact presents a difficulty however, as
higher-order coefficients are increasingly sensitive to, eventually dominated by
dynamic effects not captured by simple rigid rotation rate, such as zonal winds
and deep differential rotation. To be useful, these dynamic effects must be
accounted for.

\item If high precision or high order gravity is to be used to constrain
interior models then the planet's rotation period \emph{must} also be known to
comparable precision. Compare figures~\ref{fig:hij} and~\ref{fig:hijfixrot}.
This point is worth emphasizing since neglecting the uncertainty in rotation
period can lead to overly confident predictions.

\item No level of precision and completeness of characterization of the gravity
field and rotation state can be expected to pin down, by itself, the density at
the center of a planet. Figure~\ref{fig:hijfixrot}, yellow shaded area. To do
better than a factor of two or more will require making additional assumptions.

\item A measurement of a planet's moment of inertia factor, in
addition to and independent of the gravity field measurement, can potentially
assist with constraining the interior mass distribution, but it would have to be
a quite precise one. When only the low-order $J_2$ and $J_4$ are known and only
to rough precision, as is the case presently for Uranus and Neptune, the
distribution of the correlated MoI values is already constrained to a large
degree: { about $0.6\%$ in the case of Uranus and about $1\%$ for
Neptune}. If higher order $J_n$ become known (and recall that this implies also
precise determination of the rotation period) then the MoI is essentially fixed
and can provide no further information.
\end{enumerate}

A second goal of this work was to look at the presently available gravity field
estimate of the planets Uranus and Neptune and see what predictions can be made
about their interiors that would be, as much as possible, immune to implicit
model assumptions and to uncertainties in the equations of state {of
Hydrogen, Helium, and heavier elements}. Unsurprisingly, these predictions are
general in nature and cannot replace detailed models. Nevertheless they
illustrate the potential of more complete characterization of the gravity
fields, should one become available, to better direct such models.

\begin{enumerate}
\item As is well known, both Uranus and Neptune are much too dense to \emph{not}
include significant quantities of elements heavier than helium. In their central
regions, both planets appear to {allow} significant enrichment by
components denser than H$_2$O. To say anything more about the denser components
will require a fuller characterization of the gravity field or making more
detailed model assumptions or, very likely, both.

\item A fraction of Uranus' envelope is consistent with an adiabatic region of
H/He {at solar atmospheric abundances}. Neptune's envelope however is
not, and should be significantly metal-enriched or perhaps, somehow, He-rich.

\item An orbiter mission to better characterize the gravity field and rotation
of either or both planets would be of very high value. Even if high-order
gravity ($J_{n>6}$) cannot be reliably separated into hydrostatic and dynamic
parts, the interior barotrope (density vs.~pressure profile) could be tightly
constrained. This would allow for direct comparisons of profiles of composition
vs.~depth from formation and evolution models, opening a new era into our
understanding of Uranus and Neptune.
\end{enumerate}


\section*{}
We would like to acknowledge the support of NSF grant AST 1908615 and NASA grant
80NSSC19K1286. Resources supporting this work were provided by the \emph{lux}
supercomputer at UC Santa Cruz, funded by NSF MRI grant AST 1828315. {We
thank two anonymous reviewers for the thorough reading of the manuscript and
insightful comments.}


\appendix

\section{Sampling procedure}\label{sec:chains}
To draw a sample of $\roofs$ profiles for each planet we run the ensemble
sampler of the \code{emcee} Python package \citep{Foreman-Mackey2013} which
implements the parallel stretch-move algorithm of \citet{Goodman2010}. The goal
is a draw of a large enough sample of independent realizations of the model
parameters (eq.~\ref{eq:ppwd}) from the probability distribution dictated by the
likelihood function (eq.~\ref{eq:Jloss}), for each combination of observables
and uncertainties ($n\sub{max}$ and $\nu_n$).

\subsection{Variable transformations}
In practice, sampling is often preceded by some isomorphic transformation of the
physical model parameters into equivalent random variables whose probability
distribution is predicted to be, in some sense, smoother and thus easier for the
sampling algorithm to work with. The particular transformations are determined
by analysis or, more often, by trial-and-error or by advice or general common
wisdom suggestions. In the best cases a simple transformation, for example using
a logarithm of a nondimensionlized variable, can lead to huge gains in
efficiency. Even in the worst cases these transformations are always benign
(barring any programming bugs or math errors) so they are often left unmentioned
in publication, but in order to facilitate exact reproducibility here we give
the exact parameters used by the sampling algorithm.
Table~\ref{tab:samplingspace} defines the correspondence between the
physical-space parameters (eq.~\ref{eq:ppwd}, reproduced here for convenience)
and the sample-space parameters explored by the MCMC chains. We make no claim
that these are optimal or necessary.
\begin{equation}
\roofz = \rho(s/\rem) = \sum_{n=2}^{8}a_n(z^n - 1) + \rho_0 +
\sum_{n=1}^{2}\frac{\sigma_n}{\pi}\Bigl(\frac{\pi}{2} +
\arctan\bigl(-\nu_n(z - z_n)\bigr)\Bigr).
\end{equation}


\begin{deluxetable}{LLLL}\label{tab:samplingspace}

\tablecaption{Sampling space parameters and their priors}

\tablehead{\colhead{physical parameter (eq.~\ref{eq:ppwd})} & \colhead{prior} &
\colhead{sampling parameter} & \colhead{prior}}

\startdata
z_1 & U(0.05,0.5) & y_1=\logit(z_1) &
\propto{}(e^y)/(1 + e^y)^2\cdot{}U(\logit(0.05),\logit(0.5)) \\
\sigma_1 & U(0,\romax) & y_2=\ln(\sigma_1) &
\propto{}(e^y)\cdot{}U(-20,\ln(\romax)) \\
\nu_1 & U(20,1000) & y_3=\ln(\nu_1) &
\propto{}(e^y)\cdot{}U(\ln(20),\ln(1000)) \\
z_2 & U(0.5,0.85) & y_4=\logit(z_2) &
\propto{}(e^y)/(1 + e^y)^2\cdot{}U(\logit(0.5),\logit(0.85)) \\
\sigma_2 & U(0,\romax) & y_5=\ln(\sigma_2) &
\propto{}(e^y)\cdot{}U(-20,\ln(\romax)) \\
\nu_2 & U(20,1000) & y_6=\ln(\nu_2) &
\propto{}(e^y)\cdot{}U(\ln(20),\ln(1000)) \\
a_8,a_7,\ldots{a_2} & U(-10^7,10^7) & y_7,y_8,\ldots{y_{13}} & U(-10^7,10^7) \\
\enddata
\tablenotetext{}{$\logit(x)=\ln(x)-\ln(1-x)$}
\tablenotetext{}{transformed priors conserve probability mass: $p(x)dx=p(y)dy$}
\end{deluxetable}

\subsection{Seeds}
In theory the seeds used to initiate the walkers of the ensemble sampler are
completely unimportant, as each chain is assumed to run long enough to fully
``forget'' its initial state. A simple choice was to seed all walkers with the
simplest possible density profile consistent with the planet's mass and radius.
This would be, in our parameterization, something like
\begin{equation}
\V{y}=(0,-20,100,0,-20,100,0,0,0,0,0,0,
\frac{5}{2}\rho_0 - \frac{15}{8\pi}\frac{M}{R^3}).
\end{equation}
These values create a single quadratic in $z=r/\rem{}$ starting at $\rho=\rho_0$
at the surface and implying, for a spherical planet of radius $R$, a total mass
$M$. Starting the sampling from this bland, featureless seed the likelihood
function should produce density profiles with interesting features (i.e.
steeper gradients and sharp discontinuities) if and where they are ``preferred''
by the data. This is, in fact, exactly what happens, very slowly. The mixing
rate between walkers in our samples is very low, probably due to correlations
between parameters.

\begin{figure}[tb!]
\centering
\plotone{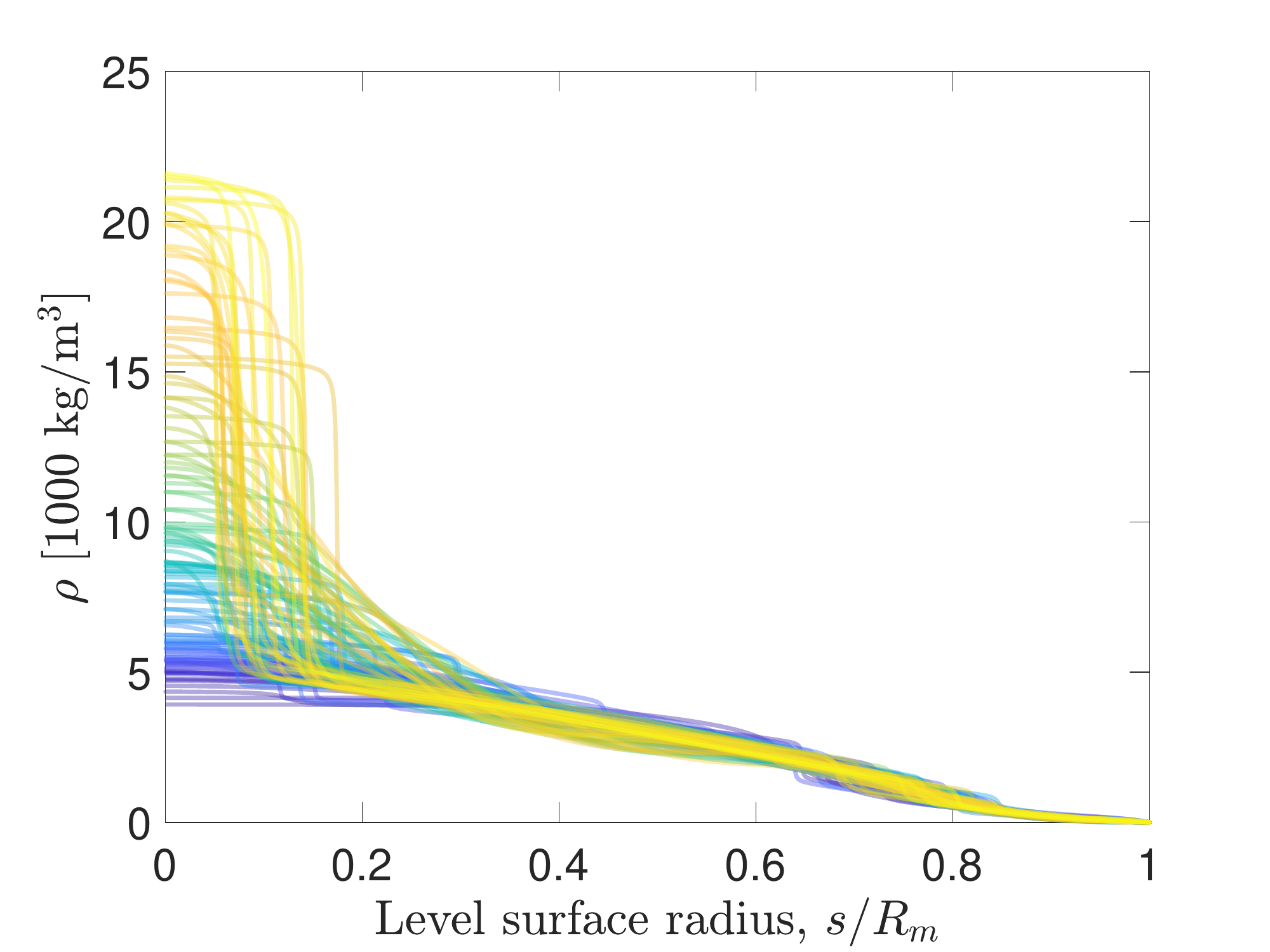}
\caption{Density profiles of a typical ensemble seed state.}
\label{fig:seedsample}
\end{figure}

In order to speed up the generation of new samples we save and reuse density
profiles (parameter values) from already generated samples. A new ensemble of
$n_w$ walkers is seeded with a random choice of $n_w$ items from a large seed
bank of diverse profiles, and may start out looking like the example in
Figure~\ref{fig:seedsample}. While this doesn't solve the problem of slow
mixing, once the chains are long enough to forget their initial state (longer
than their autocorrelation time) the final links from each chain are an
independent sample of size $n_w$.

\vspace{-4mm}
\subsection{Tempering}
To improve inter-walker mixing we use a common idea known as tempering. The
ensemble is run for a while under an artificially widened likelihood function,
designed to lower the peaks and raise the valleys in the likelihood landscape.
In our case this is equivalent to assuming very poorly known values for the
planet's observables (gravity, mass, rotation period). Under this likelihood
function walkers can readily mix and do not get stuck in local regions of high
likelihood. After a few autocorrelation times (much shorter now) the ensemble
walkers are well mixed and are spread out in parameter space following the
posterior distribution, but of the modified likelihood. When the real likelihood
is then applied, this effectively creates an ideal seed state for the ensemble.
When the walkers then continue under the real likelihood they explore their
local neighborhood more slowly, and would get stuck if they find a
high-likelihood peak. But at this point the last link from each walker is
collected and the process restarts, {repeating the cycle as many times
as necessary to produce a final sample of 1000 independent states.}

We believe but did not mathematically prove that this procedure satisfies the
requirement of detailed balance, which makes it an MCMC algorithm and guarantees
that each $\roofs{}$ appears in the final sample in proportion to its frequency
in the unknown posterior distribution. More important for the purpose of this
work, however, is that the procedure is able to efficiently explore the entire
parameter space.

\bibliography{library}{}

\begin{thebibliography}{}
\expandafter\ifx\csname natexlab\endcsname\relax\def\natexlab#1{#1}\fi
\providecommand{\url}[1]{\href{#1}{#1}}
\providecommand{\dodoi}[1]{doi:~\href{http://doi.org/#1}{\nolinkurl{#1}}}
\providecommand{\doeprint}[1]{\href{http://ascl.net/#1}{\nolinkurl{http://ascl.net/#1}}}
\providecommand{\doarXiv}[1]{\href{https://arxiv.org/abs/#1}{\nolinkurl{https://arxiv.org/abs/#1}}}

\bibitem[{Anderson \& Schubert(2007)}]{Anderson2007a}
Anderson, J.~D., \& Schubert, G. 2007, Science (80-. )., 317, 1384,
  \dodoi{10.1126/science.1144835}

\bibitem[{Bailey \& Stevenson(2021)}]{Bailey2021}
Bailey, E., \& Stevenson, D.~J. 2021, Planet. Sci. J., 2, 64,
  \dodoi{10.3847/PSJ/abd1e0}

\bibitem[{Bolton {et~al.}(2017)Bolton, Adriani, Adumitroaie, Allison, Anderson,
  Atreya, Bloxham, Brown, Connerney, DeJong, Folkner, Gautier, Grassi, Gulkis,
  Guillot, Hansen, Hubbard, Iess, Ingersoll, Janssen, Jorgensen, Kaspi, Levin,
  Li, Lunine, Miguel, Mura, Orton, Owen, Ravine, Smith, Steffes, Stone,
  Stevenson, Thorne, Waite, Durante, Ebert, Greathouse, Hue, Parisi, Szalay, \&
  Wilson}]{Bolton2017}
Bolton, S., Adriani, A., Adumitroaie, V., {et~al.} 2017, Science (80-. )., 356,
  821, \dodoi{10.1126/science.aal2108}

\bibitem[{Debras \& Chabrier(2017)}]{Debras2017}
Debras, F., \& Chabrier, G. 2017, A{\&}A, 97,
  \dodoi{10.1051/0004-6361/201731682}

\bibitem[{Durante {et~al.}(2020)Durante, Parisi, Serra, Zannoni, Notaro,
  Racioppa, Buccino, Lari, {Gomez Casajus}, Iess, Folkner, Tommei, Tortora, \&
  Bolton}]{Durante2020}
Durante, D., Parisi, M., Serra, D., {et~al.} 2020, Geophys. Res. Lett., 47,
  \dodoi{https://doi.org/10.1029/2019GL086572}

\bibitem[{Fletcher {et~al.}(2020)Fletcher, Helled, Roussos, Jones, Charnoz,
  Andr{\'{e}}, Andrews, Bannister, Bunce, Cavali{\'{e}}, Ferri, Fortney,
  Grassi, Griton, Hartogh, Hueso, Kaspi, Lamy, Masters, Melin, Moses, Mousis,
  Nettelmann, Plainaki, Schmidt, Simon, Tobie, Tortora, Tosi, \&
  Turrini}]{Fletcher2020}
Fletcher, L.~N., Helled, R., Roussos, E., {et~al.} 2020, Planet. Space Sci.,
  191, 105030, \dodoi{https://doi.org/10.1016/j.pss.2020.105030}

\bibitem[{Foreman-Mackey {et~al.}(2013)Foreman-Mackey, Hogg, Lang, \&
  Goodman}]{Foreman-Mackey2013}
Foreman-Mackey, D., Hogg, D.~W., Lang, D., \& Goodman, J. 2013, Publ. Astron.
  Soc. Pacific, 125, 306, \dodoi{10.1086/670067}

\bibitem[{Galanti \& Kaspi(2017)}]{Galanti2017}
Galanti, E., \& Kaspi, Y. 2017, Astrophys. J., 843, L25,
  \dodoi{10.3847/2041-8213/aa7aec}

\bibitem[{Galanti {et~al.}(2019)Galanti, Kaspi, Miguel, Guillot, Durante,
  Racioppa, \& Iess}]{Galanti2019}
Galanti, E., Kaspi, Y., Miguel, Y., {et~al.} 2019, Geophys. Res. Lett., 616,
  \dodoi{10.1029/2018GL078087}

\bibitem[{Gautier {et~al.}(1995)Gautier, Conrath, Owen, de~Pater, \&
  Atreya}]{Gautier1995}
Gautier, D., Conrath, B.~J., Owen, T., de~Pater, I., \& Atreya, S.~K. 1995, in
  Neptune Trit., 547--612

\bibitem[{Goodman \& Weare(2010)}]{Goodman2010}
Goodman, J., \& Weare, J. 2010, Commun. Appl. Math. Comput. Sci., 5, 65

\bibitem[{Gurnett {et~al.}(2007)Gurnett, Persoon, Kurth, Groene, Averkamp,
  Dougherty, \& Southwood}]{Gurnett2007}
Gurnett, D.~A., Persoon, A.~M., Kurth, W.~S., {et~al.} 2007, Science (80-. ).,
  316, 442, \dodoi{10.1126/science.1138562}

\bibitem[{Helled(2011)}]{Helled2011c}
Helled, R. 2011, Astrophys. J. Lett., 735, \dodoi{10.1088/2041-8205/735/1/L16}

\bibitem[{Helled {et~al.}(2011{\natexlab{a}})Helled, Anderson, Podolak, \&
  Schubert}]{Helled2011b}
Helled, R., Anderson, J.~D., Podolak, M., \& Schubert, G. 2011{\natexlab{a}},
  Astrophys. J., 726, \dodoi{10.1088/0004-637X/726/1/15}

\bibitem[{Helled {et~al.}(2010)Helled, Anderson, \& Schubert}]{Helled2010}
Helled, R., Anderson, J.~D., \& Schubert, G. 2010, Icarus, 210, 446,
  \dodoi{10.1016/j.icarus.2010.06.037}

\bibitem[{Helled {et~al.}(2011{\natexlab{b}})Helled, Anderson, Schubert, \&
  Stevenson}]{Helled2011a}
Helled, R., Anderson, J.~D., Schubert, G., \& Stevenson, D.~J.
  2011{\natexlab{b}}, Icarus, 216, 440, \dodoi{10.1016/j.icarus.2011.09.016}

\bibitem[{Helled \& Fortney(2020)}]{Helled2020b}
Helled, R., \& Fortney, J.~J. 2020, Philos. Trans. R. Soc. A Math. Phys. Eng.
  Sci., 378, \dodoi{10.1098/rsta.2019.0474}

\bibitem[{Helled {et~al.}(2015)Helled, Galanti, \& Kaspi}]{Helled2015}
Helled, R., Galanti, E., \& Kaspi, Y. 2015, Nature, \dodoi{10.1038/nature14278}

\bibitem[{Helled {et~al.}(2020)Helled, Nettelmann, \& Guillot}]{Helled2020a}
Helled, R., Nettelmann, N., \& Guillot, T. 2020, Space Sci. Rev., 216, 38,
  \dodoi{10.1007/s11214-020-00660-3}

\bibitem[{Helled {et~al.}(2009)Helled, Schubert, \& Anderson}]{Helled2009}
Helled, R., Schubert, G., \& Anderson, J.~D. 2009, Icarus, 199, 368,
  \dodoi{10.1016/j.icarus.2008.10.005}

\bibitem[{Helled \& Stevenson(2017)}]{Helled2017}
Helled, R., \& Stevenson, D.~J. 2017, Astrophys. J. Lett., 840, L4,
  \dodoi{10.3847/2041-8213/aa6d08}

\bibitem[{Hofstadter {et~al.}(2019)Hofstadter, Simon, Atreya, Banfield,
  Fortney, Hayes, Hedman, Hospodarsky, Mandt, Masters, Showalter, Soderlund,
  Turrini, Turtle, Reh, Elliott, Arora, \& Petropoulos}]{Hofstadter2019}
Hofstadter, M., Simon, A., Atreya, S., {et~al.} 2019, Planet. Space Sci., 177,
  104680, \dodoi{10.1016/j.pss.2019.06.004}

\bibitem[{Hubbard(2013)}]{Hubbard2013a}
Hubbard, W. 2013, Astrophys. J., 768, 43

\bibitem[{Hubbard(2012)}]{Hubbard2012}
Hubbard, W.~B. 2012, Astrophys. J. Lett., 756, L15,
  \dodoi{10.1088/0004-637X/768/1/43}

\bibitem[{Hubbard {et~al.}(2014)Hubbard, Schubert, Kong, \&
  Zhang}]{Hubbard2014}
Hubbard, W.~B., Schubert, G., Kong, D., \& Zhang, K. 2014, Icarus, 242, 138,
  \dodoi{10.1016/j.icarus.2014.08.014}

\bibitem[{Iess {et~al.}(2018)Iess, Folkner, Durante, Parisi, Kaspi, Galanti,
  Guillot, Hubbard, Stevenson, Anderson, Buccino, Casajus, Milani, Park,
  Racioppa, Serra, Tortora, Zannoni, Cao, Helled, Lunine, Miguel, Militzer,
  Wahl, Connerney, Levin, \& Bolton}]{Iess2018}
Iess, L., Folkner, W.~M., Durante, D., {et~al.} 2018, Nature, 555, 220,
  \dodoi{10.1038/nature25776}

\bibitem[{Iess {et~al.}(2019)Iess, Militzer, Kaspi, Nicholson, Durante,
  Racioppa, Anabtawi, Galanti, Hubbard, Mariani, Tortora, Wahl, \&
  Zannoni}]{Iess2019}
Iess, L., Militzer, B., Kaspi, Y., {et~al.} 2019, Science (80-. )., 2965,
  eaat2965, \dodoi{10.1126/science.aat2965}

\bibitem[{Jacobson(2014)}]{Jacobson2014}
Jacobson, R. 2014, Astron. J., 148, 76, \dodoi{10.1088/0004-6256/148/5/76}

\bibitem[{Jacobson(2009)}]{Jacobson2009}
Jacobson, R.~A. 2009, Astron. J., 137, 4322,
  \dodoi{10.1088/0004-6256/137/5/4322}

\bibitem[{Kaspi {et~al.}(2013)Kaspi, Showman, Hubbard, Aharonson, \&
  Helled}]{Kaspi2013}
Kaspi, Y., Showman, A.~P., Hubbard, W.~B., Aharonson, O., \& Helled, R. 2013,
  Nature, 497, 344, \dodoi{10.1038/nature12131}

\bibitem[{Kaspi {et~al.}(2017)Kaspi, Guillot, Galanti, Miguel, Helled, Hubbard,
  Militzer, Wahl, Levin, Connerney, \& Bolton}]{Kaspi2017}
Kaspi, Y., Guillot, T., Galanti, E., {et~al.} 2017, Geophys. Res. Lett., 44,
  5960, \dodoi{10.1002/2017GL073629}

\bibitem[{Kaspi {et~al.}(2018)Kaspi, Galanti, Hubbard, Stevenson, Bolton, Iess,
  Guillot, Bloxham, Connerney, Cao, Durante, Folkner, Helled, Ingersoll, Levin,
  Lunine, Miguel, Militzer, Parisi, \& Wahl}]{Kaspi2018a}
Kaspi, Y., Galanti, E., Hubbard, W.~B., {et~al.} 2018, Nature, 555, 223,
  \dodoi{10.1038/nature25793}

\bibitem[{Lindal(1992)}]{Lindal1992}
Lindal, G.~F. 1992, Astrophys. J., 103, 967, \dodoi{10.1086/116119}

\bibitem[{Mankovich \& Fortney(2020)}]{Mankovich2020a}
Mankovich, C.~R., \& Fortney, J.~J. 2020, Astrophys. J., 889, 51,
  \dodoi{10.3847/1538-4357/ab6210}

\bibitem[{Mankovich \& Fuller(2021)}]{Mankovich2021}
Mankovich, C.~R., \& Fuller, J. 2021, Nat. Astron.,
  \dodoi{10.1038/s41550-021-01448-3}

\bibitem[{Mankovich {et~al.}(2019)Mankovich, Marley, Fortney, \&
  Movshovitz}]{Mankovich2019}
Mankovich, C.~R., Marley, M.~S., Fortney, J.~J., \& Movshovitz, N. 2019,
  Astrophys. J., 871, 1, \dodoi{10.3847/1538-4357/aaf798}

\bibitem[{Marley {et~al.}(1995)Marley, Gomez, \& Podolak}]{Marley1995a}
Marley, M.~S., Gomez, P., \& Podolak, M. 1995, J. Geophys. Res., 100, 23,349,
  \dodoi{10.1029/95JE02362}

\bibitem[{Movshovitz {et~al.}(2020)Movshovitz, Fortney, Mankovich, Thorngren,
  \& Helled}]{Movshovitz2020}
Movshovitz, N., Fortney, J.~J., Mankovich, C.~R., Thorngren, D., \& Helled, R.
  2020, Astrophys. J., 891, 109, \dodoi{10.3847/1538-4357/ab71ff}

\bibitem[{Nelson {et~al.}(2018)Nelson, Ford, Buchner, Cloutier, D{\'{i}}az,
  Faria, Rajpaul, \& Rukdee}]{Nelson2018}
Nelson, B.~E., Ford, E.~B., Buchner, J., {et~al.} 2018, {Quantifying the
  Evidence for a Planet in Radial Velocity Data}.
\newblock \doarXiv{1806.04683}

\bibitem[{Nettelmann(2017)}]{Nettelmann2017a}
Nettelmann, N. 2017, A{\&}A, 139, \dodoi{10.1051/0004-6361/201731550}

\bibitem[{Nettelmann {et~al.}(2013)Nettelmann, Helled, Fortney, \&
  Redmer}]{Nettelmann2013b}
Nettelmann, N., Helled, R., Fortney, J.~J., \& Redmer, R. 2013, Planet. Space
  Sci., 77, 143, \dodoi{10.1016/j.pss.2012.06.019}

\bibitem[{Nettelmann {et~al.}(2021)Nettelmann, Movshovitz, Ni, Fortney,
  Galanti, Kaspi, Helled, Mankovich, \& Bolton}]{Nettelmann2021a}
Nettelmann, N., Movshovitz, N., Ni, D., {et~al.} 2021, Planet. Sci. J., 2, 241,
  \dodoi{10.3847/PSJ/ac390a}

\bibitem[{Neuenschwander {et~al.}(2021)Neuenschwander, Helled, Movshovitz, \&
  Fortney}]{Neuenschwander2021}
Neuenschwander, B.~A., Helled, R., Movshovitz, N., \& Fortney, J.~J. 2021,
  Astrophys. J., 910, 38, \dodoi{10.3847/1538-4357/abdfd4}

\bibitem[{Ormel {et~al.}(2021)Ormel, Vazan, \& Brouwers}]{Ormel2021}
Ormel, C.~W., Vazan, A., \& Brouwers, M.~G. 2021, Astron. Astrophys., 647,
  \dodoi{10.1051/0004-6361/202039706}

\bibitem[{Podolak \& Helled(2012)}]{Podolak2012}
Podolak, M., \& Helled, R. 2012, Astrophys. J., 759, L32,
  \dodoi{10.1088/2041-8205/759/2/L32}

\bibitem[{Podolak {et~al.}(2000)Podolak, Podolak, \& Marley}]{Podolak2000}
Podolak, M., Podolak, J., \& Marley, M. 2000, Planet. Space Sci., 48, 143,
  \dodoi{10.1016/S0032-0633(99)00088-4}

\bibitem[{Read {et~al.}(2009)Read, Dowling, \& Schubert}]{Read2009}
Read, P.~L., Dowling, T.~E., \& Schubert, G. 2009, Nature, 460, 608,
  \dodoi{10.1038/nature08194}

\bibitem[{Rymer {et~al.}(2021)Rymer, Runyon, Clyde, N{\'{u}}{\~{n}}ez,
  Nikoukar, Soderlund, Sayanagi, Hofstadter, Quick, Stern, Becker, Hedman,
  Cohen, Crary, Fortney, Vertesi, Hansen, de~Pater, Paty, Spilker, Stallard,
  Hospodarsky, Smith, Wakeford, Moran, Annex, Schenk, Ozimek, Arrieta, McNutt,
  Masters, Simon, Ensor, Apland, Bruzzi, Patthoff, Scott, Campo, Krupiarz,
  Cochrane, Gantz, Rodriguez, Gallagher, Hurley, Crowley, Abel, Provornikova,
  Turtle, Clark, Wilkes, Hunt, Roberts, Rehm, Murray, Wolfarth, Fletcher,
  Spilker, Martin, Parisi, Norkus, Izenberg, Stough, Vervack, Mandt, Stevenson,
  Kijewski, Cheng, Feldman, Allen, Prabhu, Dutta, Young, \&
  Williams}]{Rymer2021}
Rymer, A.~M., Runyon, K.~D., Clyde, B., {et~al.} 2021, Planet. Sci. J., 2, 184,
  \dodoi{10.3847/PSJ/abf654}

\bibitem[{Saumon {et~al.}(1995)Saumon, Chabrier, \& van Horn}]{Saumon1995}
Saumon, D., Chabrier, G., \& van Horn, H.~M. 1995, Astrophys. J. Suppl. v.99,
  99, 713, \dodoi{10.1086/192204}

\bibitem[{Scheibe {et~al.}(2021)Scheibe, Nettelmann, \& Redmer}]{Scheibe2021}
Scheibe, L., Nettelmann, N., \& Redmer, R. 2021, Astron. Astrophys., 650, A200,
  \dodoi{10.1051/0004-6361/202140663}

\bibitem[{Seidelmann {et~al.}(2007)Seidelmann, Archinal, A'Hearn, Conrad,
  Consolmagno, Hestroffer, Hilton, Krasinsky, Neumann, Oberst, Stooke, Tedesco,
  Tholen, Thomas, \& Williams}]{Seidelmann2007}
Seidelmann, P.~K., Archinal, B.~A., A'Hearn, M.~F., {et~al.} 2007, Celest.
  Mech. Dyn. Astron., 98, 155, \dodoi{10.1007/s10569-007-9072-y}

\bibitem[{Stevenson(2020)}]{Stevenson2020}
Stevenson, D.~J. 2020, Annu. Rev. Earth Planet. Sci., 48, 465,
  \dodoi{10.1146/annurev-earth-081619-052855}

\bibitem[{Stevenson \& Salpeter(1977)}]{Stevenson1977}
Stevenson, D.~J., \& Salpeter, E.~E. 1977, Astrophys. J. Suppl. Ser., 35, 221

\bibitem[{Stixrude {et~al.}(2021)Stixrude, Baroni, \& Grasselli}]{Stixrude2021}
Stixrude, L., Baroni, S., \& Grasselli, F. 2021, Planet. Sci. J., 2, 222,
  \dodoi{10.3847/PSJ/ac2a47}

\bibitem[{Teanby {et~al.}(2020)Teanby, Irwin, Moses, \& Helled}]{Teanby2020}
Teanby, N.~A., Irwin, P. G.~J., Moses, J.~I., \& Helled, R. 2020, Philos.
  Trans. R. Soc. A Math. Phys. Eng. Sci., 378, \dodoi{10.1098/rsta.2019.0489}

\bibitem[{Thompson(1990)}]{Thompson1990}
Thompson, S.~L. 1990, {ANEOS Analytic Equations of State for Shock Physics
  Codes Input Manual}, Tech. rep., Sandia National Laboratories, Albuquerque,
  New Mexico

\bibitem[{Tiesinga {et~al.}(2021)Tiesinga, Mohr, Newell, \&
  Taylor}]{CODATA2018}
Tiesinga, E., Mohr, P.~J., Newell, D.~B., \& Taylor, B.~N. 2021, Rev. Mod.
  Phys., 93, 7621, \dodoi{10.1103/RevModPhys.93.025010}

\bibitem[{Vazan \& Helled(2020)}]{Vazan2020}
Vazan, A., \& Helled, R. 2020, Astron. Astrophys., 633, 1,
  \dodoi{10.1051/0004-6361/201936588}

\bibitem[{Wisdom \& Hubbard(2016)}]{Wisdom2016}
Wisdom, J., \& Hubbard, W.~B. 2016, Icarus, 267, 315,
  \dodoi{10.1016/j.icarus.2015.12.030}

\end{thebibliography}
\bibliographystyle{aasjournal}
\end{document}